\documentclass[aps,pra,10pt,twocolumn,showpacs,amsmath,amssymb,floatfix,superscriptaddress]{revtex4-1}
\usepackage{bm}
\usepackage{color}
\usepackage{ulem}
\newcommand{\remy}[1]{\textcolor{black}{#1}}
\newcommand{\tomas}[1]{\textcolor{black}{#1}}

\usepackage{hyperref}
\hypersetup{colorlinks = true,	linkcolor=blue, urlcolor=blue,citecolor=blue,anchorcolor=blue}

\usepackage[sort&compress]{natbib}

\usepackage{graphicx}
\usepackage{braket}

\usepackage{multirow}

\usepackage[per-mode=symbol,separate-uncertainty]{siunitx}
\begin{document}

\title{Fast high fidelity quantum non-demolition qubit readout via a non-perturbative cross-Kerr coupling}

\author{R. Dassonneville}
\email{remy.dassonneville@ens-lyon.fr}
\affiliation{Univ. Grenoble-Alpes, CNRS, Grenoble INP, Institut N\'eel, 38000 Grenoble, France}
\author{T. Ramos}
\email{t.ramos.delrio@gmail.com}
\affiliation{Instituto  de F\'isica Fundamental, IFF-CSIC, Calle Serrano 113b, 28006 Madrid, Spain}
\affiliation{DAiTA Lab, Facultad de Estudios Interdisciplinarios, Universidad Mayor, Santiago, Chile}
\author{V. Milchakov}
\affiliation{Univ. Grenoble-Alpes, CNRS, Grenoble INP, Institut N\'eel, 38000 Grenoble, France}
\author{L. Planat}
\affiliation{Univ. Grenoble-Alpes, CNRS, Grenoble INP, Institut N\'eel, 38000 Grenoble, France}
\author{\'E. Dumur}
\affiliation{Univ. Grenoble-Alpes, CNRS, Grenoble INP, Institut N\'eel, 38000 Grenoble, France}
\author{F. Foroughi}
\affiliation{Univ. Grenoble-Alpes, CNRS, Grenoble INP, Institut N\'eel, 38000 Grenoble, France}
\author{J. Puertas}
\affiliation{Univ. Grenoble-Alpes, CNRS, Grenoble INP, Institut N\'eel, 38000 Grenoble, France}
\author{S. Leger}
\affiliation{Univ. Grenoble-Alpes, CNRS, Grenoble INP, Institut N\'eel, 38000 Grenoble, France}
\author{K. Bharadwaj}
\affiliation{Univ. Grenoble-Alpes, CNRS, Grenoble INP, Institut N\'eel, 38000 Grenoble, France}
\author{J. Delaforce}
\affiliation{Univ. Grenoble-Alpes, CNRS, Grenoble INP, Institut N\'eel, 38000 Grenoble, France}
\author{C. Naud}
\affiliation{Univ. Grenoble-Alpes, CNRS, Grenoble INP, Institut N\'eel, 38000 Grenoble, France}
\author{W. Hasch-Guichard}
\affiliation{Univ. Grenoble-Alpes, CNRS, Grenoble INP, Institut N\'eel, 38000 Grenoble, France}
\author{J. J. Garc{\'i}a-Ripoll}
\affiliation{Instituto  de F\'isica Fundamental, IFF-CSIC, Calle Serrano 113b, 28006 Madrid, Spain}
\author{N. Roch}
\affiliation{Univ. Grenoble-Alpes, CNRS, Grenoble INP, Institut N\'eel, 38000 Grenoble, France}
\author{O. Buisson}
\email{olivier.buisson@neel.cnrs.fr}
\affiliation{Univ. Grenoble-Alpes, CNRS, Grenoble INP, Institut N\'eel, 38000 Grenoble, France}

\date{\today}

\begin{abstract}
Qubit readout is an indispensable element of any quantum information processor.
In this work, we experimentally demonstrate a non-perturbative cross-Kerr coupling between a transmon and a polariton mode which enables an improved quantum non-demolition (QND) readout for superconducting qubits. The new mechanism uses the same experimental techniques as the standard QND qubit readout in the dispersive approximation, but due to its non-perturbative nature, it maximizes the speed, the single-shot fidelity and the QND properties of the readout. In addition, it minimizes the effect of unwanted decay channels such as the Purcell effect. We observed a single-shot readout fidelity of \SI{97.4}{\percent} for short \SI{50}{\nano\second} pulses, and we quantified a QND-ness of \SI{99}{\percent} for long measurement pulses with repeated single-shot readouts.
\end{abstract}

\maketitle

\section{Introduction}

In Noisy Intermediate Scale Quantum (NISQ) devices\ \cite{Preskill_2018}, measurements are usually the last step of the algorithm. Here, a high-fidelity readout is an interesting asset that reduces the overhead in error mitigation\ \cite{Li_PRX_2017} and in the characterization of gate fidelities\ \cite{Knill_PRA_2008}. However, high-fidelity quantum non-demolition (QND) single-shot measurements become a requirement once we consider scaling up quantum technologies\ \cite{DiVincenzo_FdP_2000} to large devices, using quantum error correction\ \cite{Kelly_Nature_2015,Schindler_Science_2011} and fault-tolerant quantum computation\ \cite{Bermudez_PRX_2017,Gambetta_njpQI_2017}. In this context, lowering the readout and QND-errors is as important as decreasing the single- and two-qubit gate errors below the scaling thresholds.

A fast and high-fidelity QND measurement demands a strong coupling to the measurement device combined with a good preservation of the qubit state. In trapped ion qubits, this dilemma is solved by encoding information in two long-lived states, only one of which couples to incoming radiation\ \cite{RMP_Leibfried_2003}. Fluorescence counting gives a projective measurement with errors below \SI{1}{\percent}, limited by the collection time\ \cite{Ballance_PRL_2016}.  
Cavity-QED\ \cite{Blais_PRA_2004,Nature_Volz_2011,Haroche&Raimond_2006} setups follow a different strategy. Inserting the qubit inside a cavity allows to generate a strong coupling between the qubit and the cavity electro-magnetic (EM) field but also to increase the collection efficiency.
An optical or microwave signal probes the resonator, implementing an indirect projective QND readout of the qubit polarization $\hat{\sigma}_z$\ \cite{PRApplied_Walter_2017,Nature_Volz_2011}. In these cavity-QED experiments it is very important to engineer the qubit-resonator coupling so as to maximize measurement's (i) single-shot readout fidelity, (ii) speed and (iii) QND-ness---preservation of the qubit's excited and ground state probabilities.

\begin{table*}[t]
\center
\begin{tabular}{|c|c|c|c|c|c|c|c|}
\hline\hline
Type  & Elementary & Effective QND &  QND &  Single-shot & Detection &  State-of-the-art \\
     & readout coupling &  readout coupling  &  fidelity    & readout fidelity         &  time     &  references        \\
\hline\hline
Transverse &$ \sim g_x (\hat{q}+\hat{q}^\dagger)(\hat{c}+\hat{c}^\dagger)$ &  $\sim \frac{\tomas{\alpha_{q}}{g_x}^2 }{\Delta(\Delta\tomas{+}\alpha_{q})} \hat{\sigma}_{z} \hat{c}^\dagger \hat{c}$ & Not given & \SI{99.1}{\percent}--\SI{99.6}{\percent} & \SI{48}{\nano\second}--\SI{88}{\nano\second} & \cite{PRApplied_Walter_2017} \\
\hline
Longitudinal &$ \sim g_z(t) \hat{q}^\dagger \hat{q}(\hat{c}^\dagger + \hat{c})$ & $ \sim g_{z}(t) \hat{\sigma}_{z} (\hat{c}^\dagger + \hat{c}) $ &\SI{98.4}{\percent} & \SI{98.9}{\percent} & \SI{750}{\nano\second} & \cite{arXiv_Touzard_2018} \\
\hline
Cross-Kerr & $ \sim g_{zz}(\hat{q}+\hat{q}^\dagger)^2(\hat{c}+\hat{c}^\dagger)^2$ & $ \sim g_{zz}\hat{\sigma}_{z} \hat{c}^\dag \hat{c}$ & \SI{99}{\percent} $\pm$ \SI{0.6}{\percent} &  \SI{97.4}{\percent} $\pm$ \SI{0.7}{\percent} & \SI{30}{\nano\second}--\SI{50}{\nano\second} & Present work\\
\hline\hline
\end{tabular}
\caption{State-of-the-art parameters for three different coupling types between an harmonic readout mode and a superconducting qubit. The second column shows the direct coupling terms between the qubit, described as an anharmonic oscillator with ladder operators ($\hat{q}$, $\hat{q}^\dagger$), and an harmonic read-out mode described by ($\hat{c}$, $\hat{c}^\dagger$). Column three shows the effective coupling obtained after the rotating wave approximation (RWA), and two-level system approximation for all couplings plus the dispersive approximation in the case of the transverse coupling. Notice that the present experimental work implements two non-perturbative cross-Kerr couplings of the type presented in this table since two polariton modes $\hat{c}_u$ and $\hat{c}_l$ are used for the readout [See Fig.~\ref{Schematic_qubit_readout} and Sec.~\ref{theoryModel} for more details].  
\label{tab_couplings}}
\end{table*}

To illustrate this point, we consider the ubiquitous \textit{transmon} qubit\ \cite{PRA_Koch_2007}, $\hat{H}_{q}\simeq \hbar\omega_{q} \hat{q}^\dagger\hat{q} \tomas{+} \hbar \frac{\alpha_\text{q}}{2} \hat{q}^{\dagger\,2}\hat{q}^2\simeq \frac{1}{2}\hbar {\omega}_{q}\hat{\sigma}_z$, a slightly anharmonic oscillator with frequency $\omega_{q}$, anharmonicity \tomas{$\alpha_\text{q}$}, and \tomas{ladder operator $\hat{q}$}. Three types of couplings, summarized in Table\ \ref{tab_couplings}, will be discussed.
Qubits and resonators are usually coupled via the interaction between the electric field of the qubit dipole $\hat{q}+\tomas{\hat{q}^\dag}$ and the electric field of the resonator $\hat{c}\tomas{+\hat{c}^\dag}$. This \textit{field}-\textit{field} interaction is known as the \textit{transverse coupling} and results in a term \tomas{$\sim g_x (\hat{q}+\hat{q}^\dag)(\hat{c}+\hat{c}^\dag)$} in the Hamiltonian\ \cite{Haroche&Raimond_2006,Blais_PRA_2004}. In the dispersive limit\ \cite{PRA_Koch_2007}, the qubit-cavity detuning $\Delta=\omega_{q}-\omega_{c}$ largely exceeds the coupling strength, $|g_{x}|\ll |\Delta|,\tomas{\omega_q,\omega_c}$, so that the cavity experiences an effective \textit{energy}-\textit{energy} interaction \tomas{$\sim \chi_{\rm d}\hat{\sigma}_z\hat{c}^\dagger\hat{c}$ with $\chi_{\rm d} =  \frac{\alpha_{q}{g_x}^2 }{\Delta(\Delta+\alpha_{q})}$ \cite{PRApplied_Walter_2017} } known as the dispersive or cross-Kerr interaction. It gives rise to a qubit-dependent frequency shift, mapping the state of the qubit to the signal phase probing the resonator and thus providing a good QND projective measurement\ \cite{PRApplied_Walter_2017,PRL_Jeffrey_2014}. This transverse coupling has been extensively used in most circuit-QED experiments. State-of-the-art measurement fidelities and speeds using this standard dispersive technique are summarized in the first row of Table~\ref{tab_couplings}. However, the dispersive readout is fundamentally limited by unavoidable higher order corrections to perturbation theory, which distort the qubit dynamics\ \cite{Slichter:2012bi,PRL_Sank_2016,arXiv_Lescanne_2018}, and induce additional decay channels\ \cite{Houck2008}.

Several works have investigated how to overcome these limitations, designing new quantum circuits\ \cite{PRL_Lecocq_2011, PRA_Diniz_2013,NPJ_Kerman_2013,Dumur2015,PRB_Billangeon_2015,PRB_Richer_2016,PRA_Didier_2015,arXiv_Gard_2018}. 
Implementing a coupling scheme that involves natively the energy of the qubit -- as opposed to an effective energy interaction -- resolves these limitations. Along this line, the \textit{longitudinal coupling} $ \sim g_z \hat{q}^\dag \hat{q} (\hat{c} + \hat{c}^\dag)$ is remarkable [cf. second row of Table~\ref{tab_couplings}]. It induces a qubit-dependent displacement of the cavity field $\hat{c}\remy{+\hat{c}^\dag}$ \cite{PRA_Didier_2015}. When combined with a parametric modulation $g_z(t)$ at the cavity frequency $\omega_c$, this interaction results in a faster separation of the pointer states with a QND-ness as high as $\mathcal{Q}=$ \SI{98.4}{\percent}\ \cite{arXiv_Touzard_2018,Ikonen:2018vq}.

\begin{figure}[!b]
\includegraphics[width=0.8\linewidth]{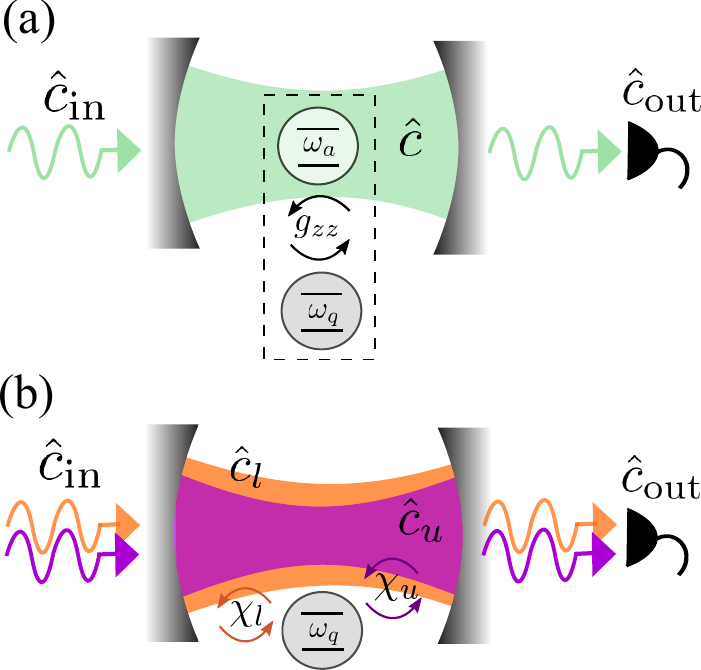}
\caption{Schematics of the circuit QED setup with the transmon molecule used for a high fidelity and fast qubit QND readout. (a) A cavity mode $\hat{c}$ is strongly and transversely coupled to an ancilla system $\hat{a}$, which in turn couples diagonally to the qubit $\hat{\sigma}_z$ as $ \sim g_{zz}\hat{\sigma}_z \hat{a}^\dag \hat{a}$. (b) The strong hybridization between cavity and ancilla is manifested by two orthogonal polariton modes $\hat{c}_u$ and $\hat{c}_l$, which couple to the qubit with a non-perturbative cross-Kerr couplings $ \sim \hat{\sigma}_z(\chi_u \hat{c}_u^\dag \hat{c}_u+ \chi_l \hat{c}_l^\dag \hat{c}_l)$ (see text). This allows us to infer the state $|g\rangle$ or $|e\rangle$ of the qubit by measuring the resonance shifts of the polaritons at the cavity transmission output.}
\label{Schematic_qubit_readout}
\end{figure}

In this work we experimentally demonstrate a new qubit-cavity coupling scheme based on a \textit{non-perturbative cross-Kerr} interaction $ \sim g_{zz}(\hat{q}+\hat{q}^\dagger)^2(\hat{c}+\hat{c}^\dagger)^2$ [cf. third row of Table~\ref{tab_couplings}]. It leads to an alternative readout mechanism for superconducting qubits. This new process is fast, has a large single-shot fidelity, maximizes the QND nature of the process, and does not require any parametric modulation. Similar non-perturbative cross-Kerr couplings have been recently proposed for the readout of a flux qubit \cite{Wang2018} and of a spin qubit \cite{Ruskov2019}. However, our experimental setup builds on ideas previously proposed in Ref.\cite{PRA_Diniz_2013} and it is realized with an artificial transmon molecule with one emergent qubit-like transmon degree of freedom and a bosonic ancilla that couples to the readout cavity [cf. Fig.\ \ref{Schematic_qubit_readout}a]. The qubit develops a Kerr-type interaction with the ancilla-cavity polariton branches [cf. Fig.\ \ref{Schematic_qubit_readout}b]. This interaction enables a detection scheme analogous to the standard dispersive measurement. Nevertheless, since our coupling is not perturbative, it does not imply any cavity-mediated excitations or decay. Moreover, the strength of the readout shift $2g_{zz}$ can be made as large as a few hundreds \SI{}{\mega\hertz}, and is independent of the qubit-cavity detuning. Thus the effect of any stray transverse coupling can be made arbitrarily small by increasing the detuning between the qubit and the cavity. This results in a very efficient single-shot QND readout of the qubit even in its first demonstration: it has a record QND-ness of \SI{99}{\percent}, a fidelity of \SI{97.4}{\percent}, and it only  requires a short measurement time of \SI{50}{\nano\second}. This readout mechanism can be combined with other paradigms of direct qubit-qubit interactions\ \cite{Barends_Nature_2008}, as an upgrade to existing quantum computing and simulation architectures.

\section{Transmon molecule inside a cavity}

In this section we give details on the physical mechanisms for the qubit readout using a non-perturbative cross-Kerr coupling. The setup demonstrating this new readout mechanism uses a transmon molecule circuit, \tomas{composed of} two transmons coupled by a parallel LC-circuit [cf.~Fig.\ \ref{sample}c], and this is inserted inside a cavity. We start by introducing the specific experimental system in Sec.~\ref{experimentalSetup}, and then, in Sec.~\ref{theoryModel}, we write down the theoretical model describing the open quantum dynamics of the system. We consider the strong coupling regime between cavity and ancilla, getting two strongly hybridized polariton modes. \tomas{A} single effective qubit then couples strongly to these two polaritons via non-perturbative cross-Kerr couplings $\chi_j$. This allows for an efficient readout of the qubit state via the transmission output of the cavity as shown below.

\subsection{Physical implementation}\label{experimentalSetup}
\begin{figure}
\includegraphics[width=8.5cm]{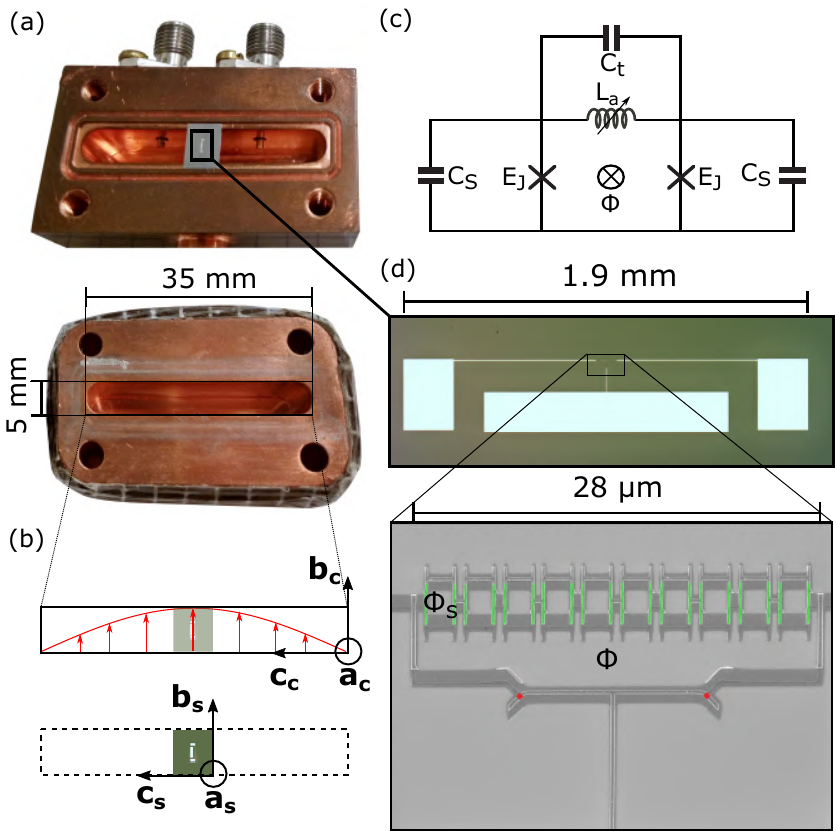}
\caption{Quantum circuit with non-perturbative cross-Kerr coupling. (a) Picture of the two parts of the Copper-OFHC 3D cavity with the input-output pin connectors. The sample is placed at the center of the cavity. (b) The electric field distribution of the first EM mode of the cavity in the center plane is sketched in red. The cavity directions ($\mathbf{a_c}$, $\mathbf{b_c}$, $\mathbf{c_c}$) and sample directions ($\mathbf{a_s}$, $\mathbf{b_s}$, $\mathbf{c_s}$) are represented. (c) Lumped element of the transmon molecule circuit. (d) Optical microscope and SEM pictures of the transmon molecule sample. The Josephson junctions are highlighted in red. The SQUID Josephson junctions implementing the coupling inductance $L_a$ are highlighted in green.}
\label{sample}
\end{figure}

The device consists of an aluminium Josephson circuit, which is deposited on an intrinsic silicon wafer and inserted in a 3D copper cavity [cf.~Fig.~\ref{sample}a]. An optical image of the molecule circuit is shown Fig.~\ref{sample}d, which implements the lumped element circuit of Fig.~\ref{sample}c. 
The molecule is realized by two identical transmon qubits with \tomas{Josephson energy $E_J$} and capacitance $C_S$, \tomas{coupled} through a parallel
LC-circuit with inductance $L_a$ and capacitance $C_t$. Here, $C_S$
represents the capacitance between an either small rectangular electrode
and the central longer one, while $C_t$ represents the capacitance
between the two small rectangular electrodes. The coupling inductor
$L_a$ \tomas{is implemented} by a chain of 10 small SQUID \tomas{loops of area} $S_{\rm SQUID}$, which are tunable by an external flux $\Phi_{\remy{S}}$ \tomas{[cf. Fig.~\ref{sample}d]}. The circuit also contains a large loop of enclosed area \remy{$A$} that is approximately $r=\remy{A}/S_{\rm SQUID} \simeq 26$ times larger than the SQUIDs. \tomas{Consequently, the flux} $\Phi = r \Phi_S$
generates a circulating current passing through both $L_a$ and the two small Josephson junctions of the transmons. As already discussed in previous work~\cite{PRL_Lecocq_2011} \remy{and also detailed in Appendix \ref{model_allFlux}}, when the applied flux satisfies $\Phi =n\Phi_0$ (with $n$ an integer \tomas{and $\Phi_0$ the magnetic flux quantum}), the dynamics of \tomas{the system effectively behaves as a single transmon qubit with cross-Kerr coupling to a slightly anharmonic ancilla} mode, described by \tomas{the Hamiltonian}
\begin{align} 
\hat{H}_{\rm mol}={}& \tomas{4}E_{C_q} \hat{n}_q^2 - 2E_J \cos(\hat{\varphi}_q)  \notag  \\
{}& + \tomas{4}E_{C_a} \hat{n}_a^2 - 2E_J \left( \cos(\hat{\varphi}_a) - \frac{L_J}{L_a (n)} \hat{\varphi}_a^2\right) \notag \\ 
{}& - \frac{E_J}{2} \hat{\varphi}_q^2 \hat{\varphi}_a^2 + {\cal O}\remy{^6}.
\label{Hamiltonian_Circuit} 
\end{align}
\tomas{Here, the phase average $\hat{\varphi}_q$ and phase difference $\hat{\varphi}_a$ between the two Josephson junctions describe the effective transmon qubit and the ancilla mode, respectively. Their conjugate charge number operators are denoted by $\hat{n}_{q}$ and $\hat{n}_{a}$. The charging energies of qubit and ancilla are given by $E_{C_{q}}=e^2/(2C_q)$ and $E_{C_{a}}=e^2/(2C_a)$, with effective capacitances $C_q=2C_S$ and $C_a=2(C_S+2C_t)$, respectively. We considered the system in the transmon regime, $E_J\gg E_{C_q}, E_{C_a}$, so that $\hat{\varphi}_q,\hat{\varphi}_a\ll 1$ and therefore expanded the coupling term between $\hat{\varphi}_q$ and $\hat{\varphi}_a$ up to fourth order in the phases}. In addition, $L_J= (\frac{\Phi_0}{2\pi})^2 \frac{1}{E_{J}}$ describes the Josephson inductance of each junction and $L_a (n)$ denotes the value of the coupling inductance for given magnetic flux $\Phi_{\tomas{S}} = \frac{n}{r} \Phi_0$. Importantly, the last term in Eq.~(\ref{Hamiltonian_Circuit}) \tomas{originates} the nonlinear \tomas{cross-Kerr} coupling between transmon qubit and ancilla as shown in the next subsection.

To measure the transmon molecule, we insert the silicon chip inside a 3D copper cavity with a volume \SI{24.5x5x35}{\milli\metre} (length $\times$ height $\times$ width) along the $\mathbf{a_c}$, $\mathbf{b_c}$ and $\mathbf{c_c}$ directions, respectively \remy{[cf.~Fig.~\ref{sample}b]}. The cavity mode considered hereafter is the fundamental TE$_{101}$ mode with the microwave electric field aligned along the $\mathbf{b_c}$ direction. \remy{All the circuit parameters of our setup are measured via spectroscopy and are summarized in Table \ref{Tab_circuit} of Appendix~\ref{app:circuit_parameters}}.

\subsection{Qubit-polaritons model}\label{theoryModel}

\tomas{To analyze the dynamics of the transmon molecule from a quantum optics point of view, it is convenient to express Eq.~(\ref{Hamiltonian_Circuit}) in the number representation and treat the qubit and ancilla modes as coupled anharmonic oscillators described by the Hamiltonian [cf.~Appendix \ref{HamiltonianNumberRep}]:}
\tomas{\begin{align}
\frac{\hat{H}_{\rm mol}}{\hbar} ={}& \omega_q \hat{q}^\dag \hat{q} \tomas{+} \frac{\alpha_q}{2} \hat{q}^\dag \hat{q}^\dag \hat{q} \hat{q}  +\omega_a \hat{a}^\dag \hat{a} + \frac{U_a}{2} \hat{a}^\dag \hat{a}^\dag \hat{a} \hat{a}\nonumber\\
{}&- \frac{g_{zz}}{2} \tomas{(\hat{a}+\hat{a}^\dag)^2 (\hat{q}+\hat{q}^\dag)^2}+{\cal O}^6.\label{Ham2}
\end{align}}
The first two terms in Eq.~(\ref{Ham2}) correspond to the Hamiltonian of a transmon \tomas{with frequency $\omega_q=4\sqrt{E_{C_q}E_J}/\hbar+\alpha_q$, anharmonicity $\alpha_q=-E_{C_q}/\hbar$, and ladder operators $\hat{q}$, $\hat{q}^\dag$. Importantly, the transmon anharmonicity $\alpha_q$ is designed to be larger than any driving in the system so that only its two lowest levels will be populated. We can thus approximate the trasmon as a two-level system or ``qubit''} with Hamiltonian $\hat{H}_q=\hbar\omega_q \hat{q}^\dag \hat{q} + \frac{\hbar\alpha_q}{2} \hat{q}^\dag \hat{q}^\dag \hat{q} \hat{q}\simeq \frac{\omega_q}{2}\hat{\sigma}_z$, where $\hat{\sigma}_z=2\hat{q}^\dag  \hat{q}-1$ corresponds to the diagonal Pauli operator.

The third and fourth terms in Eq.~(\ref{Ham2}) describe the ancilla mode with frequency \tomas{$\omega_a=4\sqrt{E_J E_{C_a} (1+ 2\frac{L_J}{L_a(n)})}/\hbar+U_a$}, anharmonicity \tomas{$U_a= -(E_{C_a}/\hbar)(1+2\frac{L_J}{L_a(n)})^{-1}$}, \tomas{and ladder operators $\hat{a}$, $\hat{a}^\dag$}. Both ancilla frequency and anharmonicity are a function of \tomas{the externally applied integer flux $\Phi=n\Phi_0$ and} we design the inductance $L_a(n)$ and capacitance $C_t$ so that the ancilla anharmonicity \tomas{$|U_a|$} is much weaker than the qubit one \tomas{$|\alpha_q|$}. In our experiments, the ancilla will be weakly populated ($\langle \hat{a}^\dag \hat{a}\rangle \lesssim 2$), allowing us to safely neglect the anharmonicity $U_a$, and regard it as a simple harmonic oscillator $\hat{H}_a = \tomas{\hbar\omega_a \hat{a}^\dag \hat{a} + \frac{\hbar U_a}{2} \hat{a}^\dag \hat{a}^\dag \hat{a} \hat{a}}\simeq \hbar \omega_a \hat{a}^\dag \hat{a}$. Interesting nonlinear and bi-stability effects arise when the ancilla is strongly populated ($\langle \hat{a}^\dag \hat{a}\rangle \gg 1$), but these effects will be discussed elsewhere. 

The last term in Eq.~(\ref{Ham2}) describes an \tomas{\textit{energy-energy}} cross-Kerr coupling between qubit and ancilla \tomas{with a strength $g_{zz} = \sqrt{\alpha_q U_a}$}. This is not only a direct consequence of the Josephson junctions non-linearity but also of the circuit symmetry \cite{PRL_Lecocq_2012}, which avoids any transverse \textit{field-field} and longitudinal \textit{field-energy} coupling to appear at a lower order than $\hat{q}^4,\hat{a}^4$ [cf.~Appendix~\ref{Asymmetry} for imperfections in the symmetry]. \tomas{Since $g_{zz}$ is much weaker than $\omega_q$, $\omega_a$ and the detuning between them $|\omega_q-\omega_a|$, we can neglect fast oscillating terms in the cross-Kerr coupling and obtain $\hat{H}_{qa}=\frac{g_{zz}}{2}(\hat{a}+\hat{a}^\dag)^2(\hat{q}+\hat{q}^\dag)^2\simeq-g_{zz}\hat{\sigma}_z\hat{a}^\dag\hat{a}-\frac{g_{zz}}{2}\hat{\sigma}_z-2g_{zz}\hat{a}^\dag\hat{a}$. In addition to the energy-energy qubit-ancilla interaction $\sim -g_{zz}\hat{\sigma}_z\hat{a}^\dag\hat{a}$, the cross-Kerr coupling also produces a renormalization of the qubit and ancilla frequencies.}

\tomas{Our final aim is to engineer a} cross-Kerr coupling between the qubit and some polariton modes \tomas{which will allow the QND readout of the qubit's state.} To obtain such effect, we strongly couple the ancilla \tomas{to a microwave cavity mode via a standard transverse interaction. The Hamiltonian reads} $\hat{H}_{\rm cav}=\hbar\omega_c \hat{c}^\dag \hat{c} + \hbar g_{ac} (\hat{a}^\dag \hat{c} + \hat{c}^\dag \hat{a})$, with $\omega_c$ the cavity frequency, \tomas{$g_{ac}\ll \omega_a,\omega_c$ the strength of the ancilla-cavity coupling and $\hat{c}$, $\hat{c}^\dag$ the cavity ladder operators.} A \tomas{precise} alignment between the sample direction $\mathbf{b_s}$ the cavity direction $\mathbf{b_c}$ is \tomas{crucial} to maximize the ancilla-cavity coupling $g_{ac}$, \tomas{while minimizing and neglecting any residual qubit-cavity coupling $g_{qc}$}. This is also guaranteed by the symmetry of the transmon molecule and of the TE$_{101}$ mode of the cavity. \tomas{Imperfections due to misalignment and a small asymmetry in the Josephson junctions are treated in Appendix \ref{Asymmetry}.} 

The total Hamiltonian of the system which includes the transmon molecule \tomas{$\hat{H}_{\rm mol}=\hat{H}_{q}+\hat{H}_{a}+\hat{H}_{qa}$} and the properly oriented cavity \tomas{$\hat{H}_{\rm cav}$} is then given by
\begin{align}
\frac{\hat{H}_{\rm tot}}{\hbar} ={}& \frac{\omega_q'}{2}\hat{\sigma}_z +\omega_a' \hat{a}^\dag \hat{a}+\omega_c \hat{c}^\dag \hat{c} \notag\\ &-g_{zz}\hat{\sigma}_z\hat{a}^\dag \hat{a}+g_{ac}(\hat{c}^\dag\hat{a}+\hat{a}^\dag\hat{c}),\label{FullHamqac}
\end{align}
\tomas{with $\omega_q'=\omega_q-g_{zz}$, and $\omega_a'=\omega_a-2g_{zz}$ the renormalized qubit and ancilla frequencies. All the parameters of our system are measured via spectroscopy in Appendix~\ref{Sec_calibration} and they are summarized in Appendix \ref{app:circuit_parameters}.}

\tomas{To strongly hybridize the cavity and ancilla modes we tune them close to resonance $|\omega_c-\omega_a'|\lesssim g_{ac}$.} This leads to two new normal modes called upper and lower polariton modes, $\hat{c}_u$ and $\hat{c}_l$, which are a linear combination of ancilla and cavity fields, $\hat{a}^\dag+\hat{a}$ and $\hat{c}^\dag+\hat{c}$. In the rotating-wave approximation (RWA), they are given by a rotation $\hat{c}_u =\cos(\theta)\hat{a}+\sin(\theta)\hat{c}$, and $\hat{c}_l =\cos(\theta)\hat{c}-\sin(\theta)\hat{a}$, where the cavity-ancilla hybridization angle reads $\tan(2\theta)= 2g_{ac}/ (\omega_a'-\omega_c)$.  In terms of these polariton modes, the total Hamiltonian takes the form [cf.~appendix \ref{Numerical_model}]
\begin{align} 
    \frac{\hat{H}_{\rm tot}}{\hbar} &= \frac{\omega_q\tomas{'}}{2}\hat{\sigma}_z + \sum_{j=u,l} \omega_j \hat{c}_j^\dag \hat{c}_j \tomas{+} \hat{\sigma}_z \sum_{j=u,l} \chi_{j}\hat{c}^\dagger_j \hat{c}_j, 
    \label{Hamiltonian_Polariton} 
\end{align}
where $\omega_u= \sin^2(\theta) \omega_c + \cos^2(\theta)\omega_a\tomas{'} + \sin(2 \theta) g_{ac} $ and $\omega_l = \cos^2(\theta)\omega_c + \sin^2(\theta)\omega_a\tomas{'}- \sin(2 \theta)g_{ac}$ are the frequencies of the upper and lower polariton modes, respectively. \tomas{Importantly,} each polariton mode is in some proportion cavity-like and therefore can be used for readout. Similarly, each polariton is also ancilla-like and thus \tomas{inherits} the \tomas{\textit{non-perturbative}} cross-Kerr coupling to the qubit. \tomas{The corresponding interaction strengths read $\chi_{u}=-g_{zz}\cos^2(\theta)$ and $\chi_{l}=-g_{zz}\sin^2(\theta)$, for the upper and lower polariton, respectively. In this way, we implement} the coupling between a qubit and a readout mode presented in the third row of Table~\ref{tab_couplings}. It is relevant to note that these cross-Kerr coupling strengths $\chi_u$ and $\chi_l$ are non-perturbative \tomas{in the sense that they} \remy{are not derived by a perturbative dispersive approximation of a transverse coupling. Thus they do} not depend on the qubit-resonator detuning but only on the hybridization angle $\theta$ and the \tomas{native} ancilla-qubit cross-Kerr coupling $g_{zz}$. 

\subsection{Conditional polariton spectroscopy}\label{conditional}

\begin{figure}
	\includegraphics[width=\columnwidth]{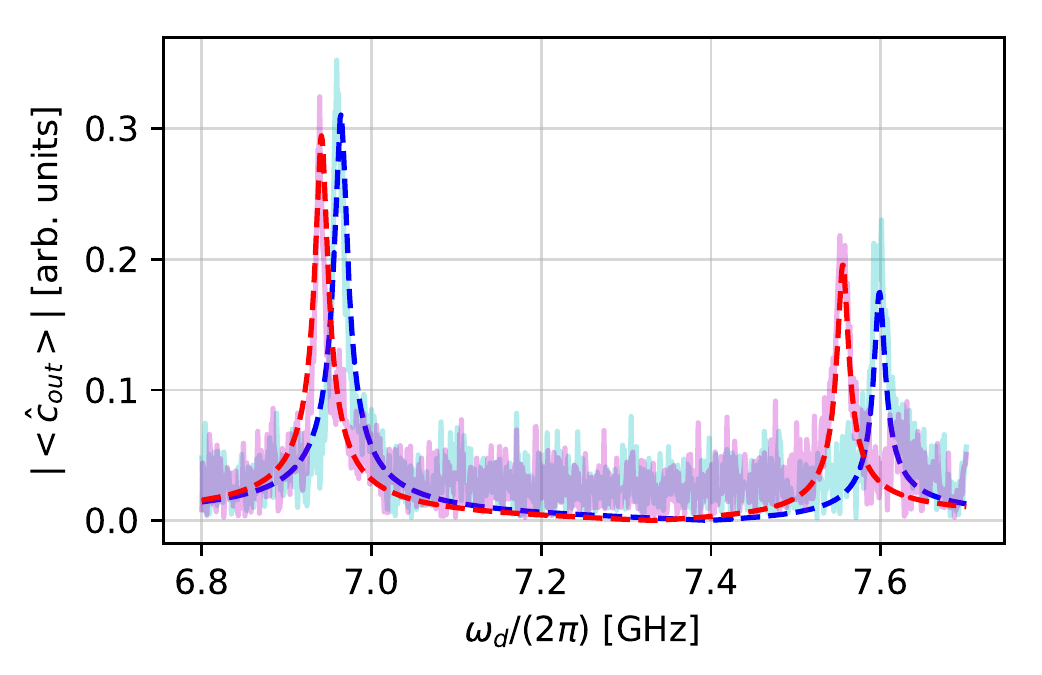}
	\caption{Polariton spectroscopy via the transmitted amplitude of the cavity as a function of the driving frequency $\omega_d$ at $\Phi=5\Phi_0$. The resonances at lower and higher frequency correspond to the lower and upper polariton modes, respectively. In addition, both polariton resonances are cross-Kerr shifted depending on the prepared qubit state (ground $|g\rangle$ in blue and excited $|e\rangle$ in red). \tomas{The highlighted red and blue lines correspond to the theoretical prediction in Eq.~(\ref{expectationC}) with the parameters given in Table~\ref{table5flux}, valid for $\Phi=5\Phi_0$.}}\label{Transmission_Measurements}
\end{figure}

Inspecting Eq.~(\ref{Hamiltonian_Polariton}) we see that, except for dissipation and dephasing effects treated in Appendix \ref{Numerical_model}, the population of the qubit remains constant during the dynamics, \tomas{$\langle \hat{\sigma}_z\rangle=\langle \hat{\sigma}_z({t_0})\rangle$}, with $t_0$ the initial time. The qubit's main effect is thus simply to shift the transition frequency of each polariton mode \tomas{and to renormalize the hybridization angle as}
\tomas{\begin{align}
&\omega_j\rightarrow \bar{\omega}_j(\langle \hat{\sigma}_z\rangle)=\omega_j+\chi_j\langle \hat{\sigma}_z\rangle,\label{shift}\\
&\theta\rightarrow\bar{\theta}(\langle \hat{\sigma}_z\rangle)=\frac{1}{2}{\rm Arctan}\left(\frac{2g_{ac}}{\omega_a'-g_{zz}\langle \hat{\sigma}_z\rangle-\omega_c}\right).\label{angle}
\end{align}}
The shift of the polariton resonances can be measured by shining a weak continuous coherent drive on the cavity and recording the amplitude of the field at the transmission output $\langle \hat{\tomas{c}}_{\rm out} \rangle_{\rm ss}$ [cf.~Fig.~\ref{Schematic_qubit_readout}]. Fig.~\ref{Transmission_Measurements} shows \tomas{a} typical spectroscopic measurement as a function of the driving frequency $\omega_d$, with the blue and red curves corresponding to the case the qubit is prepared in states $|g\rangle$ ($\langle \hat{\sigma}_z\rangle\approx -1$) and in $|e\rangle$ ($\langle \hat{\sigma}_z\rangle\approx +1$), respectively. \tomas{We clearly observe two peaks, for given qubit state, and these are well described by Lorentzian lineshapes as [cf.~Appendix~\ref{Numerical_model}]}
\begin{align}
\langle \hat{c}_{\rm out} \rangle_{\rm ss} ={}& \sin(\bar{\theta})\langle \hat{c}_u \rangle_{\rm ss}+\cos(\bar{\theta}) \langle \hat{c}_l \rangle_{\rm ss}\label{expectationC}\\
={}&\tomas{\frac{-i\Omega\sin^2(\bar{\theta})}{\kappa_u/2-i(\omega_d-\bar{\omega}_u)}+\frac{-i\Omega\cos^2(\bar{\theta})}{\kappa_l/2-i(\omega_d-\bar{\omega}_l)}}.\nonumber
\end{align}
\tomas{The resonances are centered at the upper and lower polariton frequencies $\omega_d\approx \bar{\omega}_l$ and $\omega_d\approx \bar{\omega}_u$, and their widths are given by the effective polariton decay rates $\kappa_u=\kappa_c\sin^2(\bar{\theta})+\kappa_a\cos^2(\bar{\theta})$ and $\kappa_l=\kappa_c\cos^2(\bar{\theta})+\kappa_a\sin^2(\bar{\theta})$, respectively, with $\kappa_c$ denoting the cavity decay and $\kappa_a$ the ancilla decay [cf.~Appendix~\ref{Numerical_model}]. In addition, the height of the peaks are proportional to the strength of the weak microwave drive $\Omega\ll \kappa_l,\kappa_u$.} In Fig.~\ref{Transmission_Measurements} the transmitted signal is measured using a \SI{500}{\nano\second} square microwave pulse applied immediately after preparing the qubit in $\ket{g}$ or $\ket{e}$ states. The lineshapes are fitted using Eq.~(\ref{expectationC}) and the qubit-dependent frequency shifts are clearly visible. The peaks of the lower and upper polariton branches are indeed shifted by $ \sim 2\chi_j$, up to small errors in the calibration and initial state preparation of the qubit states $\ket{g}$ and $\ket{e}$. \tomas{This effect is exploited to implement the QND qubit measurement as shown in the following.} 

\section{Single-shot quantum non-demolition measurements}
\subsection{Individual measurement records and quantum trajectories}\label{sec_trajectories}

Readout is performed using a standard microwave set-up including a high saturation-power Josephson parametric amplifier made from a SQUID chain \cite{arXiv_Planat_2018}. Next we consider the readout performance at zero flux measuring the signal transmitted through the lower polariton $j=l$. To readout the qubit state, a coherent microwave tone is applied at a frequency $\omega_{d}/2\pi=(\tomas{\bar{\omega}_l}+ 2\chi_l)/2\pi=$ 7.02\tomas{9}GHz.. The amplitude of the readout tone is $\overline{n}_{l}= \langle c_l^\dagger c_l \rangle \simeq 2$ based on a calibration using AC-Stark shift \cite{Schuster2005Stark,GambettaStarckShift}. Since the polariton resonance frequency is conditioned to the qubit state, the coherent tone is detuned by \tomas{($2\chi_l)/(2\pi) \approx -$\SI{9}{\mega\hertz},} \remy{or in resonance, when the qubit is in $\ket{g}$, or in $\ket{e}$, respectively}. Therefore the transmitted signal presents weak or large amplitude conditioned to the qubit state $\ket{g}$ and $\ket{e}$, respectively. The amplifier is operated in phase-sensitive mode leading to squeezed signal at the amplifier output. We define ${I(t)}$ and ${Q(t)}$ the in-phase and the quadrature microwave signal. Its phase has been adjusted so the information about the qubit state is only contained in ${I(t)}$. 

\begin{figure}
\includegraphics[width=8.5cm]{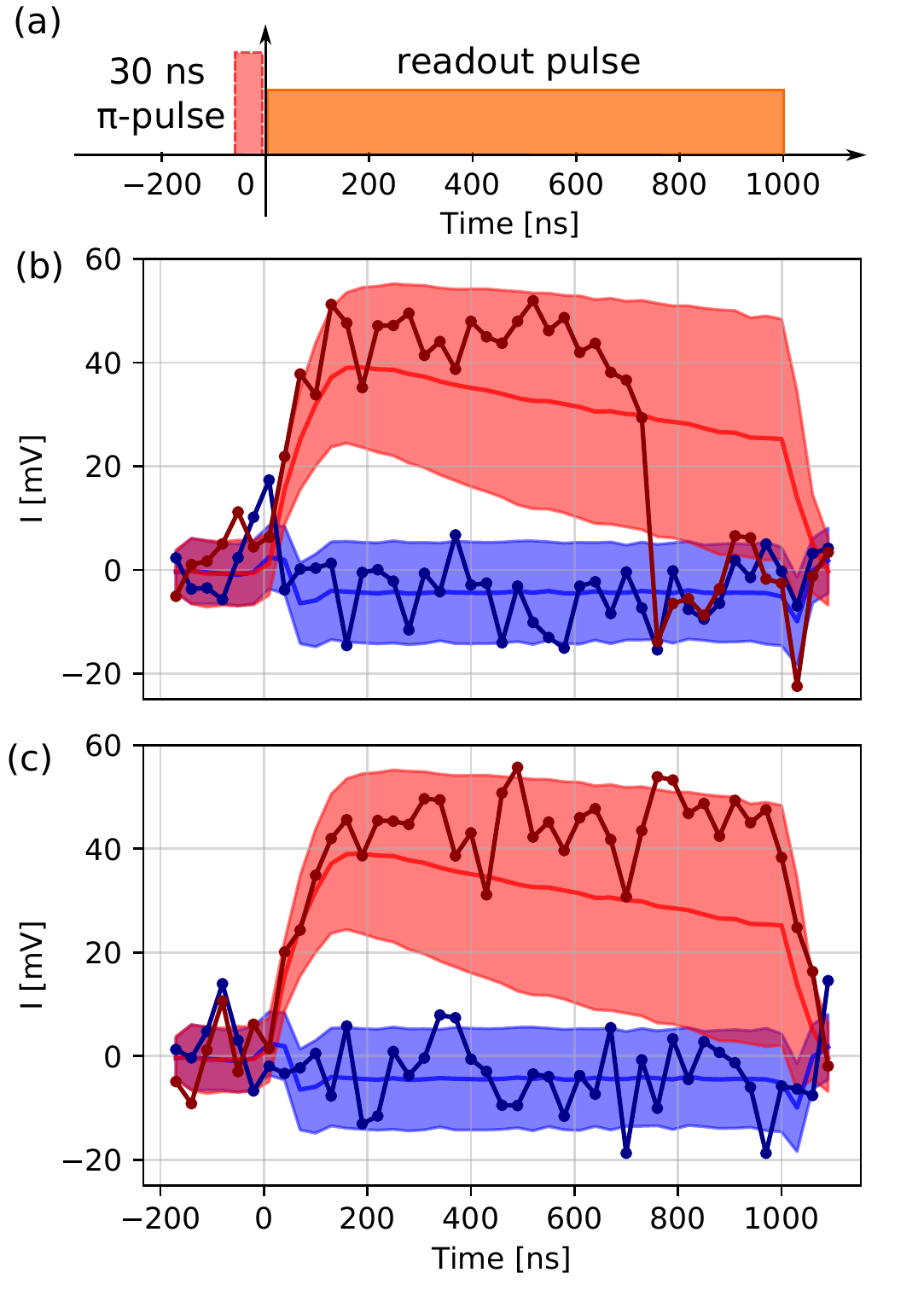}
\caption{Typical individual measurement records as a function of time, using the pulse sequence sketched in (a). We show typical quantum trajectories of the qubit in the presence (b) and absence (c) of a quantum jump. Blue and red points refer to the case the qubit is initially prepared in states $\ket{g}$ and $\ket{e}$, respectively ($t=0$). The readout pulse with amplitude $\overline{n}_{l}=2$ starts at $t=$ \SI{0}{\nano\second} and stops at $t=$ \SI{1000}{\nano\second}. Each point is measured with a \SI{30}{\nano\second} integration, corresponding to the resonator rising time $2\kappa_l^{-1}$. An average over 1000 measurement records is plotted in solid blue and red lines, as well as their standard deviation represented by corresponding shaded areas. }
\label{Q_trajectories}
\end{figure}

One thousand individual trajectories have been measured when the qubit is prepared either in $\ket{g}$ or $\ket{e}$ state. Four typical individual records are plotted in Fig. \ref{Q_trajectories}. The duration pulse is \SI{1000}{\nano\second} acquired over a larger time window (around \SI{1300}{\nano\second}). These measurement records give an insight on the real time dynamics of the qubit from single-shot trajectories. Notice that after a time of few $\kappa_l^{-1} \sim $ 15ns, the qubit state can already be inferred from a single trajectory, and that in Fig.~\ref{Q_trajectories}(b) a quantum jump \cite{PRL_Vijay_2011} of the qubit appears clearly. In addition to the individual trajectories, the mean value averaged over the one thousand trials, as well as the related standard deviation, is plotted as function of time. Due to qubit relaxation, the averaged excited state response (red solid line) decays towards the ground state response, while its corresponding standard deviation (red shaded area) grows in time. This finite qubit lifetime, can limit the distinguishability of the qubit states when the measurement itself takes a non-negligible fraction of $T_1$, highlighting the need for a fast readout. The qubit decay under drive is equal to the one measured without drive, $T_1 \simeq $ \SI{3.3}{\micro\second}, within the measurement error bars. This observation suggests a QND measurement, which we quantify in more detail in the following.

\subsection{Quantum non-demolition fidelity}\label{sec_QNDfidelity}

To check the QND-ness of the measurement, we quantify the repeatability of successive measurements. We now consider only the measurement records between time $10 \kappa_l^{-1} \sim $ \SI{150}{\nano\second} and \SI{1000}{\nano\second} to be in the steady state regime of the applied squared pulse. It corresponds to the ground state if  $I(t) < I_{th}$ or to the excited state if $I(t) > I_{th}$ with $I_{th}=$ \SI{15.5}{\milli\volt}. We define four conditional probabilities, $P_{\alpha, \beta}$, the probability to measure $\alpha$ in the first measurement and $\beta$ in the second measurement, where $\alpha,\beta=g,e$ can correspond to ground or excited states.
From these probabilities, the QND fidelity \cite{arXiv_Touzard_2018} is obtained to be
$\mathcal{Q} = \frac{P_{g, g}+ P_{e, e}}{2} =$ \SI{99}{\percent}. In $P_{e, e}= 98.3$ \SI{}{\percent}, we estimate \SI{0.7}{\percent} to be explained by relaxation during measurement, and in $P_{g, g}= 99.6$ \SI{}{\percent}, we estimate only \SI{0.02}{\percent} to be due to thermal excitation during measurement. Moreover, each probability has a statistical uncertainty, due to finite number of realizations of $\pm 0.6$ \SI{}{\percent}. These results are comparable to the QND fidelity obtained in Touzard \textit{et al} \cite{arXiv_Touzard_2018} using a parametric modulation scheme and corresponds, to the best of our knowledge, to the state-of-the-art values.

\subsection{Single-shot readout fidelity}\label{sec_ReadoutFidelity}

In the early days of circuit-QED, averaging was necessary to infer the qubit state with high fidelity. However, thanks to the advent of Josephson-based amplifier \cite{Caves1982,Yurke1996,Siddiqi2004}, high fidelity, single shot discrimination of the qubit state is now possible \cite{mallet_single-shot_2009}. Since then, works have been performed on Purcell filters and amplifiers in an attempt to increase further the readout fidelity \cite{liu_high_2014,PRL_Jeffrey_2014,krantz_single-shot_2016,bultink_active_2016}, which is now culminating at \SI{99.6}{\percent} in \SI{88}{\nano\second} \cite{PRApplied_Walter_2017}. Readout fidelity is currently limited by the balance between the time needed to discriminate the qubit state and the qubit \tomas{relaxation time} $T_1$.

\begin{figure}
\includegraphics[width=8.5cm]{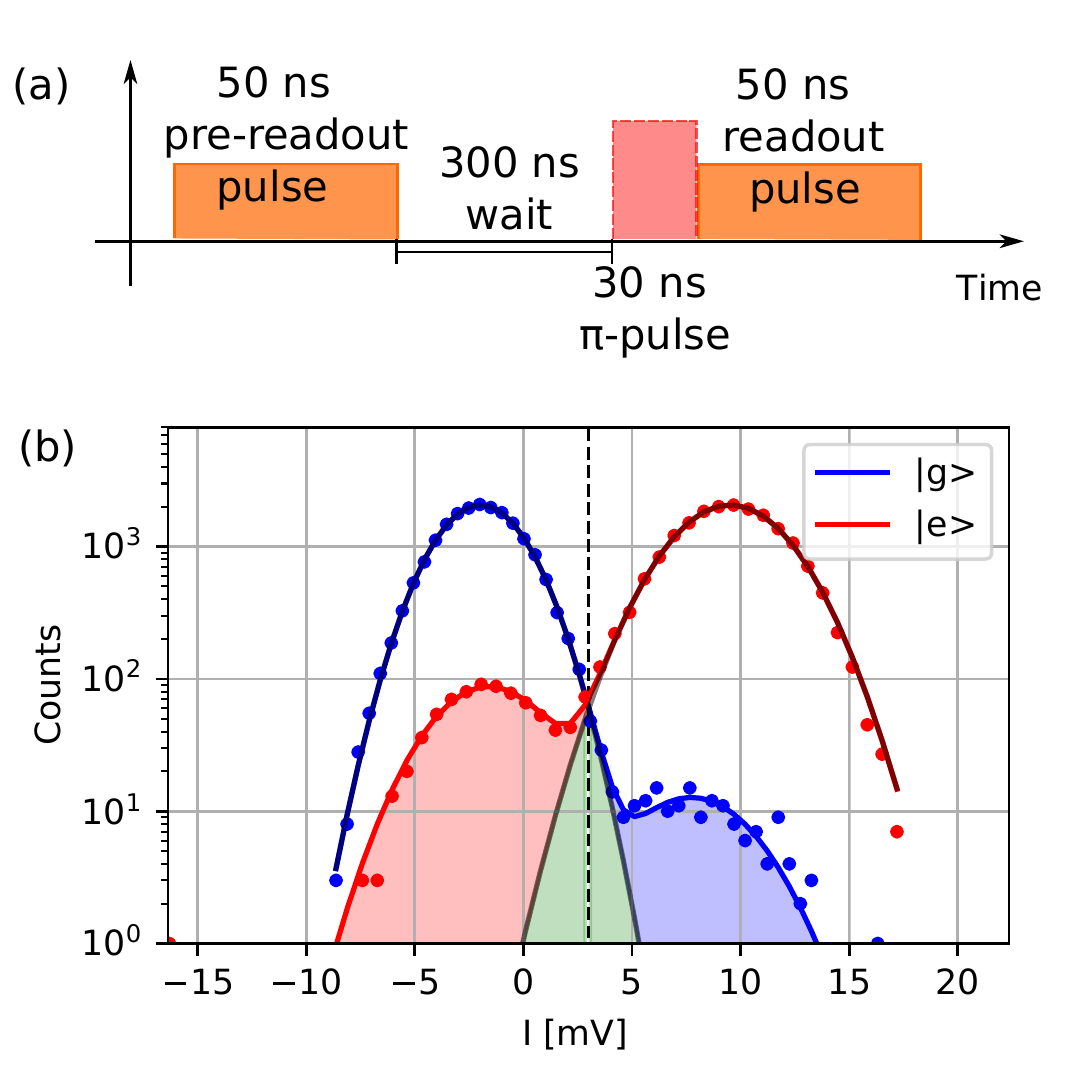}
\caption{(a) Pulse sequence sketch. (b) Histograms of \SI{50}{\nano\second} single-shot measurement for qubit prepared in ground state (blue points) and excited state (red points) with heralding. The solid blue and red line are fits with a double Gaussian model. Black line is a single Gaussian fit. The green area depicts the overlap error $\epsilon_o= $ \SI{0.8}{\percent}. The blue and red area indicates the remaining error $\epsilon_{r, g}= $ \SI{0.6}{\percent} and $\epsilon_{r, e}= $ \SI{3.9}{\percent}, respectively. It leads to a readout fidelity of \SI{97.4}{\percent}.}
\label{Histo_RO}
\end{figure}

To quantify the readout fidelity, we perform heralding \cite{Johnson2012} by first applying a \SI{50}{\nano\second} square readout pulse. In the analysis, we keep only the sequences where the qubit is found in the ground state for this first measurement. After this pulse, we wait \SI{300}{\nano\second} $ \sim 20\kappa_l^{-1}$ for the resonator to  decay back into its vacuum state before preparing the qubit in the ground or in the excited state. Then, another \SI{50}{\nano\second} square readout pulse is applied. The two measurement pulses correspond to a steady state amplitude of $\overline{n}_{l} \simeq 2$. Via the heralding procedure, we estimate a thermal equilibrium population of the excited state of \SI{2.4}{\percent}, corresponding to an effective temperature of $\sim$ \SI{80}{\milli\kelvin}.
In Fig.~\ref{Histo_RO}, histograms of $24\cdot 10^3$ single shot readouts are plotted as the function of the in-phase amplitude when the qubit is prepared in $\ket{g}$ and $\ket{e}$ states. A weight function is used to maximize the distinguishability between the two qubit states \cite{PRApplied_Walter_2017}. 
The histograms are fitted by the sum of two Gaussians (colored solid lines) as discussed in the appendix of Ref. \cite{PRApplied_Walter_2017}. The intersection of these two fitted histograms defines a threshold $I_{th}$ (vertical dash line) distinguishing the two qubit states. The readout fidelity is defined as $F = 1 - (P(e|g) + P(g|e))/2 \simeq 1 - (\epsilon_g + \epsilon_e)/2$, where $P(x|y)$ is the probability of reading out $x$ while having prepared the state $y$. In addition, $\epsilon_g$ and $\epsilon_e$ are the fraction of measured events of detecting $I \geq I_{th}$ when the qubit was prepared in g and $I \leq I_{th}$ when the qubit was prepared in e, respectively. 
Finally, we obtained a readout fidelity of $F = $ \SI{97.4}{\percent} affected by the imperfections $\epsilon_g =$ \SI{1.0}{\percent}, and $\epsilon_e= $ \SI{4.3}{\percent}.    

The following discussion is to distinguish different sources of error. One source of error is the overlap (or separation) error $\epsilon_o$, which is due to the detector noise along with the finite acquisition time. We computed from the overlap of the two main fitted Gaussians (green shaded area) an overlap error of $\epsilon_o =\epsilon_{o, g} + \epsilon_{o, e} =$ \SI{0.8}{\percent} with $\epsilon_{o, g} = \epsilon_{o, e} = $ \SI{0.4}{\percent}. For the remaining errors, $\epsilon_{r, g} = \epsilon_{g}-\epsilon_{o, g}=$ \SI{0.6}{\percent} (blue shaded area) and $\epsilon_{r, e} = \epsilon_{e}-\epsilon_{o, e}=$ \SI{3.9}{\percent} (red shaded area), we analyzed two types of sources: $\epsilon_p$, the error of false qubit state preparation and $\epsilon_t$, the transition during measurement error. In $\epsilon_{t, e}$, we expect $ \sim$ \SI{1.5}{\percent} due to relaxation during measurement. For $\epsilon_{p, e}$, we expect $ \sim $ \SI{1.4}{\percent} error due to finite $\pi$-pulse time compared to the Rabi decay time. 
We also roughly estimate $ \sim $ \SI{0.5}{\percent} error due to having prepared the f state, the second excited state of the transmon, because of the frequency spreading of the square $\pi$-pulse. The leftover errors may be attributed to a imperfect heralding procedure or possibly to measurement-induced transitions \cite{PRL_Sank_2016}, but they are within the statistical uncertainty due to finite counting of $\pm$ \SI{0.7}{\percent}. 

We believe that the readout fidelity can be further increased by implementing pulse envelop optimization such as DRAG pulse \cite{Chow2010} to have less excited state preparation error, or CLEAR pulse \cite{McClure2016} to achieve better discrimination in a shorter integration time and therefore reduce error due to relaxation during measurement.
 
\subsection{Coherence and readout quality factor}
\begin{figure}
    \centering
 \includegraphics[width=8.5cm]{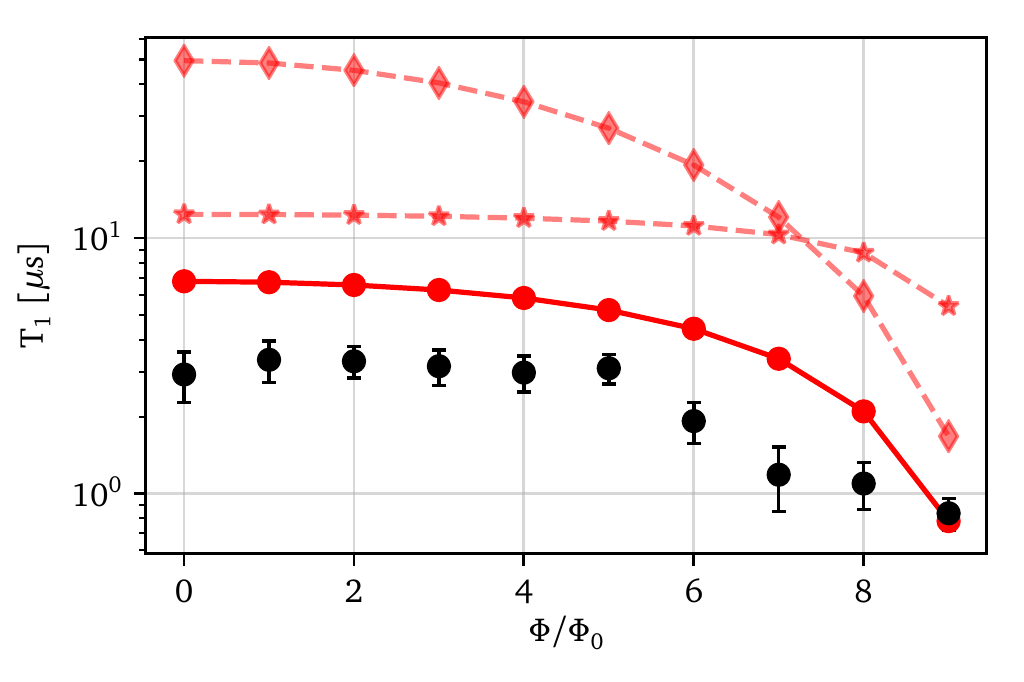}
    \caption{\remy{Qubit relaxation time $T_1$ versus flux}. Black points and error bars are the extracted values of $T_1$ from Gaussian means and standard deviations, respectively. The various types of red points correspond to computed values of Purcell-limited $T_1$, assuming a one-mode cavity, the parameters described in Appendix~\ref{app:circuit_parameters}, and \tomas{various} imperfections. For instance, the red diamond points only consider the asymmetry in \tomas{Josephson energies}, the star points only consider the misalignment, and the circle points consider both imperfections. }
    \label{fig:coherence}
\end{figure}
Both QND-ness and single-shot readout fidelity are limited by the finite $T_1$ of the qubit.  
To understand qubit lifetime limitations, we have measured its relaxation \tomas{at} several fluxes [cf.~ black points in Fig. \ref{fig:coherence}]. We found a $T_1$ ranging from \SI{3.3}{\micro\second} at zero flux to \SI{0.9}{\micro\second} at $\Phi= 9 \Phi_0 $. We identified two sources of imperfections in our system that create parasitic residual transverse coupling leading to a Purcell-limited qubit $T_1$. The first source is the asymmetry of critical current in the Josephson junctions and the second is the possible misalignment of the sample inside the cavity. The effect of these two imperfections is discussed in detail in Appendix \ref{Asymmetry}. There, we computed the Purcell-limitation due to these residual transverse couplings, and the results are shown by the various red points in Fig. \ref{fig:coherence}, where red diamond points only consider the imperfection due to asymmetry in critical current, the star points only consider the misalignment imperfection, and the circle points consider both imperfections. The overall trend of relaxation versus flux is well described by the Purcell-limited $T_1$, however further study is required to \remy{obtain better quantitative agreement and to} fully rule out other loss channels, such as dielectric loss or spurious two-level systems. 

Although our $T_1$ is limited by residual transverse couplings, it is clear that the readout shift \tomas{$(2\chi_l)/(2\pi) =-$\SI{9}{\mega\hertz}} is mainly produced by the non-perturbative cross-Kerr coupling $g_{zz}/2\pi =$ \SI{34.5}{\mega\hertz} \remy{[}cf. Fig.~\ref{polariton_modes}b\remy{]}, which does not induce qubit decay. However, to show more intuitively the benefit of the non-perturbative cross-Kerr coupling, we estimate the qubit decay as if it would be obtained solely by the usual dispersive transverse coupling between the qubit and the dominant lower polariton. For this, we consider the same readout shift \tomas{$(2\chi_l)/(2\pi)=$ -\SI{9}{\mega\hertz} (corresponding to the lower polariton in our case, cf. Fig.~\ref{polariton_modes}b),} but now \tomas{let us suppose} it is given by the dispersive approximation\tomas{, i.e.~$\chi_{\rm d}=\frac{\alpha_{q}{(g_x)}^2 }{\Delta(\Delta+\alpha_{q})}$.} With \remy{detuning $\Delta_l/(2\pi) = -754 {\rm MHz}$ and anharmonicity $\alpha_q/(2\pi) = -88 {\rm MHz}$ as measured experimentally, we then would need an hypothetical} transverse coupling \tomas{$g_x/(2\pi) \simeq$ \SI{180}{\mega\hertz}}, which would result in a Purcell-limited relaxation of $T_1 \sim \kappa_l^{-1} (\Delta_l/g_x)^2 \simeq$ \tomas{\SI{0.24}{\micro\second}}, which is one order of magnitude lower than the measured $T_1$. In
addition, if this were the case, \tomas{$|\Delta_l|/g_x \sim 4.2$}, far too low for the dispersive approximation to remain valid and would not allow QND measurements at the \SI{99}{\percent} level.  
  
Despite this limited $T_1$, we achieve a good steady state signal-to-noise ratio (SNR) per photon number \tomas{as defined in Ref.~\cite{Gambetta2008}}. Indeed, \tomas{when using the lower polariton for readout,} we obtain \remy{a readout quality factor} \remy{$Q_r= 4\chi_l^2 \kappa_l T_1/(\kappa_l^2/4 + \chi_l^2)\simeq 360$, so that} the optimal steady state SNR is given by ${\rm SNR}= \eta n Q_r$ with $n$ the photon number and $\eta$ the quantum efficiency \cite{Gambetta2008}. As a comparison, \remy{we compute from the parameters given in Refs.~\cite{PRL_Jeffrey_2014} and \cite{PRApplied_Walter_2017}, the quality factors of $Q_r = 540$ and $Q_r = 1080$, respectively.} Without limitations of the Purcell effect, it should be possible to increase our $T_1$ and maintain large values of $\kappa_{\tomas{l}}$ for fast measurements, while optimizing \tomas{$|\chi_l| \simeq \kappa_l/2$} to maximize the readout quality factor $Q_r$. In this way, we believe that one order of magnitude increase \tomas{in} $Q_r$ is within reach. Moreover, we expect \tomas{the limitation in} photon number $n$ to be less restrictive for the non-perturbative cross-Kerr coupling \tomas{compared to the standard dispersive one \cite{PRA_Koch_2007}}. \remy{Therefore, the steady state SNR may be further improved with $n$ without being restricted by the QND-ness of the readout.} Nonetheless, some \tomas{other limitations on the photon number may arise} due to the non-RWA terms of type $\sim\hat{q}^\dagger \hat{q}^\dagger \hat{a}\hat{a}$, \tomas{but these} and other related aspects will be discussed elsewhere.

\section{Conclusions and outlook}

We have developed and demonstrated an original qubit readout scheme relying on a non-perturbative cross-Kerr coupling, in contrast to the usual cross-Kerr coupling that is perturbatively obtained from the transverse coupling in the dispersive regime. Therefore, our new experimental measurement design does not suffer from cavity-mediated excitations or decay, and the strength of the readout shifts can be made large and independent of the detuning. This allows for a fast readout of the qubit, with a large single-shot fidelity, and a maximization of the QND-nature of the measurement.  
The qubit and readout performances are currently limited because of residual qubit-cavity transverse couplings. However, no fundamental reason prevents further suppression of this transverse coupling. In fact, in the future, we can obtained the same readout shifts $2\chi_j$, but with a much larger qubit-polaritons detuning, so that any residual transverse coupling produces significantly less unwanted consequences.

According to our readout error budget and to our QND-ness analysis, the measurement-induced qubit state mixing is particularly low compared to the standard literature. This could be explained by the non-perturbative nature of our cross-Kerr coupling and will be the topic of future investigations. Another appealing possibility for the future is to extend the current non-perturbative QND measurements to detect single- and multi-photon propagating fields \cite{nakamura10,besse18,ramos17,lescanne19,besse19}.

\acknowledgments
The authors thank D. Basko, D. Divincenzo, and B. Huard for fruitful discussions. The authorsthankthe referees for theirthorough review and clear remarks which helped us improvethe manuscript significantly. R.D. and S.L. acknowledge support from Fondation CFM pour la recherche. R.D., V.M. and O.B. acknowledge support from ANR REQUIEM (ANR-17-CE24-0012-01).
J.J.G.-R. and T.R. acknowledge support from project PGC2018-094792-B-I00 (MCIU/AEI/FEDER, UE) and CAM/FEDER project No. S2018/TCS-4342 (QUITEMAD-CM). T.R. further acknowledges funding from the EU Horizon 2020 program under the Marie Sk\l{}odowska-Curie grant agreement No. 798397. S.L. acknowledges the Agence Nationale de la Recherche under the  program « Investissements d'avenir » (ANR-15-IDEX-02). K.B. and J.D. acknowledge the European Union's Horizon 2020 research and innovation program under the Marie Sklodowska-Curie grant agreement No 754303. J.P. acknowledges grant from the Laboratoire d'excellence LANEF in Grenoble (ANR-10-LABX-51-01).

\appendix

\section{Experimental setup}
\label{Experimental setup}
\begin{figure}
	\includegraphics[width=8cm]{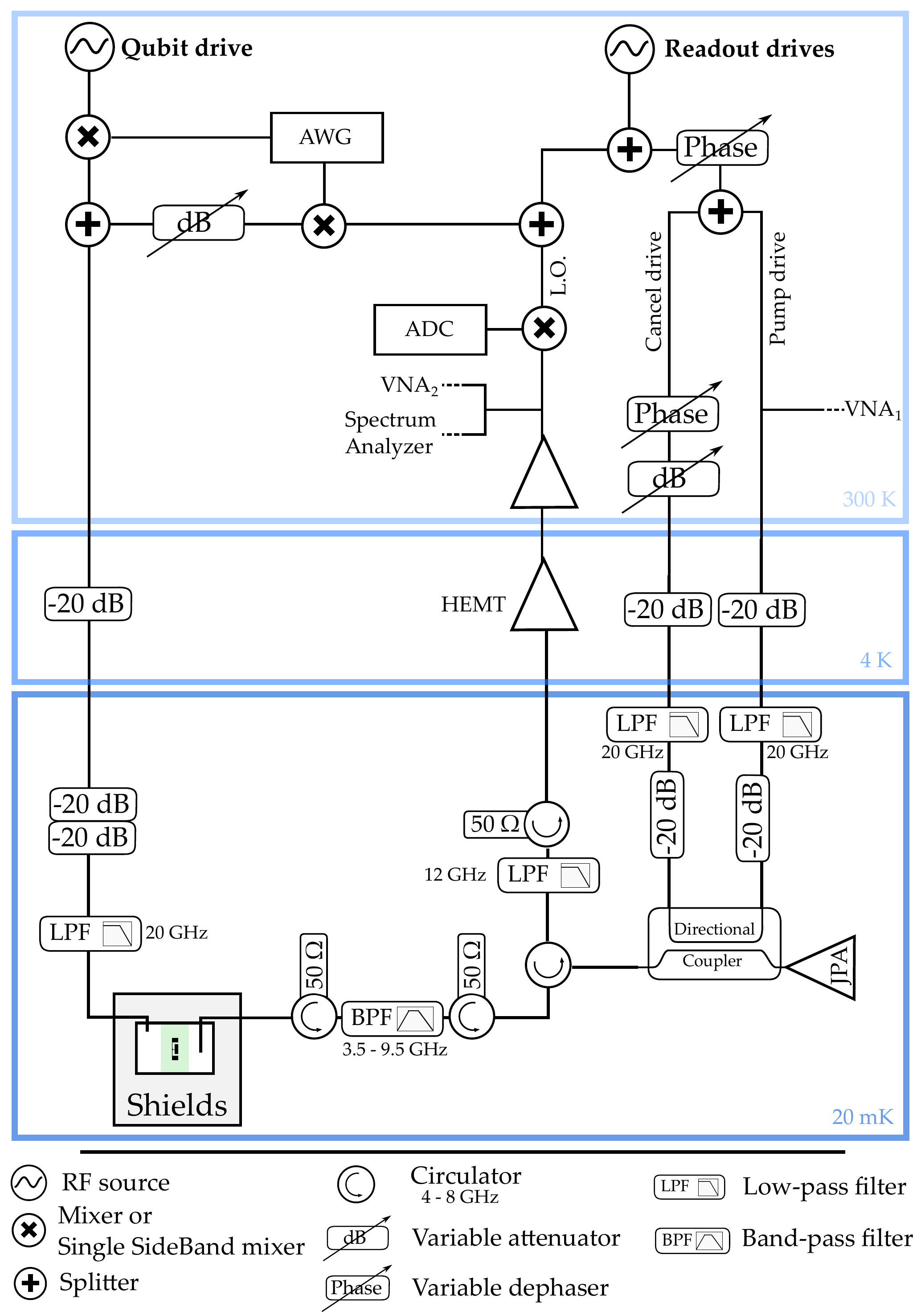}
	\caption{Schematic of the experimental setup}
	\label{Setup_figure}
\end{figure}

In this section we describe the measurement setup shown in Fig.~\ref{Setup_figure}.
Qubit and readout pulses are sent through the same input line. The transmitted signal passes through three circulators and a directional coupler before being amplified via the Josephson Parametric Amplifier (JPA). Then it passes through additional amplification stages before it is down-converted to DC voltages via an IQ mixer and digitized at 1 GS/s using an ADC. Finally, the signal is digitally integrated.

The JPA \cite{arXiv_Planat_2018} is used in the phase-sensitive regime and thus phase stability is a key feature in this setup. The pump and cancellation drives need to be tuned at the same amplitude with opposite phases \cite{PRApplied_Walter_2017,arXiv_Touzard_2018}. Moreover, the phase of the JPA also needs to be tuned to amplify the wanted quadrature. The JPA gain (\SI{20}{\decibel}) and its pump cancellation are tuned with a VNA and spectrum analyzer regardless of the sample.

\remy{\section{Superconducting quantum circuit model}\label{model_allFlux}}

\remy{In this section, we derive the circuit Hamiltonian given in Eq.~(\ref{Hamiltonian_Circuit}). We start from the classical Lagrangian $\mathrm{L}$ of the circuit, which depends on the generalized flux variables at the left and right nodes of the circuit denoted by $\Phi_1$ and $\Phi_2$, respectively [cf.~Figs.~\ref{sample}(c) and (d)]. The kinetic energy $K$, stored in the capacitances, reads
\begin{equation}
K = \frac{ C_S}{2} \dot{\Phi}_1^2 + \frac{C_S }{2} \dot{\Phi}_2^2 + \frac{C_t}{2} (\dot{\Phi}_1-\dot{\Phi}_2)^2,
\end{equation}
where $C_S$ is the capacitance of each transmon and $C_t$ the capacitance of the coupling inductor [cf.~Fig.~\ref{sample}(c)]. The potential energy $U$ is given by the Josephson energies $E_{J1}$ and $E_{J2}$ of each junction and by the inductive energy of the coupling inductance $L_a$. Explicitly, we have
\begin{align}
U =& -E_{J}(1-d_J)\cos\left(\frac{\Phi_1}{\phi_0}\right) - E_{J}(1+d_J)\cos\left(\frac{\Phi_2}{\phi_0}\right)\nonumber\\ 
+& \frac{1}{2 L_a}(\Phi_1-\Phi_2 - \Phi)^2,
\end{align}
where $d_J=(E_{J2}-E_{J1})/(E_{J2}+E_{J1})$ is a small asymmetry in the Josephson energies, $E_J=(E_{J1}+E_{J2})/2$ is the average Josephson energy, $\Phi$ is the externally applied flux, and $\phi_0 = \Phi_0/(2\pi) = \hbar/(2e)$ is reduced magnetic flux quantum. Here, $L_a$ is the coupling inductance implemented by a chain of SQUIDs and thus depends on the applied flux $\Phi$ [cf.~Figs.~\ref{sample}(d)]. }

\remy{It is convenient to introduce ``qubit'' and ``ancilla'' variables $\Phi_q$ and $\Phi_a$ as the flux average $\Phi_q = (\Phi_1 + \Phi_2)/2$ and the flux difference $\Phi_a = (\Phi_1 - \Phi_2)/2$, respectively. This allows us to write the Lagrangian of the circuit $\mathrm{L} = K-U$ as
\begin{align}
\mathrm{L}  = & C_S \dot{\Phi}_q^2 + (C_S + 2 C_t) \dot{\Phi}_a^2 \nonumber\\
+&  2 E_J \left[\cos\left(\frac{\Phi_q}{\phi_0}\right) \cos\left(\frac{\Phi_a}{\phi_0}\right) -  \frac{L_J}{(\phi_0)^2L_a }\left(\Phi_a - \Phi/2\right)^2 \right]\nonumber\\
+&2E_Jd_J\sin\left(\frac{\Phi_q}{\phi_0}\right)\sin\left(\frac{\Phi_a}{\phi_0}\right),
\end{align}
with the Josephson inductance given by $L_J=(\phi_0)^2/E_J$. We now calculate the conjugate charges $Q_{q}$ and $Q_{a}$, corresponding to the phases $\Phi_q$ and $\Phi_a$, which read
\begin{align}
Q_{q} =& \frac{\partial \mathrm{L}}{ \partial \dot{\Phi}_q } = 2 C_S \dot{\Phi}_q, \\
 Q_{a} =& \frac{\partial \mathrm{L}}{\partial \dot{\Phi}_a } = 2(C_S + 2C_t) \dot{\Phi}_a.
\end{align}
Using the Legendre transformation $H(Q_q,Q_a,\Phi_q,\Phi_a)= Q_a\dot{\Phi}_a +Q_q\dot{\Phi}_q - L$, we obtain the classical Hamiltonian of the circuit as,
\begin{align}
H = & \frac{Q_{q}^2}{2C_q}  + \frac{Q_{a}^2}{2C_a} -2E_Jd_J\sin\left(\frac{\Phi_q}{\phi_0}\right)\sin\left(\frac{\Phi_a}{\phi_0}\right)\\ 
 -&  2 E_J \left[\cos\left(\frac{\Phi_q}{\phi_0}\right) \cos\left(\frac{\Phi_a}{\phi_0}\right) -  \frac{L_J}{(\phi_0)^2L_a }\left(\Phi_a - \Phi/2\right)^2 \right],\nonumber
\end{align}
where we define the effective capacitances of the qubit and ancilla variables as $C_q = 2C_s$ and $C_a=2(C_S+2C_t)$, respectively.}

\remy{We can quantize this Hamiltonian by promoting the flux and charge variables to operators, $\Phi_j\rightarrow\hat{\Phi}_j$ and $Q_j\rightarrow\hat{Q}_j$, and imposing canonical commutation relations between them, namely $[\hat{\Phi}_j,\hat{Q}_l]=i\hbar\delta_{jl}$ with the indices $j,l$ corresponding to qubit and/or ancilla ($j,l=\lbrace q,a\rbrace$). In addition, we define dimensionless phase operators $\hat{\varphi}_j=\hat{\Phi}_j/\phi_0$ and charge number operators $\hat{n}_j=\hat{Q}_j/(2e)$, and use them to express the quantum Hamiltonian of the circuit as
\begin{align}
\hat{H}_{\rm mol} = & 4E_{C_q} \hat{n}_{q}^2 + 4E_{C_a}  \hat{n}_{a}^2 \nonumber\\
&-2E_J [\cos(\hat{\varphi}_q)\cos(\hat{\varphi}_a)  - \frac{L_J}{L_a} (\hat{\varphi}_a -\frac{\Phi}{2\phi_0})^2 ]\nonumber\\
&-2E_Jd_J\sin(\hat{\varphi}_q)\sin(\hat{\varphi}_a).
\label{Eq:VHamcos}
\end{align}
Here, we define the charging energy of qubit and ancilla as $E_{C_q} = e^2/(2C_q)$ and $E_{C_a} =e^2/ (2C_a)$, respectively. Exploiting the analogy between conjugate flux/charge operators and position/momentum operators, we can interpret the Hamiltonian (\ref{Eq:VHamcos}) as two particles with mass $m_q = \hbar^2/(8E_{C_q})=(\phi_0)^2 C_q$ and $m_a = \hbar^2/(8E_{C_a})=(\phi_0)^2 C_a$ subjected to a nonlinear two-dimensional potential $U(\hat{\varphi}_q,\hat{\varphi}_a)=-2E_J [\cos(\hat{\varphi}_q)\cos(\hat{\varphi}_a)  - \frac{L_J}{L_a} (\hat{\varphi}_a -\frac{\Phi}{2\phi_0})^2 ]-2E_Jd_J\sin(\hat{\varphi}_q)\sin(\hat{\varphi}_a)$. In the transmon regime, $E_J\gg E_{C_q},E_{C_a}$, the lowest energy bands are deep inside the sinusoidal potentials, so that we can expand the Hamiltonian (\ref{Eq:VHamcos}) in powers of the small flux $\hat{\varphi}_a, \hat{\varphi}_q \ll 1$. With corrections up to 4th order in the phases, we obtain
\begin{align}
\hat{H}_{\rm mol} ={}& 4E_{C_q} \hat{n}_{q}^2 -2E_J\cos(\hat{\varphi}_q)\nonumber\\
&+ 4 E_{C_a}  \hat{n}_{a}^2-2E_J [\cos(\hat{\varphi}_a) - \frac{L_J}{L_a} (\hat{\varphi}_a -\frac{\Phi}{2\phi_0})^2 ]\nonumber\\
&-\frac{E_J}{2}\hat{\varphi}_q^2\hat{\varphi}_a^2-2E_Jd_J\sin(\hat{\varphi}_q)\sin(\hat{\varphi}_a)+{\cal O}^6,
\label{Eq:VHamcosFULL}
\end{align}
with $\cos(\hat{\varphi}_j)=1-\hat{\varphi}_j^2/2+\hat{\varphi}_j^4/24+{\cal O}^6$ and $j=q,a$. }

\remy{The Hamiltonian in Eq.~(\ref{Hamiltonian_Circuit}) of the main text is obtained by considering an integer flux $\Phi=n\Phi_0=2\pi n\phi_0$ in Eq.~(\ref{Eq:VHamcosFULL}) and simplifying $\hat{\varphi}_a-n\pi\rightarrow \hat{\varphi}_a$ due to the cyclic property of the phase. We also rename $L_a\rightarrow L_a(n)$ to indicate the integer value of the applied flux. Finally, we neglect the interaction due to the small asymmetry in the junctions provided $d_J\ll 1$. This aspect is further discussed as a small imperfection in Appendix \ref{Asymmetry}.\\}

\section{\tomas{Circuit Hamiltonian in the number representation} }
\label{HamiltonianNumberRep}

\remy{In this Appendix we derive the Hamiltonian (\ref{Ham2}) of the main text starting from Eq.~(\ref{Hamiltonian_Circuit}). }

\remy{Since our setup works in the transmon regime of low flux, $\hat{\varphi}_q,\hat{\varphi}_a\ll 1$, we can expand the cosines in Eq.~(\ref{Hamiltonian_Circuit}), obtaining
\begin{align}
    \hat{H}_{\rm mol} ={}& 4E_{C_q}\hat{n}_q^2+\frac{E_{J_q}}{2}\hat{\varphi}_q^2+4E_{C_a}\hat{n}_a^2+\frac{E_{J_a}(n)}{2}\hat{\varphi}_a^2\nonumber\\
    {}&-\frac{E_J}{12}\left(\hat{\varphi}_q^4+\hat{\varphi}_a^4\right)-\frac{E_J}{2}\hat{\varphi}_q^2\hat{\varphi}_a^2 + {\cal O}^6,\label{expandedHam}
\end{align}
where we have defined the effective Josephson energies of qubit and ancilla as $E_{J_q}=2E_J$ and $E_{J_a}(n)=2E_J\left(1+\frac{2L_J}{L_a(n)}\right)$, respectively. To express the Hamiltonian (\ref{expandedHam}) in the number representation, we exploit the analogy between the quadratic terms in Eq.~(\ref{expandedHam}) and the Hamiltonian of independent quantum harmonic oscillators with positions $\hat{x}_j=\hat{\phi}_j$, momenta $\hat{p}_j=\hbar\hat{n}_j$, masses $m_j=\hbar^2/(8E_{C_j})$, and frequencies $\tilde{\omega}_j= \sqrt{8E_{J_j}E_{C_j}}/\hbar$, for qubit and ancilla ($j=q,a$). With these identifications, we can use the known results from the quantization of the quantum harmonic oscillator and express the phase and number operators as
\begin{align}
    \hat{\varphi}_q ={}& \left(\frac{8E_{C_q}}{E_{J_q}}\right)^{1/4}\frac{(\hat{q}+\hat{q}^\dag)}{\sqrt{2}},\label{Op1}\\
    \hat{n}_q ={}& -i\left(\frac{E_{J_q}}{8E_{C_q}}\right)^{1/4}\frac{(\hat{q}-\hat{q}^\dag)}{\sqrt{2}},\\
    \hat{\varphi}_a ={}& \left(\frac{8E_{C_a}}{E_{J_a}(n)}\right)^{1/4}\frac{(\hat{a}+\hat{a}^\dag)}{\sqrt{2}},\\ 
    \hat{n}_a ={}& -i\left(\frac{E_{J_a}(n)}{8E_{C_a}}\right)^{1/4}\frac{(\hat{a}-\hat{a}^\dag)}{\sqrt{2}},\label{Op4}
\end{align}
where $\hat{q}$, $\hat{q}^\dag$ and $\hat{a}$, $\hat{a}^\dag$ are standard ladder operators for the qubit and ancilla modes, respectively. }  

\remy{Replacing expressions (\ref{Op1})-(\ref{Op4}) into Eq.~(\ref{expandedHam}), we diagonalize the quadratic terms of the circuit Hamiltonian, allowing us to interpret the qubit and ancilla modes as two coupled anharmonic oscillators described by  
\begin{align}
\frac{\hat{H}_{\rm mol}}{\hbar}={}& \tilde{\omega}_q \hat{q}^\dag \hat{q}+\frac{\alpha_q}{12}(\hat{q}+\hat{q}^\dag)^4+\tilde{\omega}_a \hat{a}^\dag \hat{a}\label{eq:hamiltonien}\\
{}&+\frac{U_a}{12}(\hat{a}+\hat{a}^\dag)^4-\frac{g_{zz}}{2}(\hat{q}+\hat{q}^\dag)^2(\hat{a}+\hat{a}^\dag)^2.\nonumber
\end{align}
Here, the anharmonicities the qubit and ancilla are given by $\alpha_q = -E_{C_q}/\hbar$ and $U_a = -(E_{C_a}/\hbar)(1+2\frac{L_J}{L_a(n)})^{-1}$, respectively, and $g_{zz} = \sqrt{\alpha_q U_a}$ is the strength of their cross-Kerr coupling.}

\remy{We can further simplify the Hamiltonian in Eq.~(\ref{eq:hamiltonien}) by expanding the fourth order anharmonic terms proportional to $\alpha_q$ and $U_a$, and perform a rotating wave approximation (RWA), provided the anharmonicities are much smaller than the free frequencies, i.e. $\alpha_q,U_a\ll \tilde{\omega}_q, \tilde{\omega}_a$. Doing so and expressing the resulting terms in normal ordering we finally obtain the circuit Hamiltonian in Eq.~(\ref{Ham2}) of the main text, where the qubit and ancilla frequencies become renormalized by the anharmonic terms as $\omega_q = \tilde{\omega}_q+\alpha_q$, and $\omega_a = \tilde{\omega}_a+U_a$, respectively.}

\section{\tomas{Quantum optics model for decoherence and polariton spectroscopy}}\label{Numerical_model}

\remy{In this Appendix, we describe the full quantum optics model of our system and its environment, including loss sources and the coherent driving field used in the spectroscopies. We also derive the polariton Hamiltonian in Eq.~(\ref{Hamiltonian_Polariton}), and the cavity transmission amplitude in Eq.~(\ref{expectationC}), which models the spectroscopic measurements.}

\remy{Our experimental setup consists of a transmon molecule circuit coupled to a microwave cavity mode as described by the Hamiltonian in Eq.~(\ref{FullHamqac}) of the main text. Under realistic experimental conditions qubit, ancilla and cavity modes are not perfectly isolated from their environment and they undergo dissipation and decoherence. As a consequence, the state of the system is represented by a density matrix $\hat{\rho}(t)$, whose dynamics can be well described in a Master equation formalism as,
\begin{align}
\frac{d\hat{\rho}}{dt} ={}& -\frac{i}{\hbar}[\hat{H}_{\rm tot}+\hat{H}_{\rm drive},\hat{\rho}] + \kappa_c {\cal D}[\hat{c}]\hat{\rho} + \kappa_a {\cal D}[\hat{a}]\hat{\rho}\nonumber\\
{}&+ \kappa_q {\cal D}[\hat{\sigma}^-]\hat{\rho} + 2\gamma_q {\cal D}[\hat{\sigma}^+ \hat{\sigma}^-]\hat{\rho}.\label{simplifiedMaster}
\end{align}
Here, the coherent part of the dynamics is governed by the Hamiltonian $\hat{H}_{\rm tot}$ in Eq.~(\ref{FullHamqac}) and by $\hat{H}_{\rm drive}=\hbar \Omega(\hat{c}e^{i\omega_d t}+ \hat{c}^\dag e^{-i\omega_d t})$, which describes a coherent driving field of strength $\Omega$ and frequency $\omega_d$ acting on the cavity mode $\hat{c}$. In addition, photon decay of the cavity mode is described by the Lindblad term $\kappa_c {\cal D}[\hat{c}]\hat{\rho}$, where $\kappa_c$ is the cavity decay rate and ${\cal D}[\hat{x}]\hat{\rho}=\hat{x}\hat{\rho}\hat{x}^\dag-(\hat{x}^\dag\hat{x}\hat{\rho}+\hat{\rho}\hat{x}^\dag\hat{x})/2$. Similarly, $\kappa_a$ is the decay rate of the ancilla mode, and $\kappa_q$ the decay rate of the qubit. We also include pure dephasing of the qubit with rate $\gamma_q$. The relaxation and pure dephasing times of the qubit are then given by $T_1=1/\kappa_q$, and $T_2^\ast=1/\gamma_q$, respectively.}

\remy{In our experiments the cavity and ancilla are strongly coupled and close to resonance $|\omega_a'-\omega_c|\lesssim g_{ac}$, so that these two modes become strongly hybridized into upper and lower polariton modes given by $\hat{c}_u =\cos(\theta)\hat{a}+\sin(\theta)\hat{c}$, and $\hat{c}_l =\cos(\theta)\hat{c}-\sin(\theta)\hat{a}$, respectively, with $\tan(2\theta)= 2g_{ac}/ (\omega_a'-\omega_c)$. Re-expressing the master equation (\ref{simplifiedMaster}) in terms of these polaritons, we obtain 
\begin{align}
\frac{d\hat{\rho}}{dt} ={}& -\frac{i}{\hbar}[\hat{H}_{\rm tot}+\hat{H}_{\rm drive},\hat{\rho}] + \sum_{j=u,l}\kappa_j {\cal D}[\hat{c}_j]\hat{\rho}\label{simplifiedMaster2}\\
{}&+\kappa_q {\cal D}[\hat{\sigma}^-]\hat{\rho} + 2\gamma_q {\cal D}[\hat{\sigma}^+ \hat{\sigma}^-]\hat{\rho},\nonumber
\end{align}
where $\hat{H}_{\rm tot}$ is given in Eq.~(\ref{Hamiltonian_Polariton}) of the main text, and the coherent drive on the polariton modes is decribed by $\hat{H}_{\rm drive}=\sum_{j=u,l}\Omega_j(\hat{c}_je^{i\omega_d t}+\hat{c}_j^\dag e^{-i\omega_d t})$ with $\Omega_l=\Omega\cos(\theta)$ and $\Omega_u=\Omega\sin(\theta)$ the effective driving strengths. In addition, the decay effective decay rates of upper and lower polariton read $\kappa_u=\kappa_c\sin^2(\theta)+\kappa_a\cos^2(\theta)$ and $\kappa_l=\kappa_c\cos^2(\theta)+\kappa_a\sin^2(\theta)$, respectively. Importantly, to derive these effective expressions and the master equation (\ref{simplifiedMaster2}), we have neglected fast oscillating terms in a RWA provided $g_{zz},\kappa_u, \kappa_l\ll \omega_u, \omega_l, |\omega_u-\omega_l|$, where $\omega_u = \sin^2(\theta) \omega_c + \cos^2(\theta)\omega_a' + \sin(2 \theta) g_{ac} $ and $\omega_l = \cos^2(\theta)\omega_c + \sin^2(\theta)\omega_a'- \sin(2 \theta)g_{ac}$ are the effective polariton resonance frequencies. We also require a low occupation of the polariton modes, which is ensured in our experiments by having a weak driving strength $\Omega_j\ll \kappa_j$.}

\remy{To end this Appendix, we show how to derive Eq.~(\ref{expectationC}) of the main text, which models the shape of the polariton resonances observed in the spectroscopic measurements of this article [cf.~Sec.~\ref{conditional} and Appendix~\ref{Sec_calibration}]. We perform the spectroscopy by shining a weak coherent drive on the cavity as described by the master equation (\ref{simplifiedMaster2}), and then measuring the amplitude of the cavity field leaking through its transmission output $\langle\hat{\xi}_{\rm out}\rangle$. The input-output relation, $\hat{\xi}_{\rm out}(t)=\hat{\xi}_{\rm in}(t)+\sqrt{\kappa_c^{\rm out}}\hat{c}$ \cite{ramos_correlated_2018,QuantumNoise}, allows us to calculate this output field from the knowledge of the internal dynamics of cavity mode $\hat{c}$, the input noise $\hat{\xi}_{\rm in}(t)$, and the cavity decay on the transmission output $\kappa_c^{\rm out}\leq \kappa_c$. Taking averages and assuming vacuum input noise, we find that the normalized cavity output field reads 
\begin{align}
    \langle \hat{c}_{\rm out}\rangle=\frac{\langle \hat{\xi}_{\rm out}\rangle}{\sqrt{\kappa_c^{\rm out}}}=\langle\hat{c}\rangle=\sin(\theta)\langle\hat{c}_{u}\rangle+\cos(\theta)\langle\hat{c}_{l}\rangle.\label{outExp}
\end{align}
Importantly, the polariton averages $\langle \hat{c}_u\rangle$ and $\langle \hat{c}_l\rangle$ can be calculated from Eq.~(\ref{simplifiedMaster2}). Since the qubit couples to the polaritons via a cross-Kerr coupling only $\sim \sum_j\chi_j\hat{\sigma}_z\hat{c}_j^\dag \hat{c}_j$, the master equation (\ref{simplifiedMaster2}) predicts that the qubit occupation $\langle \hat{\sigma}_z \rangle$ will remain constant during a dynamics much shorter than the qubit coherence times $t\ll T_1,T_2^\ast$. Experimentally, we perform the measurements in time scales shorter than $T_1,T_2^\ast$, so that the main effect of the qubit is simply to shift the resonance frequency of the polaritons $\omega_j\rightarrow \bar{\omega}_j=\omega_j+\chi_j\langle \hat{\sigma}_z \rangle$ and to renormalize the hybridization angle $\theta\rightarrow \bar{\theta}$,  conditioned on the initial state of the qubit, as shown in Eqs.~(\ref{shift})-(\ref{angle}) of the main text. Putting all these together, we neglect $\kappa_q$ and $\gamma_q$ in Eq.~(\ref{simplifiedMaster2}) and assume a constant $\langle \hat{\sigma}_z \rangle$, so that the dynamics of the polaritons reduces simply to two independent driven-dissipative harmonic oscillators, whose steady state expectation values read
\begin{align}
    \langle\hat{c}_{j}\rangle_{\rm ss} = \frac{-i\Omega_j}{\kappa_j/2-i(\omega_d-\bar{\omega}_j)},\qquad j=u,l.\label{expectationPolariton}
\end{align}
Finally, if we replace Eq.~(\ref{expectationPolariton}) into Eq.~(\ref{outExp}) and use the renormalized angle $\bar{\theta}$ in Eq.~(\ref{angle}), we obtain Eq.~(\ref{expectationC}) of the main text.}

\section{Imperfections}
\label{Asymmetry}

\tomas{In this Appendix, we analyze the two main} sources of imperfections that can lead to a non-zero \tomas{transverse couplings between qubit and polariton modes, and} thus limit the readout performance. \tomas{At the end of the Appendix, we also comment on the estimations of the Purcell limited qubit relaxation times $T_1$ shown in Fig.~\ref{fig:coherence}.} 

The first \tomas{source of imperfection for the readout} is the Josephson junction asymmetry $d_J = (E_{J_2}-E_{J_1})/(E_{J_1}+E_{J_2})$ in the transmon molecule circuit, which is experimentally challenging to fully suppress it. To \tomas{estimate the effect of this imperfection, we evaluate the interaction term $\hat{H}^{\rm asy}_{qa}=-2E_J d_J \sin(\hat{\varphi}_q)\sin(\hat{\varphi}_a)$, which was neglected so far from the full Hamiltonian in Eq.~(\ref{Eq:VHamcosFULL})}. \tomas{Notice that $E_J$ denotes} the mean Josephson energy of the two Josephson junctions. At first order, this new term corresponds to a tranverse coupling between the qubit and the ancilla $\tomas{\hat{H}^{\rm asy}_{qa}=\hbar}g_{qa} (\hat{q}+\hat{q}^\dagger)(\hat{a}+\hat{a}^\dagger)$, \remy{where the coupling $g_{qa} = -\frac{d_J}{2}  \sqrt{\frac{\tilde{\omega}_q \tilde{\omega}_a}{1+2L_J/L_a(n)}}$ can be calculated using the identifications of Appendix~\ref{HamiltonianNumberRep}}. \tomas{In order to exprimentally characterize $d_J$,} we measured the room temperature resistances between each pad of the sample. These resistances have contributions from the Josephson junction resistances $R_{J_1}$, $R_{J_2}$, the resistance of the array of SQUIDs and resistances from the connecting wires. The wire resistances are estimated via measurement of wires-only test structures on a dedicated test-chip fabricated during the same process.
In the end, we solve a set of 3 equations with 3 unknowns and found an asymmetry $d_J=(R_{J_2}-R_{J_1})/(R_{J_1}+R_{J_2})= $ \SI{1.3}{\percent}, giving $\tomas{|g_{qa}|}/2\pi = $ \remy{\SI{26.1}{\mega\hertz}} at zero applied flux.

The second \tomas{source of} imperfection is a misalignment of the sample inside the 3D cavity, creating a direct transverse coupling between the qubit and the cavity $\tomas{\hat{H}^{\rm asy}_{qc}=\hbar}g_{qc} (\hat{q}+\hat{q}^\dagger)(\hat{c}+\hat{c}^\dagger)$. Considering the size of the cavity groove and of the sample, we estimate a misalignment angle up to $\theta_{\rm m} = \pm 5$ deg. Assuming that the ratio between transverse couplings $g_{qc}/g_{ac}$ is roughly given by $\tan(\theta_{\rm m})$, we estimate \tomas{that the} qubit-cavity transverse coupling \tomas{is bounded by} $\vert g_{qc} \vert /2\pi \lesssim 25\tomas{.8}{\rm MHz}$. In Fig.~\ref{fig:coherence}, we took the worst case scenario of $|g_{qc}|/2\pi = 25\tomas{.8}{\rm MHz}\ll g_{ac}$.

\tomas{Regarding the analysis of the qubit relaxation times in Fig.~\ref{fig:coherence}, we can analytically} estimate the Purcell limited $T_1$ via the decay rates of the cavity $\kappa_{\tomas{c}}$ and the ancilla $\tomas{\kappa}_a$ as $T_1=1/\Gamma_P$ with $\Gamma_P = \kappa_{\tomas{c}} (\frac{g_{qc}}{\Delta_{qc}})^2 + \tomas{\kappa}_a (\frac{g_{qa}}{\Delta_{qa}})^2 $. \tomas{Here, $\Delta_{qc}$ and $\Delta_{qa}$ are the detunings of the qubit with respect to cavity and ancilla, respectively.} For a more precise computation of the Purcell-limited $T_1$ in Fig.~\ref{fig:coherence}, we numerically diagonalize the total Hamiltonian as described in Appendix \ref{general_diagonalization} and then compute the Purcell rate as $\Gamma_P = \kappa_c \vert \bra{\psi_g} \hat{c} \ket{\psi_e} \vert^2 + \kappa_a \vert \bra{\psi_g} \hat{a} \ket{\psi_e} \vert^2 $ where $\ket{\psi_g}$ and $\ket{\psi_e}$ are the dressed eigenstates of the system corresponding to the ground and excited state of the qubit, respectively. The red diamond points in Fig.~\ref{fig:coherence} only consider imperfections from the asymmetry in \tomas{the Josephson energy of the junctions}, the star points only consider the misalignment \tomas{between cavity and qubit}, and the circle points consider both imperfections.

\section{System characterization}\label{Sec_calibration}

\tomas{In this Appendix, we detail the spectroscopic methods we used to experimentally characterize all the parameters of our system.  First, in Sec.~\ref{general_diagonalization} we give details on the numerical diagonalization used to fit the spectroscopic data valid at any value of the applied flux $\Phi$. Then, in Sec.~\ref{spectroscopy} we show the results of the single- and two-tone spectroscopy, allowing us to characterize the resonance frequencies of the system. In Sec.~\ref{polariton_tunability} we extract the ancilla-cavity coupling $g_{ac}$ and the flux dependence of the cross-Kerr couplings between qubit and polariton modes $\chi_j$. Finally, in Sec.~\ref{app:circuit_parameters} we summarize all the parameters of our experimental setup.} 

\subsection{Numerical diagonalization of the Hamiltonian valid at all flux}
\label{general_diagonalization}

\tomas{The theoretical model discussed in the main text and in Appendix~\ref{Numerical_model} accounts for the full interaction between the transmon molecule and the microwave cavity mode, but it is restricted to integer values of the flux only $\Phi=n\Phi_0$. Nevertheless, a complete spectroscopy of the system requires studying the transition frequencies and couplings of the system as a function of all possible values of the flux, including non-integer fluxes $\Phi\neq n\Phi_0$.}

\begin{figure}
\includegraphics[width=8.5cm]{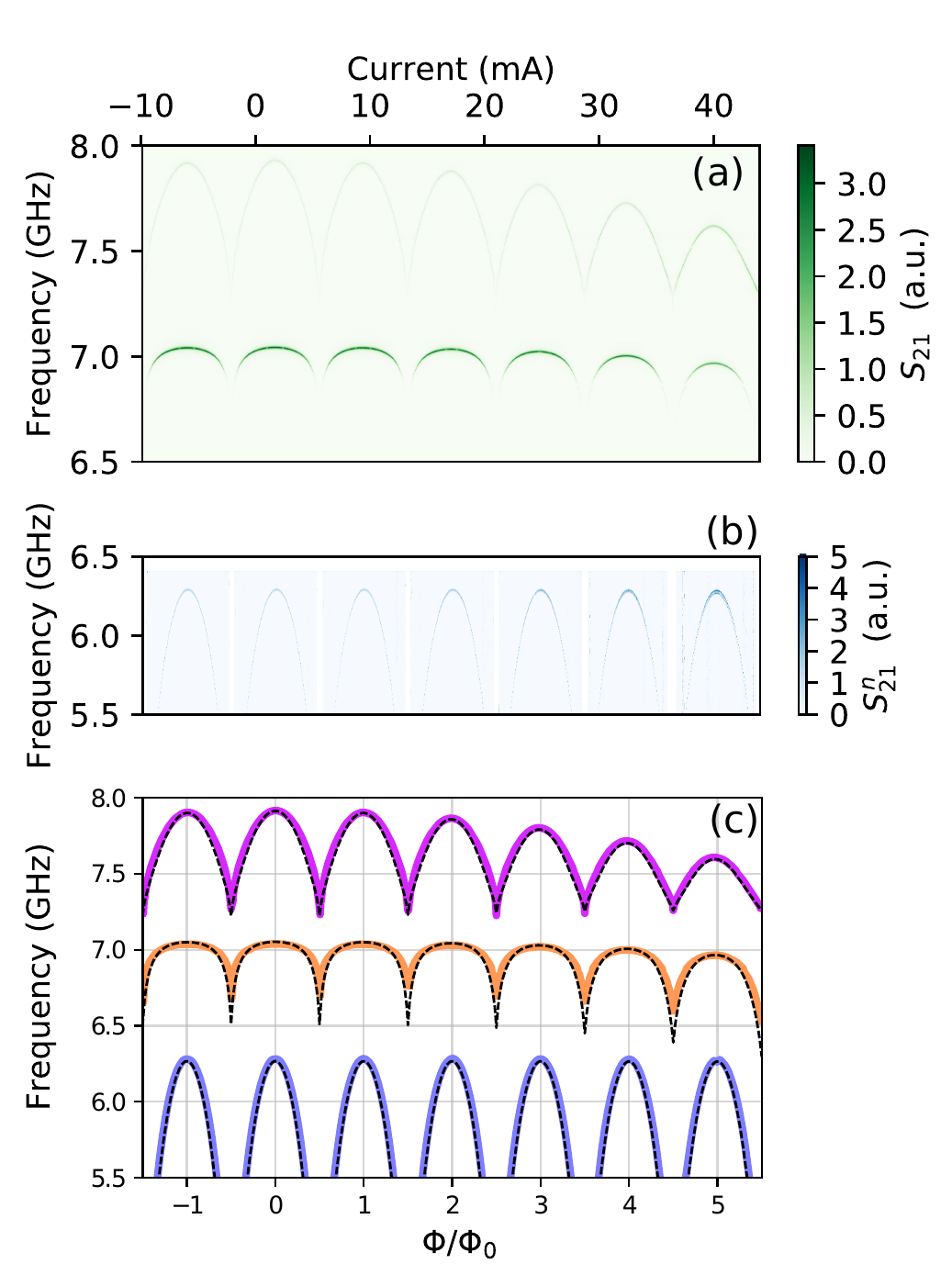}
\caption{
(a) Single-tone transmission $S_{21}$ measurements as function of \tomas{driving} frequency and flux (coil current). 
(b) Two-tone measurement, where the corresponding transmission $S_{21}^n$ is normalized by its value without second tone. \remy{(c)} Extracted resonan\tomas{ce} frequencies \tomas{of qubit $\omega_q'$ (in blue), lower polariton $\bar{\omega}_l$ (in orange), and upper polariton $\bar{\omega}_u$ (in purple) as a funtion of the applied flux $\Phi/\Phi_0$. The dashed black lines correspond to the theoretical predictions from the numerical diagonalization of the circuit model in Sec.~\ref{general_diagonalization}.}}
\label{Spectroscopy_polariton&Qubit}
\end{figure}

A theoretical model of the system at all flux is obtained by the total Hamiltonian $\hat{H}_{\rm tot} = \hat{H}_{\rm mol} + \hat{H}_{\rm cav}$, where $\hat{H}_{\rm mol}$ corresponds to the general circuit Hamiltonian \tomas{in Eq.~(\ref{Eq:VHamcosFULL})} and $\hat{H}_{\rm cav}=\hbar\omega_c \hat{c}^\dag \hat{c} + \hbar g_{ac} (\hat{a}^\dag \hat{c} + \hat{c}^\dag \hat{a})$ is the standard Hamiltonian including cavity and coupling. \tomas{When expanding the Hamiltonian (\ref{Eq:VHamcosFULL}) up to fourth order in $\hat{\varphi}_q,\hat{\varphi}_q\ll 1$, additional coupling terms appear on order $\sim\hat{\varphi}_q\hat{\varphi}_a$, $\sim\hat{\varphi}_a^3$, and $\sim\hat{\varphi}_q^2 \hat{\varphi}_a$ due to the non-integer values of the flux $\Phi\neq n\Phi_0$, and due to asymmetries in the Josephson junctions \cite{PRL_Lecocq_2011}.} Anyways, we numerically diagonalize this general Hamiltonian in the number representation using $8$ states in qubit, ancilla, and cavity, and for different values of the applied flux $\Phi$. \tomas{The results are used below to fit the single- and two-tone spectroscopy measurements shown in Fig.~\ref{Spectroscopy_polariton&Qubit}(c). Notice that} around frustration points, \tomas{where the low flux expansion of the Hamiltonian} becomes less valid, the predicted eigenenergies are still fitted within \SI{2}{\percent} errors.

\subsection{Qubit-polaritons spectroscopy}\label{spectroscopy}

Fig.~\ref{Spectroscopy_polariton&Qubit}(a) presents the single tone spectroscopy performed by measuring the cavity transmission versus magnetic flux $\Phi$ and \tomas{driving} frequency. The two resonant polariton modes are observed as two maximal transmission peak that strongly vary with $\Phi$. It demonstrates a direct coupling to the traveling microwave signal. The bare cavity resonant frequency $\omega_\text{c}/2\pi =$ \SI{7.169}{\giga\hertz} of the fundamental mode has been measured at \SI{4}{\kelvin} but it is no longer visible at this frequency. Indeed because of its strong hybridization with the ancilla mode, the cavity is now split into the two polariton modes. From the cavity they inherit their direct coupling to traveling microwave signal and from the ancilla they get a flux dependence. The two polariton frequencies vary rapidly in flux with a period given by flux quantization in the large circuit loop. In addition a slow variation is superimposed and this affects differently to the two modes. 

The two polariton modes present a non linear response inherited from the ancilla anharmonicity. When the input microwave power is large, the polariton dynamics shows a bi-stability behaviour. This regime is beyond the scope of this article \tomas{and will be treated elsewhere. Here,} we focus on the linear regime \tomas{of low input power}.

No qubit resonance is directly detected via single-tone spectroscopy. Therefore two-tone spectroscopy is needed to reveal it. One tone is swept between \SI{5.5}{\giga\hertz} and \SI{6.4}{\giga\hertz} in the vicinity of the qubit resonance. The second tone measures the transmission signal at the resonant frequency of one of the polariton modes. This two-tone spectroscopy reveals the qubit flux dependence [cf. Fig. \ref{Spectroscopy_polariton&Qubit}b]. We observed a flux dependence periodic in $\Phi$ but without any superimposed slow variation.
 
\tomas{We extract the resonance frequencies of the two polariton modes and the qubit from the single- and two-tone spectroscopy and we plot the results in} Fig. \ref{Spectroscopy_polariton&Qubit}(c) as function of flux $\Phi$. They are well fitted by the numerical model \tomas{discussed in Sec.~\ref{general_diagonalization}}, which nicely describes the flux variation of the \tomas{resonance frequencies of the} qubit \tomas{and} the two polariton modes. \remy{Using two-tone spectroscopy with an increasing Rabi drive to observe the two-photon transition from ground to second-excited state \cite{Schreier2008}, we extracted} the qubit anharmonicity to be $\alpha_q/2\pi =\tomas{-}$ \SI{88}{\mega\hertz}. 

\subsection{Polaritons tunability}\label{polariton_tunability}

Interestingly, the different flux working points allow to tune the ancilla-cavity hybridization angle without affecting the qubit frequency [cf. Fig. \ref{Spectroscopy_polariton&Qubit}]. Therefore, we can tune in-situ the parameters $\bar{\omega}_j$ and $\chi_j$, \tomas{which determine the Hamiltonian of our system} in Eq.~(\ref{Hamiltonian_Polariton}). 

\begin{figure}
	\includegraphics[width=7.5cm]{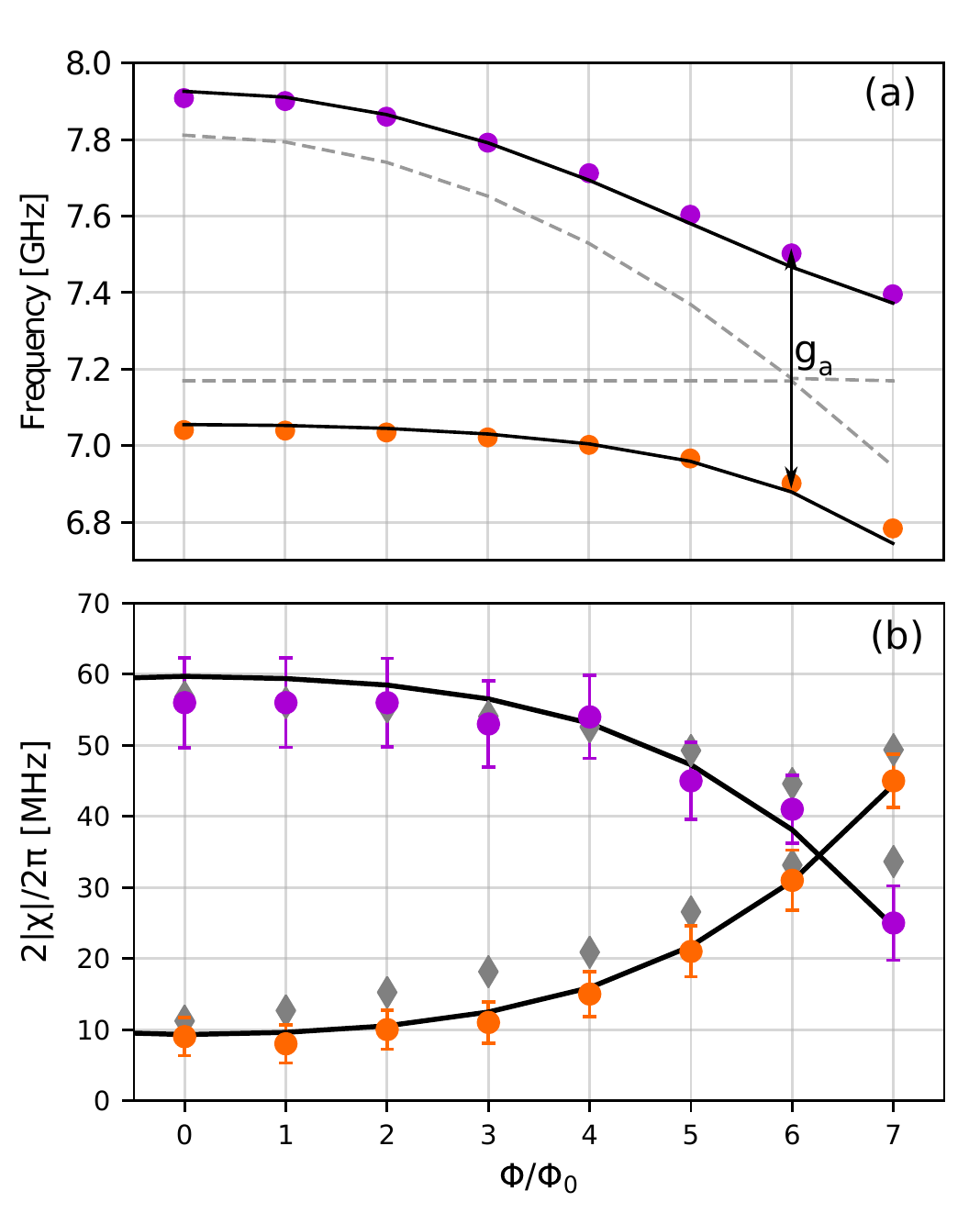}
	\caption{(a) The lower (orange) and upper (purple) polariton resonant frequencies as function of integer quantum flux. They are fitted (black lines) using the numerical model discussed in Appendix \ref{model_allFlux}. The grey dashed lines correspond to the bare cavity and bare ancilla frequencies. An avoided crossing between ancilla and cavity can thus be seen. (b) Cross-Kerr strengths between qubit and lower (orange) and upper (purple) polaritons. Black lines are the expected cross-Kerr coupling using $\chi_{l}= \tomas{-}g_{zz} \sin^2(\theta)$ and $\chi_{u}= \tomas{-}g_{zz} \cos^2(\theta)$ with $g_{zz}/(2\pi)=$ \tomas{34.5} MHz. The grey diamonds are simulated points computed using Black Box Quantization \cite{PRL_Nigg_2012} with EM simulation.}
	\label{polariton_modes}
\end{figure}

In Fig.~\ref{polariton_modes}(a), the two polaritons resonance frequencies are plotted versus the integer flux quantum $n$. They are quantitatively described by the lower and upper polariton modes $\hat{c}_l$ and $\hat{c}_u$ previously discussed. Here we set the cavity frequency to the value measured at $4\ \text{K}$ and the ancilla \tomas{frequency, when the qubit is prepared in the ground state $\bar{\omega}_a=\omega_a'+g_{zz}$,} is extracted from the expression $\bar{\omega}_a=\bar{\omega}_l + \bar{\omega}_u-\omega_c$. On resonance ($\bar{\omega}_a=\omega_c$), the two polaritons are maximally hybridized. We measure $g_{a\tomas{c}}/2\pi=$ \SI{295}{\mega\hertz} from the anti-level crossing. The hybridization weights $ \sin^2(\bar{\theta})$ and $\cos^2(\bar{\theta})$ between cavity and ancilla are then fitted. At zero flux, the upper polariton mode is mainly ancilla-like while the lower polariton is mainly cavity-like. When the cavity and ancilla are resonant, the hybridization weight is \SI{50}{\percent}. The large value of ancilla-cavity transverse coupling $g_{ac}$ has been designed in order to insure a strong hybridization over a large flux window.

Each polariton resonance is shifted by the cross-Kerr coupling strength $2\chi_j$ conditioned on the qubit state. The cross-Kerr coupling between the qubit and the two polariton modes are plotted in Fig.~\ref{polariton_modes}(b) as a function of integer flux quantum. A single tone spectroscopy is performed around the polariton resonances \tomas{$\bar{\omega}_{l}(n,\langle \hat{\sigma}_z\rangle)$ and $\bar{\omega}_{u}(n,\langle \hat{\sigma}_z\rangle)$ -- which differ for each polariton, for each value of flux $n$, and for each qubit occupation $\langle \hat{\sigma}_z\rangle$ (depending if the $\pi$-pulse is applied or not).}  
Because of relaxation, these experiments are performed in the time domain with a \SI{30}{\nano\second} $\pi$-pulse immediately followed by a \SI{500}{\nano\second} readout pulse. The cross-Kerr coupling is quantitatively described by $2\chi_j$ as predicted by the effective polariton model. We measured large readout shifts $(2\chi_j)/(2\pi)$ from \tomas{-}$9$ to \tomas{-}\SI{57}{\mega\hertz} thanks to the non-perturbative cross-Kerr coupling. These readout shifts are neither limited by the validity of the dispersive approximation nor by the multi-level aspects of the transmon. For instance, in Ref.~\cite{PRApplied_Walter_2017} the effective coupling for readout has been optimized and is reported to be \tomas{$\chi_{\rm d}=\frac{\alpha_{q}{g_x}^2 }{\Delta(\Delta+\alpha_{q})}=-2\pi\cdot$ \SI{7.9}{\mega\hertz}.} This is on the order or below of what we can achieve with the present setup without doing an intense optimization of our parameters. Interestingly, at zero flux, the upper polariton, which is further detuned from the qubit than the lower polariton, has a stronger readout shift than the lower polariton.\\

\subsection{Circuit parameters}\label{app:circuit_parameters}

\tomas{In the following we summarize how we experimentally determine all the parameters of our setup. All the resulting quantities are displayed in Tables~\ref{Tab_freq}-\ref{Tab_circuit}.}

\remy{First, the mode frequencies $\omega_q'$, $\bar{\omega}_l$ and $\bar{\omega}_u$} are obtained from spectroscopies at different applied fluxes, at a temperature of \SI{20}{\milli\kelvin}, \tomas{and with the qubit prepared in the ground state} [cf.~Fig.~\ref{Spectroscopy_polariton&Qubit}]. \tomas{On the other hand, the cavity frequency $\omega_c$ is obtained from spectroscopy at \SI{4}{\kelvin}. \tomas{From these quantities we determine the ancilla frequency, when the qubit is prepared in the ground state $\bar{\omega}_a=\omega_a'+g_{zz}$, using the formula $\bar{\omega}_a = \bar{\omega}_l + \bar{\omega}_u - \omega_c$. In the first row of Table \ref{Tab_freq} we show the values of these frequencies at zero flux.} In addition, the ancilla-cavity coupling $g_{ac}$ is fitted from the spectroscopy of the polariton resonances at different flux [cf.~Fig.~\ref{Spectroscopy_polariton&Qubit} and Fig.~\ref{polariton_modes}(a)]. The polariton cross-Kerr couplings $\chi_l$ and $\chi_u$ are measured directly from the two-tone spectroscopy for given flux as shown in Fig.~\ref{polariton_modes}(b), and the ancilla-qubit cross-Kerr coupling $g_{zz}$ is fitted from the global flux dependence of this plot. The qubit anharmonicity $\alpha_q$ is measured using standard methods of two-tone spectroscopy \remy{with an increasing rabi drive to observe the two-photon transition from ground to second-excited state \cite{Schreier2008}}. Finally, the ancilla anharmonicity is estimated as $U_a=g_{zz}^2/\alpha_q$, according to the circuit model in Appendix~\ref{HamiltonianNumberRep}. The values of all the above quantities at zero flux are shown in the second row of Table \ref{Tab_freq}.} 

In Table \ref{Tab_lifetime}, \tomas{we detail the coherence times and decay of the various modes at zero flux. We measure the polariton decay rates $\kappa_l$ and $\kappa_u$ from the widths of the polariton resonances at \SI{20}{\milli\kelvin} [cf.~Eq.~\ref{expectationC}]. Subsequently, we determine the cavity and ancilla decay, $\kappa_c$ and $\kappa_a$, from the hybridization angle $\bar{\theta}=(1/2){\rm Arctan}(2g_{ac}/[\bar{\omega}_a-\bar{\omega}_c])$ and the inverse relations $\kappa_c =-\frac{\sin^2(\theta)}{\cos(2\theta)}\kappa_u + \frac{\cos^2(\theta)}{\cos(2\theta)}\kappa_l$ and $\kappa_a =\frac{\cos^2(\theta)}{\cos(2\theta)}\kappa_u - \frac{\sin^2(\theta)}{\cos(2\theta)}\kappa_l$. The results are shown in Table.~\ref{Tab_lifetime}. To have direct access to the cavity decay (without hybridization into polaritons) we also performed transmission spectroscopy at \SI{4.2}{\kelvin}. Indeed, at this temperature, the aluminium of the transmon molecule circuit is not superconducting. From the resonance width we obtained $\kappa_c^{\rm 4K}/(2\pi)= 19.6 {\rm MHz}$, which is slightly larger than reported in Table.~\ref{Tab_lifetime} at \SI{20}{\milli\kelvin}, probably due to extra losses in the metal and the dielectric. Finally, we measured the qubit decay time $T_1$ and dephasing time $T_2$ at \SI{20}{\milli\kelvin} via relaxation and Ramsey experiments, respectively.} 

\tomas{The ancilla frequency and decay depend strongly on flux because of the SQUIDs. Therefore, in Table \ref{table5flux} we state the corresponding values at non-zero flux, $\Phi=5\Phi_0$, which we use in the theoretical prediction of Fig.~\ref{Transmission_Measurements} All the rest of the parameters are the same as in Tables \ref{Tab_freq} and \ref{Tab_lifetime}.} 

In Table~\ref{Tab_circuit} we display the microscopic parameters describing the transmon molecule circuit. \remy{The asymmetry $d_J$ is measured from room temperature resistance measurement [cf.~Appendix \ref{Asymmetry}]}. \tomas{All the other parameters are derived using the expressions from the circuit model in Appendices \ref{model_allFlux} and \ref{HamiltonianNumberRep}, which relate the circuit parameters to the measurable frequencies, anharmonicities, and couplings in Table \ref{Tab_freq}. Explicitly, we use the formulas: $E_{C_q} = -\hbar\alpha_q$, $\tilde{\omega}_q=\omega_q'-\alpha_q + g_{zz}$, $E_J = (\hbar\tilde{\omega}_q)^2/(16E_{C_q})$, $C_q = e^2/(2E_{C_q})$, $C_S = C_q/2$, $L_J = (\phi_0)^2/E_J$, $\tilde{\omega}_a=\bar{\omega}_a-U_a + g_{zz}$, $E_{C_a} = \sqrt{-\hbar U_a (\hbar\tilde{\omega}_a)^2/(16E_J)}$, $L_a = 2L_J/[E_{C_a}/(-\hbar U_a) -1 ]$, $C_a = e^2/(2E_{C_a})$, $C_t = (C_a-2C_S)/4$, and the critical current of the Josephson junctions read $I_C = E_J/\phi_0$. The resulting values are shown in Table~\ref{Tab_circuit} with $3$ significant digits. They are consistent with the parameters obtained from the numerical fit of Fig. \ref{Spectroscopy_polariton&Qubit}c and also from estimations based on HFSS simulation and room temperature resistance measurements of the transmon Josephson junctions and SQUIDs chain.}

\begin{table}[b!]
	\begin{center}
		\begin{tabular}{c c c c c c c}
			\hline
			$\omega_{q}\tomas{'}/2\pi$ & 	$\bar{\omega}_{a}/2\pi$  & 	$\omega_{c}/2\pi$  & 	$\bar{\omega}_{l}/2\pi$  & 	$\bar{\omega}_{u}/2\pi$  & {} &\multirow{2}{*}{(\SI{}{\giga\hertz})} \\
			6.284 & 7.780 & 7.169 & 7.038 & 7.911 &\\
			\hline 
		\end{tabular} 
	\\
	\begin{tabular}{c c c c c c c}
		$g_{zz}/2\pi$ & 	$g_{a\tomas{c}}/2\pi$  & 	$\tomas{\chi_l}/2\pi$  & 	$\tomas{\chi_u}/2\pi$ & $\alpha_q/2\pi$& \tomas{$U_a/2\pi$} &\multirow{2}{*}{(\SI{}{\mega\hertz})}  \\
		34.5 & 295 & \tomas{-}4.5 & \tomas{-}28.5& \tomas{-}88 & \tomas{-13.5} &\\
		\hline 
	\end{tabular} 
		\caption{\tomas{Transition} frequencies, \tomas{anharmonicities, and coupling} strengths at zero flux \tomas{$\Phi=0$ and $T=20 {\rm mK}$}.}
		\label{Tab_freq}
	\end{center}  
\end{table} 

\begin{table}[b!]
	\begin{center}
			\begin{tabular}{c c c c c c c}
				\hline
				 $T_1$ & $T_2$ & $\kappa_l/2\pi$ & $\kappa_u/2\pi$& $\kappa_c/2\pi$ & $\tomas{\kappa_a}/2\pi$ & $\tomas{\bar{\theta}}$  \\ 
				 3.3 \SI{}{\micro\second} & 3.2 \SI{}{\micro\second}& \tomas{11.8}\SI{}{\mega\hertz} & \tomas{7.1}\SI{}{\mega\hertz} & \tomas{12.7}\SI{}{\mega\hertz} & \tomas{6.2}\SI{}{\mega\hertz} & \tomas{\SI{0.384}{\radian}}\\
				\hline 
		\end{tabular} 
		\caption{Coherence \tomas{times} and decay \tomas{rates} \tomas{at zero flux $\Phi=0$ and $T=20 {\rm mK}$.}}
		\label{Tab_lifetime}
	\end{center}
\end{table} 

\begin{table}[b!]
	\begin{center}
		\begin{tabular}{c c c c c}
			\hline
		$\tomas{\bar{\omega}_a/2\pi}$ & $\remy{\bar{\omega}_l/2\pi}$& $\remy{\bar{\omega}_u/2\pi}$& $\tomas{\chi_l/2\pi}$ & $\tomas{\chi_u/2\pi}$ \\
			$\tomas{7.396 {\rm GHz}}$ & \tomas{6.966 {\rm GHz}} & \tomas{7.599 {\rm GHz}}& $\tomas{-11.1 {\rm MHz}}$ & $\tomas{-23.4 {\rm MHz}}$\\
			\hline 
		\end{tabular} 
	\\
	\begin{tabular}{c c c c}
		$\tomas{\kappa_a/2\pi}$ & $\tomas{\kappa_l/2\pi}$ & $\tomas{\kappa_u/2\pi}$ & $\tomas{\bar{\theta}}$  \\
		 $\tomas{11.2 {\rm MHz}}$ & $\tomas{12.1 {\rm MHz}}$ & $\tomas{11.6 {\rm MHz}}$ & $\tomas{\SI{0.602}{\radian}}$ \\
		\hline 
	\end{tabular} 
		\caption{\tomas{Parameters at non-zero flux $\Phi=5\Phi_0$ and $T=20 {\rm mK}$. All the rest of the parameters do not strongly depend on flux and they are the same as in Tables \ref{Tab_freq} and \ref{Tab_lifetime}}.}
		\label{table5flux}
	\end{center}  
\end{table} 

\begin{table}[b!]
	\begin{center}
	\remy{
			\begin{tabular}{c c c c c c}
				\hline
				 $I_C$ (\SI{}{\nano\ampere})& 
				 \tomas{$L_J$ (\SI{}{\nano\henry})}&
				 $L_a$ (\SI{}{\nano\henry})& $C_S$ (\SI{}{\femto\farad})& $C_t$ (\SI{}{\femto\farad})& $d_J$ (\SI{}{\percent})\\ 
				 58.\tomas{6} &
				 \tomas{5.63}
				 &\tomas{5.32} & \tomas{110} & \tomas{59.6} & 1.3 \\
				\hline 
		\end{tabular} 
		\\
			\begin{tabular}{c c c c}
		$E_J/(2\pi\hbar)$ & 	$E_{Cq}/(2\pi\hbar)$  & 	$E_{Ca}/(2\pi\hbar)$  & \multirow{2}{*}{\tomas{(\SI{}{\mega\hertz})}}  \\
		29\tomas{200} & \tomas{88} & \tomas{42.2} & \\
		\hline 
	\end{tabular} 
		\caption{Circuit parameters at zero flux $\Phi=0$ and $T=20 {\rm mK}$.} 
		\label{Tab_circuit} }
	\end{center}
\end{table} 

\bibliographystyle{apsrev4-1}

\begin{thebibliography}{58}%
	\makeatletter
	\providecommand \@ifxundefined [1]{%
		\@ifx{#1\undefined}
	}%
	\providecommand \@ifnum [1]{%
		\ifnum #1\expandafter \@firstoftwo
		\else \expandafter \@secondoftwo
		\fi
	}%
	\providecommand \@ifx [1]{%
		\ifx #1\expandafter \@firstoftwo
		\else \expandafter \@secondoftwo
		\fi
	}%
	\providecommand \natexlab [1]{#1}%
	\providecommand \enquote  [1]{``#1''}%
	\providecommand \bibnamefont  [1]{#1}%
	\providecommand \bibfnamefont [1]{#1}%
	\providecommand \citenamefont [1]{#1}%
	\providecommand \href@noop [0]{\@secondoftwo}%
	\providecommand \href [0]{\begingroup \@sanitize@url \@href}%
	\providecommand \@href[1]{\@@startlink{#1}\@@href}%
	\providecommand \@@href[1]{\endgroup#1\@@endlink}%
	\providecommand \@sanitize@url [0]{\catcode `\\12\catcode `\$12\catcode
		`\&12\catcode `\#12\catcode `\^12\catcode `\_12\catcode `\%12\relax}%
	\providecommand \@@startlink[1]{}%
	\providecommand \@@endlink[0]{}%
	\providecommand \url  [0]{\begingroup\@sanitize@url \@url }%
	\providecommand \@url [1]{\endgroup\@href {#1}{\urlprefix }}%
	\providecommand \urlprefix  [0]{URL }%
	\providecommand \Eprint [0]{\href }%
	\providecommand \doibase [0]{http://dx.doi.org/}%
	\providecommand \selectlanguage [0]{\@gobble}%
	\providecommand \bibinfo  [0]{\@secondoftwo}%
	\providecommand \bibfield  [0]{\@secondoftwo}%
	\providecommand \translation [1]{[#1]}%
	\providecommand \BibitemOpen [0]{}%
	\providecommand \bibitemStop [0]{}%
	\providecommand \bibitemNoStop [0]{.\EOS\space}%
	\providecommand \EOS [0]{\spacefactor3000\relax}%
	\providecommand \BibitemShut  [1]{\csname bibitem#1\endcsname}%
	\let\auto@bib@innerbib\@empty
	\bibitem [{\citenamefont {Preskill}(2018)}]{Preskill_2018}%
	\BibitemOpen
	\bibfield  {author} {\bibinfo {author} {\bibfnamefont {John}\ \bibnamefont
			{Preskill}},\ }\bibfield  {title} {\enquote {\bibinfo {title} {Quantum
				{C}omputing in the {NISQ} era and beyond},}\ }\href {\doibase
		10.22331/q-2018-08-06-79} {\bibfield  {journal} {\bibinfo  {journal}
			{{Quantum}}\ }\textbf {\bibinfo {volume} {2}},\ \bibinfo {pages} {79}
		(\bibinfo {year} {2018})}\BibitemShut {NoStop}%
	\bibitem [{\citenamefont {Li}\ and\ \citenamefont
		{Benjamin}(2017)}]{Li_PRX_2017}%
	\BibitemOpen
	\bibfield  {author} {\bibinfo {author} {\bibfnamefont {Ying}\ \bibnamefont
			{Li}}\ and\ \bibinfo {author} {\bibfnamefont {Simon~C.}\ \bibnamefont
			{Benjamin}},\ }\bibfield  {title} {\enquote {\bibinfo {title} {Efficient
				variational quantum simulator incorporating active error minimization},}\
	}\href {\doibase 10.1103/PhysRevX.7.021050} {\bibfield  {journal} {\bibinfo
			{journal} {Phys. Rev. X}\ }\textbf {\bibinfo {volume} {7}},\ \bibinfo {pages}
		{021050} (\bibinfo {year} {2017})}\BibitemShut {NoStop}%
	\bibitem [{\citenamefont {Knill}\ \emph {et~al.}(2008)\citenamefont {Knill},
		\citenamefont {Leibfried}, \citenamefont {Reichle}, \citenamefont {Britton},
		\citenamefont {Blakestad}, \citenamefont {Jost}, \citenamefont {Langer},
		\citenamefont {Ozeri}, \citenamefont {Seidelin},\ and\ \citenamefont
		{Wineland}}]{Knill_PRA_2008}%
	\BibitemOpen
	\bibfield  {author} {\bibinfo {author} {\bibfnamefont {E.}~\bibnamefont
			{Knill}}, \bibinfo {author} {\bibfnamefont {D.}~\bibnamefont {Leibfried}},
		\bibinfo {author} {\bibfnamefont {R.}~\bibnamefont {Reichle}}, \bibinfo
		{author} {\bibfnamefont {J.}~\bibnamefont {Britton}}, \bibinfo {author}
		{\bibfnamefont {R.~B.}\ \bibnamefont {Blakestad}}, \bibinfo {author}
		{\bibfnamefont {J.~D.}\ \bibnamefont {Jost}}, \bibinfo {author}
		{\bibfnamefont {C.}~\bibnamefont {Langer}}, \bibinfo {author} {\bibfnamefont
			{R.}~\bibnamefont {Ozeri}}, \bibinfo {author} {\bibfnamefont
			{S.}~\bibnamefont {Seidelin}}, \ and\ \bibinfo {author} {\bibfnamefont
			{D.~J.}\ \bibnamefont {Wineland}},\ }\bibfield  {title} {\enquote {\bibinfo
			{title} {Randomized benchmarking of quantum gates},}\ }\href {\doibase
		10.1103/PhysRevA.77.012307} {\bibfield  {journal} {\bibinfo  {journal} {Phys.
				Rev. A}\ }\textbf {\bibinfo {volume} {77}},\ \bibinfo {pages} {012307}
		(\bibinfo {year} {2008})}\BibitemShut {NoStop}%
	\bibitem [{\citenamefont {{Divincenzo}}(2000)}]{DiVincenzo_FdP_2000}%
	\BibitemOpen
	\bibfield  {author} {\bibinfo {author} {\bibfnamefont {David~P.}\
			\bibnamefont {{Divincenzo}}},\ }\bibfield  {title} {\enquote {\bibinfo
			{title} {{The Physical Implementation of Quantum Computation}},}\ }\href
	{\doibase 10.1002/1521-3978(200009)48:9/11<771::AID-PROP771>3.0.CO;2-E}
	{\bibfield  {journal} {\bibinfo  {journal} {Fortschritte der Physik}\
		}\textbf {\bibinfo {volume} {48}},\ \bibinfo {pages} {771--783} (\bibinfo
		{year} {2000})},\ \Eprint {http://arxiv.org/abs/quant-ph/0002077}
	{arXiv:quant-ph/0002077 [quant-ph]} \BibitemShut {NoStop}%
	\bibitem [{\citenamefont {{Kelly}}\ \emph {et~al.}(2015)\citenamefont
		{{Kelly}}, \citenamefont {{Barends}}, \citenamefont {{Fowler}}, \citenamefont
		{{Megrant}}, \citenamefont {{Jeffrey}}, \citenamefont {{White}},
		\citenamefont {{Sank}}, \citenamefont {{Mutus}}, \citenamefont {{Campbell}},
		\citenamefont {{Chen}}, \citenamefont {{Chen}}, \citenamefont {{Chiaro}},
		\citenamefont {{Dunsworth}}, \citenamefont {{Hoi}}, \citenamefont {{Neill}},
		\citenamefont {{O'Malley}}, \citenamefont {{Quintana}}, \citenamefont
		{{Roushan}}, \citenamefont {{Vainsencher}}, \citenamefont {{Wenner}},
		\citenamefont {{Cleland}},\ and\ \citenamefont
		{{Martinis}}}]{Kelly_Nature_2015}%
	\BibitemOpen
	\bibfield  {author} {\bibinfo {author} {\bibfnamefont {J.}~\bibnamefont
			{{Kelly}}}, \bibinfo {author} {\bibfnamefont {R.}~\bibnamefont {{Barends}}},
		\bibinfo {author} {\bibfnamefont {A.~G.}\ \bibnamefont {{Fowler}}}, \bibinfo
		{author} {\bibfnamefont {A.}~\bibnamefont {{Megrant}}}, \bibinfo {author}
		{\bibfnamefont {E.}~\bibnamefont {{Jeffrey}}}, \bibinfo {author}
		{\bibfnamefont {T.~C.}\ \bibnamefont {{White}}}, \bibinfo {author}
		{\bibfnamefont {D.}~\bibnamefont {{Sank}}}, \bibinfo {author} {\bibfnamefont
			{J.~Y.}\ \bibnamefont {{Mutus}}}, \bibinfo {author} {\bibfnamefont
			{B.}~\bibnamefont {{Campbell}}}, \bibinfo {author} {\bibfnamefont
			{Y.}~\bibnamefont {{Chen}}}, \bibinfo {author} {\bibfnamefont
			{Z.}~\bibnamefont {{Chen}}}, \bibinfo {author} {\bibfnamefont
			{B.}~\bibnamefont {{Chiaro}}}, \bibinfo {author} {\bibfnamefont
			{A.}~\bibnamefont {{Dunsworth}}}, \bibinfo {author} {\bibfnamefont {I.-C.}\
			\bibnamefont {{Hoi}}}, \bibinfo {author} {\bibfnamefont {C.}~\bibnamefont
			{{Neill}}}, \bibinfo {author} {\bibfnamefont {P.~J.~J.}\ \bibnamefont
			{{O'Malley}}}, \bibinfo {author} {\bibfnamefont {C.}~\bibnamefont
			{{Quintana}}}, \bibinfo {author} {\bibfnamefont {P.}~\bibnamefont
			{{Roushan}}}, \bibinfo {author} {\bibfnamefont {A.}~\bibnamefont
			{{Vainsencher}}}, \bibinfo {author} {\bibfnamefont {J.}~\bibnamefont
			{{Wenner}}}, \bibinfo {author} {\bibfnamefont {A.~N.}\ \bibnamefont
			{{Cleland}}}, \ and\ \bibinfo {author} {\bibfnamefont {J.~M.}\ \bibnamefont
			{{Martinis}}},\ }\bibfield  {title} {\enquote {\bibinfo {title} {{State
					preservation by repetitive error detection in a superconducting quantum
					circuit}},}\ }\href {\doibase 10.1038/nature14270} {\bibfield  {journal}
		{\bibinfo  {journal} {Nature}\ }\textbf {\bibinfo {volume} {519}},\ \bibinfo
		{pages} {66--69} (\bibinfo {year} {2015})}\BibitemShut {NoStop}%
	\bibitem [{\citenamefont {Schindler}\ \emph {et~al.}(2011)\citenamefont
		{Schindler}, \citenamefont {Barreiro}, \citenamefont {Monz}, \citenamefont
		{Nebendahl}, \citenamefont {Nigg}, \citenamefont {Chwalla}, \citenamefont
		{Hennrich},\ and\ \citenamefont {Blatt}}]{Schindler_Science_2011}%
	\BibitemOpen
	\bibfield  {author} {\bibinfo {author} {\bibfnamefont {Philipp}\ \bibnamefont
			{Schindler}}, \bibinfo {author} {\bibfnamefont {Julio~T.}\ \bibnamefont
			{Barreiro}}, \bibinfo {author} {\bibfnamefont {Thomas}\ \bibnamefont {Monz}},
		\bibinfo {author} {\bibfnamefont {Volckmar}\ \bibnamefont {Nebendahl}},
		\bibinfo {author} {\bibfnamefont {Daniel}\ \bibnamefont {Nigg}}, \bibinfo
		{author} {\bibfnamefont {Michael}\ \bibnamefont {Chwalla}}, \bibinfo {author}
		{\bibfnamefont {Markus}\ \bibnamefont {Hennrich}}, \ and\ \bibinfo {author}
		{\bibfnamefont {Rainer}\ \bibnamefont {Blatt}},\ }\bibfield  {title}
	{\enquote {\bibinfo {title} {Experimental repetitive quantum error
				correction},}\ }\href {\doibase 10.1126/science.1203329} {\bibfield
		{journal} {\bibinfo  {journal} {Science}\ }\textbf {\bibinfo {volume}
			{332}},\ \bibinfo {pages} {1059--1061} (\bibinfo {year} {2011})}\BibitemShut
	{NoStop}%
	\bibitem [{\citenamefont {Bermudez}\ \emph {et~al.}(2017)\citenamefont
		{Bermudez}, \citenamefont {Xu}, \citenamefont {Nigmatullin}, \citenamefont
		{O'Gorman}, \citenamefont {Negnevitsky}, \citenamefont {Schindler},
		\citenamefont {Monz}, \citenamefont {Poschinger}, \citenamefont {Hempel},
		\citenamefont {Home}, \citenamefont {Schmidt-Kaler}, \citenamefont {Biercuk},
		\citenamefont {Blatt}, \citenamefont {Benjamin},\ and\ \citenamefont
		{M\"uller}}]{Bermudez_PRX_2017}%
	\BibitemOpen
	\bibfield  {author} {\bibinfo {author} {\bibfnamefont {A.}~\bibnamefont
			{Bermudez}}, \bibinfo {author} {\bibfnamefont {X.}~\bibnamefont {Xu}},
		\bibinfo {author} {\bibfnamefont {R.}~\bibnamefont {Nigmatullin}}, \bibinfo
		{author} {\bibfnamefont {J.}~\bibnamefont {O'Gorman}}, \bibinfo {author}
		{\bibfnamefont {V.}~\bibnamefont {Negnevitsky}}, \bibinfo {author}
		{\bibfnamefont {P.}~\bibnamefont {Schindler}}, \bibinfo {author}
		{\bibfnamefont {T.}~\bibnamefont {Monz}}, \bibinfo {author} {\bibfnamefont
			{U.~G.}\ \bibnamefont {Poschinger}}, \bibinfo {author} {\bibfnamefont
			{C.}~\bibnamefont {Hempel}}, \bibinfo {author} {\bibfnamefont
			{J.}~\bibnamefont {Home}}, \bibinfo {author} {\bibfnamefont {F.}~\bibnamefont
			{Schmidt-Kaler}}, \bibinfo {author} {\bibfnamefont {M.}~\bibnamefont
			{Biercuk}}, \bibinfo {author} {\bibfnamefont {R.}~\bibnamefont {Blatt}},
		\bibinfo {author} {\bibfnamefont {S.}~\bibnamefont {Benjamin}}, \ and\
		\bibinfo {author} {\bibfnamefont {M.}~\bibnamefont {M\"uller}},\ }\bibfield
	{title} {\enquote {\bibinfo {title} {Assessing the progress of trapped-ion
				processors towards fault-tolerant quantum computation},}\ }\href {\doibase
		10.1103/PhysRevX.7.041061} {\bibfield  {journal} {\bibinfo  {journal} {Phys.
				Rev. X}\ }\textbf {\bibinfo {volume} {7}},\ \bibinfo {pages} {041061}
		(\bibinfo {year} {2017})}\BibitemShut {NoStop}%
	\bibitem [{\citenamefont {Gambetta}\ \emph {et~al.}(2017)\citenamefont
		{Gambetta}, \citenamefont {Chow},\ and\ \citenamefont
		{Steffen}}]{Gambetta_njpQI_2017}%
	\BibitemOpen
	\bibfield  {author} {\bibinfo {author} {\bibfnamefont {Jay~M}\ \bibnamefont
			{Gambetta}}, \bibinfo {author} {\bibfnamefont {Jerry~M}\ \bibnamefont
			{Chow}}, \ and\ \bibinfo {author} {\bibfnamefont {Matthias}\ \bibnamefont
			{Steffen}},\ }\bibfield  {title} {\enquote {\bibinfo {title} {Building
				logical qubits in a superconducting quantum computing system},}\ }\href
	{\doibase 10.1038/s41534-016-0004-0} {\bibfield  {journal} {\bibinfo
			{journal} {npj Quantum Information}\ }\textbf {\bibinfo {volume} {3}},\
		\bibinfo {pages} {2} (\bibinfo {year} {2017})}\BibitemShut {NoStop}%
	\bibitem [{\citenamefont {Leibfried}\ \emph {et~al.}(2003)\citenamefont
		{Leibfried}, \citenamefont {Blatt}, \citenamefont {Monroe},\ and\
		\citenamefont {Wineland}}]{RMP_Leibfried_2003}%
	\BibitemOpen
	\bibfield  {author} {\bibinfo {author} {\bibfnamefont {D.}~\bibnamefont
			{Leibfried}}, \bibinfo {author} {\bibfnamefont {R.}~\bibnamefont {Blatt}},
		\bibinfo {author} {\bibfnamefont {C.}~\bibnamefont {Monroe}}, \ and\ \bibinfo
		{author} {\bibfnamefont {D.}~\bibnamefont {Wineland}},\ }\bibfield  {title}
	{\enquote {\bibinfo {title} {Quantum dynamics of single trapped ions},}\
	}\href {\doibase 10.1103/RevModPhys.75.281} {\bibfield  {journal} {\bibinfo
			{journal} {Rev. Mod. Phys.}\ }\textbf {\bibinfo {volume} {75}},\ \bibinfo
		{pages} {281--324} (\bibinfo {year} {2003})}\BibitemShut {NoStop}%
	\bibitem [{\citenamefont {Ballance}\ \emph {et~al.}(2016)\citenamefont
		{Ballance}, \citenamefont {Harty}, \citenamefont {Linke}, \citenamefont
		{Sepiol},\ and\ \citenamefont {Lucas}}]{Ballance_PRL_2016}%
	\BibitemOpen
	\bibfield  {author} {\bibinfo {author} {\bibfnamefont {C.~J.}\ \bibnamefont
			{Ballance}}, \bibinfo {author} {\bibfnamefont {T.~P.}\ \bibnamefont {Harty}},
		\bibinfo {author} {\bibfnamefont {N.~M.}\ \bibnamefont {Linke}}, \bibinfo
		{author} {\bibfnamefont {M.~A.}\ \bibnamefont {Sepiol}}, \ and\ \bibinfo
		{author} {\bibfnamefont {D.~M.}\ \bibnamefont {Lucas}},\ }\bibfield  {title}
	{\enquote {\bibinfo {title} {High-fidelity quantum logic gates using
				trapped-ion hyperfine qubits},}\ }\href {\doibase
		10.1103/PhysRevLett.117.060504} {\bibfield  {journal} {\bibinfo  {journal}
			{Phys. Rev. Lett.}\ }\textbf {\bibinfo {volume} {117}},\ \bibinfo {pages}
		{060504} (\bibinfo {year} {2016})}\BibitemShut {NoStop}%
	\bibitem [{\citenamefont {Blais}\ \emph {et~al.}(2004)\citenamefont {Blais},
		\citenamefont {Huang}, \citenamefont {Wallraff}, \citenamefont {Girvin},\
		and\ \citenamefont {Schoelkopf}}]{Blais_PRA_2004}%
	\BibitemOpen
	\bibfield  {author} {\bibinfo {author} {\bibfnamefont {Alexandre}\
			\bibnamefont {Blais}}, \bibinfo {author} {\bibfnamefont {Ren-Shou}\
			\bibnamefont {Huang}}, \bibinfo {author} {\bibfnamefont {Andreas}\
			\bibnamefont {Wallraff}}, \bibinfo {author} {\bibfnamefont {S.~M.}\
			\bibnamefont {Girvin}}, \ and\ \bibinfo {author} {\bibfnamefont {R.~J.}\
			\bibnamefont {Schoelkopf}},\ }\bibfield  {title} {\enquote {\bibinfo {title}
			{Cavity quantum electrodynamics for superconducting electrical circuits: An
				architecture for quantum computation},}\ }\href {\doibase
		10.1103/PhysRevA.69.062320} {\bibfield  {journal} {\bibinfo  {journal} {Phys.
				Rev. A}\ }\textbf {\bibinfo {volume} {69}},\ \bibinfo {pages} {062320}
		(\bibinfo {year} {2004})}\BibitemShut {NoStop}%
	\bibitem [{\citenamefont {Volz}\ \emph {et~al.}(2011)\citenamefont {Volz},
		\citenamefont {Gehr}, \citenamefont {Dubois}, \citenamefont {Est{\`e}ve},\
		and\ \citenamefont {Reichel}}]{Nature_Volz_2011}%
	\BibitemOpen
	\bibfield  {author} {\bibinfo {author} {\bibfnamefont {J{\"u}rgen}\
			\bibnamefont {Volz}}, \bibinfo {author} {\bibfnamefont {Roger}\ \bibnamefont
			{Gehr}}, \bibinfo {author} {\bibfnamefont {Guilhem}\ \bibnamefont {Dubois}},
		\bibinfo {author} {\bibfnamefont {J{\'e}r{\^o}me}\ \bibnamefont
			{Est{\`e}ve}}, \ and\ \bibinfo {author} {\bibfnamefont {Jakob}\ \bibnamefont
			{Reichel}},\ }\bibfield  {title} {\enquote {\bibinfo {title} {Measurement of
				the internal state of a single atom without energy exchange},}\ }\href
	{https://www.nature.com/articles/nature10225} {\bibfield  {journal} {\bibinfo
			{journal} {Nature}\ }\textbf {\bibinfo {volume} {475}},\ \bibinfo {pages}
		{210} (\bibinfo {year} {2011})}\BibitemShut {NoStop}%
	\bibitem [{\citenamefont {Haroche}\ and\ \citenamefont
		{Raimond}(2006)}]{Haroche&Raimond_2006}%
	\BibitemOpen
	\bibfield  {author} {\bibinfo {author} {\bibfnamefont {Serge}\ \bibnamefont
			{Haroche}}\ and\ \bibinfo {author} {\bibfnamefont {J.-M.}\ \bibnamefont
			{Raimond}},\ }\href@noop {} {\emph {\bibinfo {title} {Exploring the quantum:
				atoms, cavities, and photons}}}\ (\bibinfo  {publisher} {Oxford University
		Press},\ \bibinfo {address} {Oxford},\ \bibinfo {year} {2006})\BibitemShut
	{NoStop}%
	\bibitem [{\citenamefont {Walter}\ \emph {et~al.}(2017)\citenamefont {Walter},
		\citenamefont {Kurpiers}, \citenamefont {Gasparinetti}, \citenamefont
		{Magnard}, \citenamefont {Poto\ifmmode~\check{c}\else \v{c}\fi{}nik},
		\citenamefont {Salath\'e}, \citenamefont {Pechal}, \citenamefont {Mondal},
		\citenamefont {Oppliger}, \citenamefont {Eichler},\ and\ \citenamefont
		{Wallraff}}]{PRApplied_Walter_2017}%
	\BibitemOpen
	\bibfield  {author} {\bibinfo {author} {\bibfnamefont {T.}~\bibnamefont
			{Walter}}, \bibinfo {author} {\bibfnamefont {P.}~\bibnamefont {Kurpiers}},
		\bibinfo {author} {\bibfnamefont {S.}~\bibnamefont {Gasparinetti}}, \bibinfo
		{author} {\bibfnamefont {P.}~\bibnamefont {Magnard}}, \bibinfo {author}
		{\bibfnamefont {A.}~\bibnamefont {Poto\ifmmode~\check{c}\else
				\v{c}\fi{}nik}}, \bibinfo {author} {\bibfnamefont {Y.}~\bibnamefont
			{Salath\'e}}, \bibinfo {author} {\bibfnamefont {M.}~\bibnamefont {Pechal}},
		\bibinfo {author} {\bibfnamefont {M.}~\bibnamefont {Mondal}}, \bibinfo
		{author} {\bibfnamefont {M.}~\bibnamefont {Oppliger}}, \bibinfo {author}
		{\bibfnamefont {C.}~\bibnamefont {Eichler}}, \ and\ \bibinfo {author}
		{\bibfnamefont {A.}~\bibnamefont {Wallraff}},\ }\bibfield  {title} {\enquote
		{\bibinfo {title} {Rapid high-fidelity single-shot dispersive readout of
				superconducting qubits},}\ }\href {\doibase 10.1103/PhysRevApplied.7.054020}
	{\bibfield  {journal} {\bibinfo  {journal} {Phys. Rev. Applied}\ }\textbf
		{\bibinfo {volume} {7}},\ \bibinfo {pages} {054020} (\bibinfo {year}
		{2017})}\BibitemShut {NoStop}%
	\bibitem [{\citenamefont {Touzard}\ \emph {et~al.}(2019)\citenamefont
		{Touzard}, \citenamefont {Kou}, \citenamefont {Frattini}, \citenamefont
		{Sivak}, \citenamefont {Puri}, \citenamefont {Grimm}, \citenamefont
		{Frunzio}, \citenamefont {Shankar},\ and\ \citenamefont
		{Devoret}}]{arXiv_Touzard_2018}%
	\BibitemOpen
	\bibfield  {author} {\bibinfo {author} {\bibfnamefont {S.}~\bibnamefont
			{Touzard}}, \bibinfo {author} {\bibfnamefont {A.}~\bibnamefont {Kou}},
		\bibinfo {author} {\bibfnamefont {N.{\hspace{0.167em}}E.}\ \bibnamefont
			{Frattini}}, \bibinfo {author} {\bibfnamefont {V.{\hspace{0.167em}}V.}\
			\bibnamefont {Sivak}}, \bibinfo {author} {\bibfnamefont {S.}~\bibnamefont
			{Puri}}, \bibinfo {author} {\bibfnamefont {A.}~\bibnamefont {Grimm}},
		\bibinfo {author} {\bibfnamefont {L.}~\bibnamefont {Frunzio}}, \bibinfo
		{author} {\bibfnamefont {S.}~\bibnamefont {Shankar}}, \ and\ \bibinfo
		{author} {\bibfnamefont {M.{\hspace{0.167em}}H.}\ \bibnamefont {Devoret}},\
	}\bibfield  {title} {\enquote {\bibinfo {title} {Gated conditional
				displacement readout of superconducting qubits},}\ }\href
	{https://doi.org/10.1103/physrevlett.122.080502} {\bibfield  {journal}
		{\bibinfo  {journal} {Physical Review Letters}\ }\textbf {\bibinfo {volume}
			{122}} (\bibinfo {year} {2019})}\BibitemShut {NoStop}%
	\bibitem [{\citenamefont {Koch}\ \emph {et~al.}(2007)\citenamefont {Koch},
		\citenamefont {Yu}, \citenamefont {Gambetta}, \citenamefont {Houck},
		\citenamefont {Schuster}, \citenamefont {Majer}, \citenamefont {Blais},
		\citenamefont {Devoret}, \citenamefont {Girvin},\ and\ \citenamefont
		{Schoelkopf}}]{PRA_Koch_2007}%
	\BibitemOpen
	\bibfield  {author} {\bibinfo {author} {\bibfnamefont {Jens}\ \bibnamefont
			{Koch}}, \bibinfo {author} {\bibfnamefont {Terri~M.}\ \bibnamefont {Yu}},
		\bibinfo {author} {\bibfnamefont {Jay}\ \bibnamefont {Gambetta}}, \bibinfo
		{author} {\bibfnamefont {A.~A.}\ \bibnamefont {Houck}}, \bibinfo {author}
		{\bibfnamefont {D.~I.}\ \bibnamefont {Schuster}}, \bibinfo {author}
		{\bibfnamefont {J.}~\bibnamefont {Majer}}, \bibinfo {author} {\bibfnamefont
			{Alexandre}\ \bibnamefont {Blais}}, \bibinfo {author} {\bibfnamefont {M.~H.}\
			\bibnamefont {Devoret}}, \bibinfo {author} {\bibfnamefont {S.~M.}\
			\bibnamefont {Girvin}}, \ and\ \bibinfo {author} {\bibfnamefont {R.~J.}\
			\bibnamefont {Schoelkopf}},\ }\bibfield  {title} {\enquote {\bibinfo {title}
			{Charge-insensitive qubit design derived from the cooper pair box},}\ }\href
	{\doibase 10.1103/PhysRevA.76.042319} {\bibfield  {journal} {\bibinfo
			{journal} {Phys. Rev. A}\ }\textbf {\bibinfo {volume} {76}},\ \bibinfo
		{pages} {042319} (\bibinfo {year} {2007})}\BibitemShut {NoStop}%
	\bibitem [{\citenamefont {Jeffrey}\ \emph {et~al.}(2014)\citenamefont
		{Jeffrey}, \citenamefont {Sank}, \citenamefont {Mutus}, \citenamefont
		{White}, \citenamefont {Kelly}, \citenamefont {Barends}, \citenamefont
		{Chen}, \citenamefont {Chen}, \citenamefont {Chiaro}, \citenamefont
		{Dunsworth}, \citenamefont {Megrant}, \citenamefont {O'Malley}, \citenamefont
		{Neill}, \citenamefont {Roushan}, \citenamefont {Vainsencher}, \citenamefont
		{Wenner}, \citenamefont {Cleland},\ and\ \citenamefont
		{Martinis}}]{PRL_Jeffrey_2014}%
	\BibitemOpen
	\bibfield  {author} {\bibinfo {author} {\bibfnamefont {Evan}\ \bibnamefont
			{Jeffrey}}, \bibinfo {author} {\bibfnamefont {Daniel}\ \bibnamefont {Sank}},
		\bibinfo {author} {\bibfnamefont {J.~Y.}\ \bibnamefont {Mutus}}, \bibinfo
		{author} {\bibfnamefont {T.~C.}\ \bibnamefont {White}}, \bibinfo {author}
		{\bibfnamefont {J.}~\bibnamefont {Kelly}}, \bibinfo {author} {\bibfnamefont
			{R.}~\bibnamefont {Barends}}, \bibinfo {author} {\bibfnamefont
			{Y.}~\bibnamefont {Chen}}, \bibinfo {author} {\bibfnamefont {Z.}~\bibnamefont
			{Chen}}, \bibinfo {author} {\bibfnamefont {B.}~\bibnamefont {Chiaro}},
		\bibinfo {author} {\bibfnamefont {A.}~\bibnamefont {Dunsworth}}, \bibinfo
		{author} {\bibfnamefont {A.}~\bibnamefont {Megrant}}, \bibinfo {author}
		{\bibfnamefont {P.~J.~J.}\ \bibnamefont {O'Malley}}, \bibinfo {author}
		{\bibfnamefont {C.}~\bibnamefont {Neill}}, \bibinfo {author} {\bibfnamefont
			{P.}~\bibnamefont {Roushan}}, \bibinfo {author} {\bibfnamefont
			{A.}~\bibnamefont {Vainsencher}}, \bibinfo {author} {\bibfnamefont
			{J.}~\bibnamefont {Wenner}}, \bibinfo {author} {\bibfnamefont {A.~N.}\
			\bibnamefont {Cleland}}, \ and\ \bibinfo {author} {\bibfnamefont {John~M.}\
			\bibnamefont {Martinis}},\ }\bibfield  {title} {\enquote {\bibinfo {title}
			{Fast accurate state measurement with superconducting qubits},}\ }\href
	{\doibase 10.1103/PhysRevLett.112.190504} {\bibfield  {journal} {\bibinfo
			{journal} {Phys. Rev. Lett.}\ }\textbf {\bibinfo {volume} {112}},\ \bibinfo
		{pages} {190504} (\bibinfo {year} {2014})}\BibitemShut {NoStop}%
	\bibitem [{\citenamefont {Slichter}\ \emph {et~al.}(2012)\citenamefont
		{Slichter}, \citenamefont {Vijay}, \citenamefont {Weber}, \citenamefont
		{Boutin}, \citenamefont {Boissonneault}, \citenamefont {Gambetta},
		\citenamefont {Blais},\ and\ \citenamefont {Siddiqi}}]{Slichter:2012bi}%
	\BibitemOpen
	\bibfield  {author} {\bibinfo {author} {\bibfnamefont {D~H}\ \bibnamefont
			{Slichter}}, \bibinfo {author} {\bibfnamefont {R}~\bibnamefont {Vijay}},
		\bibinfo {author} {\bibfnamefont {S~J}\ \bibnamefont {Weber}}, \bibinfo
		{author} {\bibfnamefont {S}~\bibnamefont {Boutin}}, \bibinfo {author}
		{\bibfnamefont {M}~\bibnamefont {Boissonneault}}, \bibinfo {author}
		{\bibfnamefont {J~M}\ \bibnamefont {Gambetta}}, \bibinfo {author}
		{\bibfnamefont {A}~\bibnamefont {Blais}}, \ and\ \bibinfo {author}
		{\bibfnamefont {I}~\bibnamefont {Siddiqi}},\ }\bibfield  {title} {\enquote
		{\bibinfo {title} {{Measurement-Induced Qubit State Mixing in Circuit QED
					from Up-Converted Dephasing Noise}},}\ }\href
	{https://link.aps.org/doi/10.1103/PhysRevLett.109.153601} {\bibfield
		{journal} {\bibinfo  {journal} {Physical Review Letters}\ }\textbf {\bibinfo
			{volume} {109}},\ \bibinfo {pages} {153601--5} (\bibinfo {year}
		{2012})}\BibitemShut {NoStop}%
	\bibitem [{\citenamefont {Sank}\ \emph {et~al.}(2016)\citenamefont {Sank},
		\citenamefont {Chen}, \citenamefont {Khezri}, \citenamefont {Kelly},
		\citenamefont {Barends}, \citenamefont {Campbell}, \citenamefont {Chen},
		\citenamefont {Chiaro}, \citenamefont {Dunsworth}, \citenamefont {Fowler},
		\citenamefont {Jeffrey}, \citenamefont {Lucero}, \citenamefont {Megrant},
		\citenamefont {Mutus}, \citenamefont {Neeley}, \citenamefont {Neill},
		\citenamefont {O'Malley}, \citenamefont {Quintana}, \citenamefont {Roushan},
		\citenamefont {Vainsencher}, \citenamefont {White}, \citenamefont {Wenner},
		\citenamefont {Korotkov},\ and\ \citenamefont {Martinis}}]{PRL_Sank_2016}%
	\BibitemOpen
	\bibfield  {author} {\bibinfo {author} {\bibfnamefont {Daniel}\ \bibnamefont
			{Sank}}, \bibinfo {author} {\bibfnamefont {Zijun}\ \bibnamefont {Chen}},
		\bibinfo {author} {\bibfnamefont {Mostafa}\ \bibnamefont {Khezri}}, \bibinfo
		{author} {\bibfnamefont {J.}~\bibnamefont {Kelly}}, \bibinfo {author}
		{\bibfnamefont {R.}~\bibnamefont {Barends}}, \bibinfo {author} {\bibfnamefont
			{B.}~\bibnamefont {Campbell}}, \bibinfo {author} {\bibfnamefont
			{Y.}~\bibnamefont {Chen}}, \bibinfo {author} {\bibfnamefont {B.}~\bibnamefont
			{Chiaro}}, \bibinfo {author} {\bibfnamefont {A.}~\bibnamefont {Dunsworth}},
		\bibinfo {author} {\bibfnamefont {A.}~\bibnamefont {Fowler}}, \bibinfo
		{author} {\bibfnamefont {E.}~\bibnamefont {Jeffrey}}, \bibinfo {author}
		{\bibfnamefont {E.}~\bibnamefont {Lucero}}, \bibinfo {author} {\bibfnamefont
			{A.}~\bibnamefont {Megrant}}, \bibinfo {author} {\bibfnamefont
			{J.}~\bibnamefont {Mutus}}, \bibinfo {author} {\bibfnamefont
			{M.}~\bibnamefont {Neeley}}, \bibinfo {author} {\bibfnamefont
			{C.}~\bibnamefont {Neill}}, \bibinfo {author} {\bibfnamefont {P.~J.~J.}\
			\bibnamefont {O'Malley}}, \bibinfo {author} {\bibfnamefont {C.}~\bibnamefont
			{Quintana}}, \bibinfo {author} {\bibfnamefont {P.}~\bibnamefont {Roushan}},
		\bibinfo {author} {\bibfnamefont {A.}~\bibnamefont {Vainsencher}}, \bibinfo
		{author} {\bibfnamefont {T.}~\bibnamefont {White}}, \bibinfo {author}
		{\bibfnamefont {J.}~\bibnamefont {Wenner}}, \bibinfo {author} {\bibfnamefont
			{Alexander~N.}\ \bibnamefont {Korotkov}}, \ and\ \bibinfo {author}
		{\bibfnamefont {John~M.}\ \bibnamefont {Martinis}},\ }\bibfield  {title}
	{\enquote {\bibinfo {title} {Measurement-induced state transitions in a
				superconducting qubit: Beyond the rotating wave approximation},}\ }\href
	{\doibase 10.1103/PhysRevLett.117.190503} {\bibfield  {journal} {\bibinfo
			{journal} {Phys. Rev. Lett.}\ }\textbf {\bibinfo {volume} {117}},\ \bibinfo
		{pages} {190503} (\bibinfo {year} {2016})}\BibitemShut {NoStop}%
	\bibitem [{\citenamefont {Lescanne}\ \emph
		{et~al.}(2019{\natexlab{a}})\citenamefont {Lescanne}, \citenamefont {Verney},
		\citenamefont {Ficheux}, \citenamefont {Devoret}, \citenamefont {Huard},
		\citenamefont {Mirrahimi},\ and\ \citenamefont
		{Leghtas}}]{arXiv_Lescanne_2018}%
	\BibitemOpen
	\bibfield  {author} {\bibinfo {author} {\bibfnamefont {Raphaël}\
			\bibnamefont {Lescanne}}, \bibinfo {author} {\bibfnamefont {Lucas}\
			\bibnamefont {Verney}}, \bibinfo {author} {\bibfnamefont {Quentin}\
			\bibnamefont {Ficheux}}, \bibinfo {author} {\bibfnamefont {Michel~H.}\
			\bibnamefont {Devoret}}, \bibinfo {author} {\bibfnamefont {Benjamin}\
			\bibnamefont {Huard}}, \bibinfo {author} {\bibfnamefont {Mazyar}\
			\bibnamefont {Mirrahimi}}, \ and\ \bibinfo {author} {\bibfnamefont {Zaki}\
			\bibnamefont {Leghtas}},\ }\bibfield  {title} {\enquote {\bibinfo {title}
			{Escape of a driven quantum josephson circuit into unconfined states},}\
	}\href {https://doi.org/10.1103/physrevapplied.11.014030} {\bibfield
		{journal} {\bibinfo  {journal} {Physical Review Applied}\ }\textbf {\bibinfo
			{volume} {11}} (\bibinfo {year} {2019}{\natexlab{a}})}\BibitemShut {NoStop}%
	\bibitem [{\citenamefont {Houck}\ \emph {et~al.}(2008)\citenamefont {Houck},
		\citenamefont {Schreier}, \citenamefont {Johnson}, \citenamefont {Chow},
		\citenamefont {Koch}, \citenamefont {Gambetta}, \citenamefont {Schuster},
		\citenamefont {Frunzio}, \citenamefont {Devoret}, \citenamefont {Girvin},\
		and\ \citenamefont {Schoelkopf}}]{Houck2008}%
	\BibitemOpen
	\bibfield  {author} {\bibinfo {author} {\bibfnamefont {A.~A.}\ \bibnamefont
			{Houck}}, \bibinfo {author} {\bibfnamefont {J.~A.}\ \bibnamefont {Schreier}},
		\bibinfo {author} {\bibfnamefont {B.~R.}\ \bibnamefont {Johnson}}, \bibinfo
		{author} {\bibfnamefont {J.~M.}\ \bibnamefont {Chow}}, \bibinfo {author}
		{\bibfnamefont {Jens}\ \bibnamefont {Koch}}, \bibinfo {author} {\bibfnamefont
			{J.~M.}\ \bibnamefont {Gambetta}}, \bibinfo {author} {\bibfnamefont {D.~I.}\
			\bibnamefont {Schuster}}, \bibinfo {author} {\bibfnamefont {L.}~\bibnamefont
			{Frunzio}}, \bibinfo {author} {\bibfnamefont {M.~H.}\ \bibnamefont
			{Devoret}}, \bibinfo {author} {\bibfnamefont {S.~M.}\ \bibnamefont {Girvin}},
		\ and\ \bibinfo {author} {\bibfnamefont {R.~J.}\ \bibnamefont {Schoelkopf}},\
	}\bibfield  {title} {\enquote {\bibinfo {title} {Controlling the spontaneous
				emission of a superconducting transmon qubit},}\ }\href
	{https://doi.org/10.1103/PhysRevLett.101.080502} {\bibfield  {journal}
		{\bibinfo  {journal} {Physical Review Letters}\ }\textbf {\bibinfo {volume}
			{101}} (\bibinfo {year} {2008})}\BibitemShut {NoStop}%
	\bibitem [{\citenamefont {Lecocq}\ \emph {et~al.}(2011)\citenamefont {Lecocq},
		\citenamefont {Claudon}, \citenamefont {Buisson},\ and\ \citenamefont
		{Milman}}]{PRL_Lecocq_2011}%
	\BibitemOpen
	\bibfield  {author} {\bibinfo {author} {\bibfnamefont {F.}~\bibnamefont
			{Lecocq}}, \bibinfo {author} {\bibfnamefont {J.}~\bibnamefont {Claudon}},
		\bibinfo {author} {\bibfnamefont {O.}~\bibnamefont {Buisson}}, \ and\
		\bibinfo {author} {\bibfnamefont {P.}~\bibnamefont {Milman}},\ }\bibfield
	{title} {\enquote {\bibinfo {title} {Nonlinear coupling between the two
				oscillation modes of a dc squid},}\ }\href
	{https://doi.org/10.1103/PhysRevLett.107.197002} {\bibfield  {journal}
		{\bibinfo  {journal} {Phys. Rev. Lett.}\ }\textbf {\bibinfo {volume} {107}},\
		\bibinfo {pages} {197002} (\bibinfo {year} {2011})}\BibitemShut {NoStop}%
	\bibitem [{\citenamefont {Diniz}\ \emph {et~al.}(2013)\citenamefont {Diniz},
		\citenamefont {Dumur}, \citenamefont {Buisson},\ and\ \citenamefont
		{Auff\`eves}}]{PRA_Diniz_2013}%
	\BibitemOpen
	\bibfield  {author} {\bibinfo {author} {\bibfnamefont {I.}~\bibnamefont
			{Diniz}}, \bibinfo {author} {\bibfnamefont {E.}~\bibnamefont {Dumur}},
		\bibinfo {author} {\bibfnamefont {O.}~\bibnamefont {Buisson}}, \ and\
		\bibinfo {author} {\bibfnamefont {A.}~\bibnamefont {Auff\`eves}},\ }\bibfield
	{title} {\enquote {\bibinfo {title} {Ultrafast quantum nondemolition
				measurements based on a diamond-shaped artificial atom},}\ }\href {\doibase
		10.1103/PhysRevA.87.033837} {\bibfield  {journal} {\bibinfo  {journal} {Phys.
				Rev. A}\ }\textbf {\bibinfo {volume} {87}},\ \bibinfo {pages} {033837}
		(\bibinfo {year} {2013})}\BibitemShut {NoStop}%
	\bibitem [{\citenamefont {Kerman}(2013)}]{NPJ_Kerman_2013}%
	\BibitemOpen
	\bibfield  {author} {\bibinfo {author} {\bibfnamefont {Andrew~J}\
			\bibnamefont {Kerman}},\ }\bibfield  {title} {\enquote {\bibinfo {title}
			{Quantum information processing using quasiclassical electromagnetic
				interactions between qubits and electrical resonators},}\ }\href {\doibase
		10.1088/1367-2630/15/12/123011} {\bibfield  {journal} {\bibinfo  {journal}
			{New J. Phys.}\ }\textbf {\bibinfo {volume} {15}},\ \bibinfo {pages} {123011}
		(\bibinfo {year} {2013})}\BibitemShut {NoStop}%
	\bibitem [{\citenamefont {Dumur}\ \emph {et~al.}(2015)\citenamefont {Dumur},
		\citenamefont {K\"ung}, \citenamefont {Feofanov}, \citenamefont {Weissl},
		\citenamefont {Roch}, \citenamefont {Naud}, \citenamefont {Guichard},\ and\
		\citenamefont {Buisson}}]{Dumur2015}%
	\BibitemOpen
	\bibfield  {author} {\bibinfo {author} {\bibfnamefont {{\'{E}}.}~\bibnamefont
			{Dumur}}, \bibinfo {author} {\bibfnamefont {B.}~\bibnamefont {K\"ung}},
		\bibinfo {author} {\bibfnamefont {A.~K.}\ \bibnamefont {Feofanov}}, \bibinfo
		{author} {\bibfnamefont {T.}~\bibnamefont {Weissl}}, \bibinfo {author}
		{\bibfnamefont {N.}~\bibnamefont {Roch}}, \bibinfo {author} {\bibfnamefont
			{C.}~\bibnamefont {Naud}}, \bibinfo {author} {\bibfnamefont {W.}~\bibnamefont
			{Guichard}}, \ and\ \bibinfo {author} {\bibfnamefont {O.}~\bibnamefont
			{Buisson}},\ }\bibfield  {title} {\enquote {\bibinfo {title} {V-shaped
				superconducting artificial atom based on two inductively coupled
				transmons},}\ }\href {https://doi.org/10.1103/PhysRevB.92.020515} {\bibfield
		{journal} {\bibinfo  {journal} {Physical Review B}\ }\textbf {\bibinfo
			{volume} {92}} (\bibinfo {year} {2015})}\BibitemShut {NoStop}%
	\bibitem [{\citenamefont {Billangeon}\ \emph {et~al.}(2015)\citenamefont
		{Billangeon}, \citenamefont {Tsai},\ and\ \citenamefont
		{Nakamura}}]{PRB_Billangeon_2015}%
	\BibitemOpen
	\bibfield  {author} {\bibinfo {author} {\bibfnamefont {P.-M.}\ \bibnamefont
			{Billangeon}}, \bibinfo {author} {\bibfnamefont {J.~S.}\ \bibnamefont
			{Tsai}}, \ and\ \bibinfo {author} {\bibfnamefont {Y.}~\bibnamefont
			{Nakamura}},\ }\bibfield  {title} {\enquote {\bibinfo {title}
			{Circuit-qed-based scalable architectures for quantum information processing
				with superconducting qubits},}\ }\href {\doibase 10.1103/PhysRevB.91.094517}
	{\bibfield  {journal} {\bibinfo  {journal} {Phys. Rev. B}\ }\textbf {\bibinfo
			{volume} {91}},\ \bibinfo {pages} {094517} (\bibinfo {year}
		{2015})}\BibitemShut {NoStop}%
	\bibitem [{\citenamefont {Richer}\ and\ \citenamefont
		{DiVincenzo}(2016)}]{PRB_Richer_2016}%
	\BibitemOpen
	\bibfield  {author} {\bibinfo {author} {\bibfnamefont {Susanne}\ \bibnamefont
			{Richer}}\ and\ \bibinfo {author} {\bibfnamefont {David}\ \bibnamefont
			{DiVincenzo}},\ }\bibfield  {title} {\enquote {\bibinfo {title} {Circuit
				design implementing longitudinal coupling: A scalable scheme for
				superconducting qubits},}\ }\href {\doibase 10.1103/PhysRevB.93.134501}
	{\bibfield  {journal} {\bibinfo  {journal} {Phys. Rev. B}\ }\textbf {\bibinfo
			{volume} {93}},\ \bibinfo {pages} {134501} (\bibinfo {year}
		{2016})}\BibitemShut {NoStop}%
	\bibitem [{\citenamefont {Didier}\ \emph {et~al.}(2015)\citenamefont {Didier},
		\citenamefont {Bourassa},\ and\ \citenamefont {Blais}}]{PRA_Didier_2015}%
	\BibitemOpen
	\bibfield  {author} {\bibinfo {author} {\bibfnamefont {Nicolas}\ \bibnamefont
			{Didier}}, \bibinfo {author} {\bibfnamefont {J\'er\^ome}\ \bibnamefont
			{Bourassa}}, \ and\ \bibinfo {author} {\bibfnamefont {Alexandre}\
			\bibnamefont {Blais}},\ }\bibfield  {title} {\enquote {\bibinfo {title} {Fast
				quantum nondemolition readout by parametric modulation of longitudinal
				qubit-oscillator interaction},}\ }\href {\doibase
		10.1103/PhysRevLett.115.203601} {\bibfield  {journal} {\bibinfo  {journal}
			{Phys. Rev. Lett.}\ }\textbf {\bibinfo {volume} {115}},\ \bibinfo {pages}
		{203601} (\bibinfo {year} {2015})}\BibitemShut {NoStop}%
	\bibitem [{\citenamefont {{Gard}}\ \emph {et~al.}(2018)\citenamefont {{Gard}},
		\citenamefont {{Jacobs}}, \citenamefont {{Aumentado}},\ and\ \citenamefont
		{{Simmonds}}}]{arXiv_Gard_2018}%
	\BibitemOpen
	\bibfield  {author} {\bibinfo {author} {\bibfnamefont {Bryan~T.}\
			\bibnamefont {{Gard}}}, \bibinfo {author} {\bibfnamefont {Kurt}\ \bibnamefont
			{{Jacobs}}}, \bibinfo {author} {\bibfnamefont {Jos{\'e}}\ \bibnamefont
			{{Aumentado}}}, \ and\ \bibinfo {author} {\bibfnamefont {Raymond~W.}\
			\bibnamefont {{Simmonds}}},\ }\href@noop {} {\enquote {\bibinfo {title}
			{{Fast, High-Fidelity, Quantum Non-demolition Readout of a Superconducting
					Qubit Using a Transverse Coupling}},}\ } (\bibinfo {year} {2018}),\ \bibinfo
	{note} {preprint},\ \Eprint {http://arxiv.org/abs/1809.02597}
	{arXiv:1809.02597} \BibitemShut {NoStop}%
	\bibitem [{\citenamefont {Ikonen}\ \emph {et~al.}(2019)\citenamefont {Ikonen},
		\citenamefont {Goetz}, \citenamefont {Ilves}, \citenamefont {Keränen},
		\citenamefont {Gunyho}, \citenamefont {Partanen}, \citenamefont {Tan},
		\citenamefont {Hazra}, \citenamefont {Grönberg}, \citenamefont {Vesterinen},
		\citenamefont {Simbierowicz}, \citenamefont {Hassel},\ and\ \citenamefont
		{Möttönen}}]{Ikonen:2018vq}%
	\BibitemOpen
	\bibfield  {author} {\bibinfo {author} {\bibfnamefont {Joni}\ \bibnamefont
			{Ikonen}}, \bibinfo {author} {\bibfnamefont {Jan}\ \bibnamefont {Goetz}},
		\bibinfo {author} {\bibfnamefont {Jesper}\ \bibnamefont {Ilves}}, \bibinfo
		{author} {\bibfnamefont {Aarne}\ \bibnamefont {Keränen}}, \bibinfo {author}
		{\bibfnamefont {Andras~M.}\ \bibnamefont {Gunyho}}, \bibinfo {author}
		{\bibfnamefont {Matti}\ \bibnamefont {Partanen}}, \bibinfo {author}
		{\bibfnamefont {Kuan~Y.}\ \bibnamefont {Tan}}, \bibinfo {author}
		{\bibfnamefont {Dibyendu}\ \bibnamefont {Hazra}}, \bibinfo {author}
		{\bibfnamefont {Leif}\ \bibnamefont {Grönberg}}, \bibinfo {author}
		{\bibfnamefont {Visa}\ \bibnamefont {Vesterinen}}, \bibinfo {author}
		{\bibfnamefont {Slawomir}\ \bibnamefont {Simbierowicz}}, \bibinfo {author}
		{\bibfnamefont {Juha}\ \bibnamefont {Hassel}}, \ and\ \bibinfo {author}
		{\bibfnamefont {Mikko}\ \bibnamefont {Möttönen}},\ }\bibfield  {title}
	{\enquote {\bibinfo {title} {Qubit measurement by multichannel driving},}\
	}\href {https://doi.org/10.1103/physrevlett.122.080503} {\bibfield  {journal}
		{\bibinfo  {journal} {Physical Review Letters}\ }\textbf {\bibinfo {volume}
			{122}} (\bibinfo {year} {2019})}\BibitemShut {NoStop}%
	\bibitem [{\citenamefont {Wang}\ \emph {et~al.}()\citenamefont {Wang},
		\citenamefont {Miranowicz},\ and\ \citenamefont {Nori}}]{Wang2018}%
	\BibitemOpen
	\bibfield  {author} {\bibinfo {author} {\bibfnamefont {Xin}\ \bibnamefont
			{Wang}}, \bibinfo {author} {\bibfnamefont {Adam}\ \bibnamefont {Miranowicz}},
		\ and\ \bibinfo {author} {\bibfnamefont {Franco}\ \bibnamefont {Nori}},\
	}\bibfield  {title} {\enquote {\bibinfo {title} {Ideal quantum nondemolition
				readout of a flux qubit without purcell limitations},}\ }\href@noop {} {\
	}\Eprint {http://arxiv.org/abs/1811.09048v2} {1811.09048v2} \BibitemShut
	{NoStop}%
	\bibitem [{\citenamefont {Ruskov}\ and\ \citenamefont
		{Tahan}(2019)}]{Ruskov2019}%
	\BibitemOpen
	\bibfield  {author} {\bibinfo {author} {\bibfnamefont {Rusko}\ \bibnamefont
			{Ruskov}}\ and\ \bibinfo {author} {\bibfnamefont {Charles}\ \bibnamefont
			{Tahan}},\ }\bibfield  {title} {\enquote {\bibinfo {title} {Quantum-limited
				measurement of spin qubits via curvature couplings to a cavity},}\ }\href
	{\doibase 10.1103/physrevb.99.245306} {\bibfield  {journal} {\bibinfo
			{journal} {Physical Review B}\ }\textbf {\bibinfo {volume} {99}} (\bibinfo
		{year} {2019}),\ 10.1103/physrevb.99.245306}\BibitemShut {NoStop}%
	\bibitem [{\citenamefont {Barends}\ \emph {et~al.}(2014)\citenamefont
		{Barends}, \citenamefont {Kelly}, \citenamefont {Megrant}, \citenamefont
		{Veitia}, \citenamefont {Sank}, \citenamefont {Jeffrey}, \citenamefont
		{White}, \citenamefont {Mutus}, \citenamefont {Fowler}, \citenamefont
		{Campbell}, \citenamefont {Chen}, \citenamefont {Chen}, \citenamefont
		{Chiaro}, \citenamefont {Dunsworth}, \citenamefont {Neill}, \citenamefont
		{O’Malley}, \citenamefont {Roushan}, \citenamefont {Vainsencher},
		\citenamefont {Wenner}, \citenamefont {Korotkov}, \citenamefont {Cleland},\
		and\ \citenamefont {Martinis}}]{Barends_Nature_2008}%
	\BibitemOpen
	\bibfield  {author} {\bibinfo {author} {\bibfnamefont {R.}~\bibnamefont
			{Barends}}, \bibinfo {author} {\bibfnamefont {J.}~\bibnamefont {Kelly}},
		\bibinfo {author} {\bibfnamefont {A.}~\bibnamefont {Megrant}}, \bibinfo
		{author} {\bibfnamefont {A.}~\bibnamefont {Veitia}}, \bibinfo {author}
		{\bibfnamefont {D.}~\bibnamefont {Sank}}, \bibinfo {author} {\bibfnamefont
			{E.}~\bibnamefont {Jeffrey}}, \bibinfo {author} {\bibfnamefont {T.~C.}\
			\bibnamefont {White}}, \bibinfo {author} {\bibfnamefont {J.}~\bibnamefont
			{Mutus}}, \bibinfo {author} {\bibfnamefont {A.~G.}\ \bibnamefont {Fowler}},
		\bibinfo {author} {\bibfnamefont {B.}~\bibnamefont {Campbell}}, \bibinfo
		{author} {\bibfnamefont {Y.}~\bibnamefont {Chen}}, \bibinfo {author}
		{\bibfnamefont {Z.}~\bibnamefont {Chen}}, \bibinfo {author} {\bibfnamefont
			{B.}~\bibnamefont {Chiaro}}, \bibinfo {author} {\bibfnamefont
			{A}~\bibnamefont {Dunsworth}}, \bibinfo {author} {\bibfnamefont
			{C}~\bibnamefont {Neill}}, \bibinfo {author} {\bibfnamefont {P}~\bibnamefont
			{O’Malley}}, \bibinfo {author} {\bibfnamefont {P.}~\bibnamefont {Roushan}},
		\bibinfo {author} {\bibfnamefont {A.}~\bibnamefont {Vainsencher}}, \bibinfo
		{author} {\bibfnamefont {J.}~\bibnamefont {Wenner}}, \bibinfo {author}
		{\bibfnamefont {A.~N.}\ \bibnamefont {Korotkov}}, \bibinfo {author}
		{\bibfnamefont {A.~N.}\ \bibnamefont {Cleland}}, \ and\ \bibinfo {author}
		{\bibfnamefont {J.~M.}\ \bibnamefont {Martinis}},\ }\bibfield  {title}
	{\enquote {\bibinfo {title} {Superconducting quantum circuits at the surface
				code threshold for fault tolerance},}\ }\href
	{https://www.nature.com/articles/nature13171} {\bibfield  {journal} {\bibinfo
			{journal} {Nature}\ }\textbf {\bibinfo {volume} {508}} (\bibinfo {year}
		{2014})}\BibitemShut {NoStop}%
	\bibitem [{\citenamefont {Lecocq}\ \emph {et~al.}(2012)\citenamefont {Lecocq},
		\citenamefont {Pop}, \citenamefont {Matei}, \citenamefont {Dumur},
		\citenamefont {Feofanov}, \citenamefont {Naud}, \citenamefont {Guichard},\
		and\ \citenamefont {Buisson}}]{PRL_Lecocq_2012}%
	\BibitemOpen
	\bibfield  {author} {\bibinfo {author} {\bibfnamefont {F.}~\bibnamefont
			{Lecocq}}, \bibinfo {author} {\bibfnamefont {I.~M.}\ \bibnamefont {Pop}},
		\bibinfo {author} {\bibfnamefont {I.}~\bibnamefont {Matei}}, \bibinfo
		{author} {\bibfnamefont {E.}~\bibnamefont {Dumur}}, \bibinfo {author}
		{\bibfnamefont {A.~K.}\ \bibnamefont {Feofanov}}, \bibinfo {author}
		{\bibfnamefont {C.}~\bibnamefont {Naud}}, \bibinfo {author} {\bibfnamefont
			{W.}~\bibnamefont {Guichard}}, \ and\ \bibinfo {author} {\bibfnamefont
			{O.}~\bibnamefont {Buisson}},\ }\bibfield  {title} {\enquote {\bibinfo
			{title} {Coherent frequency conversion in a superconducting artificial atom
				with two internal degrees of freedom},}\ }\href {\doibase
		10.1103/PhysRevLett.108.107001} {\bibfield  {journal} {\bibinfo  {journal}
			{Phys. Rev. Lett.}\ }\textbf {\bibinfo {volume} {108}},\ \bibinfo {pages}
		{107001} (\bibinfo {year} {2012})}\BibitemShut {NoStop}%
	\bibitem [{\citenamefont {Planat}\ \emph {et~al.}(2019)\citenamefont {Planat},
		\citenamefont {Dassonneville}, \citenamefont {Mart{\'{\i}}nez}, \citenamefont
		{Foroughi}, \citenamefont {Buisson}, \citenamefont {Hasch-Guichard},
		\citenamefont {Naud}, \citenamefont {Vijay}, \citenamefont {Murch},\ and\
		\citenamefont {Roch}}]{arXiv_Planat_2018}%
	\BibitemOpen
	\bibfield  {author} {\bibinfo {author} {\bibfnamefont {Luca}\ \bibnamefont
			{Planat}}, \bibinfo {author} {\bibfnamefont {R{\'{e}}my}\ \bibnamefont
			{Dassonneville}}, \bibinfo {author} {\bibfnamefont {Javier~Puertas}\
			\bibnamefont {Mart{\'{\i}}nez}}, \bibinfo {author} {\bibfnamefont {Farshad}\
			\bibnamefont {Foroughi}}, \bibinfo {author} {\bibfnamefont {Olivier}\
			\bibnamefont {Buisson}}, \bibinfo {author} {\bibfnamefont {Wiebke}\
			\bibnamefont {Hasch-Guichard}}, \bibinfo {author} {\bibfnamefont
			{C{\'{e}}cile}\ \bibnamefont {Naud}}, \bibinfo {author} {\bibfnamefont
			{R.}~\bibnamefont {Vijay}}, \bibinfo {author} {\bibfnamefont {Kater}\
			\bibnamefont {Murch}}, \ and\ \bibinfo {author} {\bibfnamefont {Nicolas}\
			\bibnamefont {Roch}},\ }\bibfield  {title} {\enquote {\bibinfo {title}
			{Understanding the saturation power of josephson parametric amplifiers made
				from {SQUID} arrays},}\ }\href
	{https://doi.org/10.1103/physrevapplied.11.034014} {\bibfield  {journal}
		{\bibinfo  {journal} {Physical Review Applied}\ }\textbf {\bibinfo {volume}
			{11}} (\bibinfo {year} {2019})}\BibitemShut {NoStop}%
	\bibitem [{\citenamefont {Schuster}\ \emph {et~al.}(2005)\citenamefont
		{Schuster}, \citenamefont {Wallraff}, \citenamefont {Blais}, \citenamefont
		{Frunzio}, \citenamefont {Huang}, \citenamefont {Majer}, \citenamefont
		{Girvin},\ and\ \citenamefont {Schoelkopf}}]{Schuster2005Stark}%
	\BibitemOpen
	\bibfield  {author} {\bibinfo {author} {\bibfnamefont {D.~I.}\ \bibnamefont
			{Schuster}}, \bibinfo {author} {\bibfnamefont {A.}~\bibnamefont {Wallraff}},
		\bibinfo {author} {\bibfnamefont {A.}~\bibnamefont {Blais}}, \bibinfo
		{author} {\bibfnamefont {L.}~\bibnamefont {Frunzio}}, \bibinfo {author}
		{\bibfnamefont {R.-S.}\ \bibnamefont {Huang}}, \bibinfo {author}
		{\bibfnamefont {J.}~\bibnamefont {Majer}}, \bibinfo {author} {\bibfnamefont
			{S.~M.}\ \bibnamefont {Girvin}}, \ and\ \bibinfo {author} {\bibfnamefont
			{R.~J.}\ \bibnamefont {Schoelkopf}},\ }\bibfield  {title} {\enquote {\bibinfo
			{title} {ac stark shift and dephasing of a superconducting qubit strongly
				coupled to a cavity field},}\ }\href
	{https://doi.org/10.1103/physrevlett.94.123602} {\bibfield  {journal}
		{\bibinfo  {journal} {Physical Review Letters}\ }\textbf {\bibinfo {volume}
			{94}} (\bibinfo {year} {2005})}\BibitemShut {NoStop}%
	\bibitem [{\citenamefont {Gambetta}\ \emph {et~al.}(2006)\citenamefont
		{Gambetta}, \citenamefont {Blais}, \citenamefont {Schuster}, \citenamefont
		{Wallraff}, \citenamefont {Frunzio}, \citenamefont {Majer}, \citenamefont
		{Devoret}, \citenamefont {Girvin},\ and\ \citenamefont
		{Schoelkopf}}]{GambettaStarckShift}%
	\BibitemOpen
	\bibfield  {author} {\bibinfo {author} {\bibfnamefont {Jay}\ \bibnamefont
			{Gambetta}}, \bibinfo {author} {\bibfnamefont {Alexandre}\ \bibnamefont
			{Blais}}, \bibinfo {author} {\bibfnamefont {D.~I.}\ \bibnamefont {Schuster}},
		\bibinfo {author} {\bibfnamefont {A.}~\bibnamefont {Wallraff}}, \bibinfo
		{author} {\bibfnamefont {L.}~\bibnamefont {Frunzio}}, \bibinfo {author}
		{\bibfnamefont {J.}~\bibnamefont {Majer}}, \bibinfo {author} {\bibfnamefont
			{M.~H.}\ \bibnamefont {Devoret}}, \bibinfo {author} {\bibfnamefont {S.~M.}\
			\bibnamefont {Girvin}}, \ and\ \bibinfo {author} {\bibfnamefont {R.~J.}\
			\bibnamefont {Schoelkopf}},\ }\bibfield  {title} {\enquote {\bibinfo {title}
			{Qubit-photon interactions in a cavity: Measurement-induced dephasing and
				number splitting},}\ }\href {https://doi.org/10.1103/physreva.74.042318}
	{\bibfield  {journal} {\bibinfo  {journal} {Physical Review A}\ }\textbf
		{\bibinfo {volume} {74}} (\bibinfo {year} {2006})}\BibitemShut {NoStop}%
	\bibitem [{\citenamefont {Vijay}\ \emph {et~al.}(2011)\citenamefont {Vijay},
		\citenamefont {Slichter},\ and\ \citenamefont {Siddiqi}}]{PRL_Vijay_2011}%
	\BibitemOpen
	\bibfield  {author} {\bibinfo {author} {\bibfnamefont {R.}~\bibnamefont
			{Vijay}}, \bibinfo {author} {\bibfnamefont {D.~H.}\ \bibnamefont {Slichter}},
		\ and\ \bibinfo {author} {\bibfnamefont {I.}~\bibnamefont {Siddiqi}},\
	}\bibfield  {title} {\enquote {\bibinfo {title} {Observation of quantum jumps
				in a superconducting artificial atom},}\ }\href {\doibase
		10.1103/PhysRevLett.106.110502} {\bibfield  {journal} {\bibinfo  {journal}
			{Phys. Rev. Lett.}\ }\textbf {\bibinfo {volume} {106}},\ \bibinfo {pages}
		{110502} (\bibinfo {year} {2011})}\BibitemShut {NoStop}%
	\bibitem [{\citenamefont {Caves}(1982)}]{Caves1982}%
	\BibitemOpen
	\bibfield  {author} {\bibinfo {author} {\bibfnamefont {Carlton~M.}\
			\bibnamefont {Caves}},\ }\bibfield  {title} {\enquote {\bibinfo {title}
			{Quantum limits on noise in linear amplifiers},}\ }\href {\doibase
		10.1103/physrevd.26.1817} {\bibfield  {journal} {\bibinfo  {journal}
			{Physical Review D}\ }\textbf {\bibinfo {volume} {26}},\ \bibinfo {pages}
		{1817--1839} (\bibinfo {year} {1982})}\BibitemShut {NoStop}%
	\bibitem [{\citenamefont {Yurke}\ \emph {et~al.}(1996)\citenamefont {Yurke},
		\citenamefont {Roukes}, \citenamefont {Movshovich},\ and\ \citenamefont
		{Pargellis}}]{Yurke1996}%
	\BibitemOpen
	\bibfield  {author} {\bibinfo {author} {\bibfnamefont {B.}~\bibnamefont
			{Yurke}}, \bibinfo {author} {\bibfnamefont {M.~L.}\ \bibnamefont {Roukes}},
		\bibinfo {author} {\bibfnamefont {R.}~\bibnamefont {Movshovich}}, \ and\
		\bibinfo {author} {\bibfnamefont {A.~N.}\ \bibnamefont {Pargellis}},\
	}\bibfield  {title} {\enquote {\bibinfo {title} {A low-noise series-array
				josephson junction parametric amplifier},}\ }\href {\doibase
		10.1063/1.116845} {\bibfield  {journal} {\bibinfo  {journal} {Applied Physics
				Letters}\ }\textbf {\bibinfo {volume} {69}},\ \bibinfo {pages} {3078--3080}
		(\bibinfo {year} {1996})}\BibitemShut {NoStop}%
	\bibitem [{\citenamefont {Siddiqi}\ \emph {et~al.}(2004)\citenamefont
		{Siddiqi}, \citenamefont {Vijay}, \citenamefont {Pierre}, \citenamefont
		{Wilson}, \citenamefont {Metcalfe}, \citenamefont {Rigetti}, \citenamefont
		{Frunzio},\ and\ \citenamefont {Devoret}}]{Siddiqi2004}%
	\BibitemOpen
	\bibfield  {author} {\bibinfo {author} {\bibfnamefont {I.}~\bibnamefont
			{Siddiqi}}, \bibinfo {author} {\bibfnamefont {R.}~\bibnamefont {Vijay}},
		\bibinfo {author} {\bibfnamefont {F.}~\bibnamefont {Pierre}}, \bibinfo
		{author} {\bibfnamefont {C.~M.}\ \bibnamefont {Wilson}}, \bibinfo {author}
		{\bibfnamefont {M.}~\bibnamefont {Metcalfe}}, \bibinfo {author}
		{\bibfnamefont {C.}~\bibnamefont {Rigetti}}, \bibinfo {author} {\bibfnamefont
			{L.}~\bibnamefont {Frunzio}}, \ and\ \bibinfo {author} {\bibfnamefont
			{M.~H.}\ \bibnamefont {Devoret}},\ }\bibfield  {title} {\enquote {\bibinfo
			{title} {{RF}-driven josephson bifurcation amplifier for quantum
				measurement},}\ }\href {https://doi.org/10.1103/PhysRevLett.93.207002}
	{\bibfield  {journal} {\bibinfo  {journal} {Physical Review Letters}\
		}\textbf {\bibinfo {volume} {93}} (\bibinfo {year} {2004})}\BibitemShut
	{NoStop}%
	\bibitem [{\citenamefont {Mallet}\ \emph {et~al.}(2009)\citenamefont {Mallet},
		\citenamefont {Ong}, \citenamefont {Palacios-Laloy}, \citenamefont {Nguyen},
		\citenamefont {Bertet}, \citenamefont {Vion},\ and\ \citenamefont
		{Esteve}}]{mallet_single-shot_2009}%
	\BibitemOpen
	\bibfield  {author} {\bibinfo {author} {\bibfnamefont {François}\
			\bibnamefont {Mallet}}, \bibinfo {author} {\bibfnamefont {Florian~R.}\
			\bibnamefont {Ong}}, \bibinfo {author} {\bibfnamefont {Agustin}\ \bibnamefont
			{Palacios-Laloy}}, \bibinfo {author} {\bibfnamefont {François}\ \bibnamefont
			{Nguyen}}, \bibinfo {author} {\bibfnamefont {Patrice}\ \bibnamefont
			{Bertet}}, \bibinfo {author} {\bibfnamefont {Denis}\ \bibnamefont {Vion}}, \
		and\ \bibinfo {author} {\bibfnamefont {Daniel}\ \bibnamefont {Esteve}},\
	}\bibfield  {title} {\enquote {\bibinfo {title} {Single-shot qubit readout in
				circuit quantum electrodynamics},}\ }\href {\doibase 10.1038/nphys1400}
	{\bibfield  {journal} {\bibinfo  {journal} {Nature Phys.}\ }\textbf {\bibinfo
			{volume} {5}},\ \bibinfo {pages} {791} (\bibinfo {year} {2009})}\BibitemShut
	{NoStop}%
	\bibitem [{\citenamefont {Liu}\ \emph {et~al.}(2014)\citenamefont {Liu},
		\citenamefont {Srinivasan}, \citenamefont {Hover}, \citenamefont {Zhu},
		\citenamefont {McDermott},\ and\ \citenamefont {Houck}}]{liu_high_2014}%
	\BibitemOpen
	\bibfield  {author} {\bibinfo {author} {\bibfnamefont {Yanbing}\ \bibnamefont
			{Liu}}, \bibinfo {author} {\bibfnamefont {Srikanth~J.}\ \bibnamefont
			{Srinivasan}}, \bibinfo {author} {\bibfnamefont {D.}~\bibnamefont {Hover}},
		\bibinfo {author} {\bibfnamefont {Shaojiang}\ \bibnamefont {Zhu}}, \bibinfo
		{author} {\bibfnamefont {R.}~\bibnamefont {McDermott}}, \ and\ \bibinfo
		{author} {\bibfnamefont {A.~A.}\ \bibnamefont {Houck}},\ }\bibfield  {title}
	{\enquote {\bibinfo {title} {High fidelity readout of a transmon qubit using
				a superconducting low-inductance undulatory galvanometer microwave
				amplifier},}\ }\href {\doibase 10.1088/1367-2630/16/11/113008} {\bibfield
		{journal} {\bibinfo  {journal} {New J. Phys.}\ }\textbf {\bibinfo {volume}
			{16}},\ \bibinfo {pages} {113008} (\bibinfo {year} {2014})}\BibitemShut
	{NoStop}%
	\bibitem [{\citenamefont {Krantz}\ \emph {et~al.}(2016)\citenamefont {Krantz},
		\citenamefont {Bengtsson}, \citenamefont {Simoen}, \citenamefont
		{Gustavsson}, \citenamefont {Shumeiko}, \citenamefont {Oliver}, \citenamefont
		{Wilson}, \citenamefont {Delsing},\ and\ \citenamefont
		{Bylander}}]{krantz_single-shot_2016}%
	\BibitemOpen
	\bibfield  {author} {\bibinfo {author} {\bibfnamefont {Philip}\ \bibnamefont
			{Krantz}}, \bibinfo {author} {\bibfnamefont {Andreas}\ \bibnamefont
			{Bengtsson}}, \bibinfo {author} {\bibfnamefont {Michaël}\ \bibnamefont
			{Simoen}}, \bibinfo {author} {\bibfnamefont {Simon}\ \bibnamefont
			{Gustavsson}}, \bibinfo {author} {\bibfnamefont {Vitaly}\ \bibnamefont
			{Shumeiko}}, \bibinfo {author} {\bibfnamefont {W.~D.}\ \bibnamefont
			{Oliver}}, \bibinfo {author} {\bibfnamefont {C.~M.}\ \bibnamefont {Wilson}},
		\bibinfo {author} {\bibfnamefont {Per}\ \bibnamefont {Delsing}}, \ and\
		\bibinfo {author} {\bibfnamefont {Jonas}\ \bibnamefont {Bylander}},\
	}\bibfield  {title} {\enquote {\bibinfo {title} {Single-shot read-out of a
				superconducting qubit using a josephson parametric oscillator},}\ }\href
	{\doibase 10.1038/ncomms11417} {\bibfield  {journal} {\bibinfo  {journal}
			{Nature Comm.}\ }\textbf {\bibinfo {volume} {7}},\ \bibinfo {pages} {11417}
		(\bibinfo {year} {2016})}\BibitemShut {NoStop}%
	\bibitem [{\citenamefont {Bultink}\ \emph {et~al.}(2016)\citenamefont
		{Bultink}, \citenamefont {Rol}, \citenamefont {O’Brien}, \citenamefont
		{Fu}, \citenamefont {Dikken}, \citenamefont {Dickel}, \citenamefont
		{Vermeulen}, \citenamefont {de~Sterke}, \citenamefont {Bruno}, \citenamefont
		{Schouten},\ and\ \citenamefont {DiCarlo}}]{bultink_active_2016}%
	\BibitemOpen
	\bibfield  {author} {\bibinfo {author} {\bibfnamefont {C. C.}\ \bibnamefont
			{Bultink}}, \bibinfo {author} {\bibfnamefont {M. A.}\ \bibnamefont {Rol}},
		\bibinfo {author} {\bibfnamefont {T. E.}\ \bibnamefont {O’Brien}},
		\bibinfo {author} {\bibfnamefont {X.}~\bibnamefont {Fu}}, \bibinfo {author}
		{\bibfnamefont {B. C. S.}\ \bibnamefont {Dikken}}, \bibinfo {author}
		{\bibfnamefont {C.}~\bibnamefont {Dickel}}, \bibinfo {author} {\bibfnamefont
			{R. F. L.}\ \bibnamefont {Vermeulen}}, \bibinfo {author} {\bibfnamefont
			{J. C.}\ \bibnamefont {de~Sterke}}, \bibinfo {author} {\bibfnamefont
			{A.}~\bibnamefont {Bruno}}, \bibinfo {author} {\bibfnamefont {R. N.}\
			\bibnamefont {Schouten}}, \ and\ \bibinfo {author} {\bibfnamefont
			{L.}~\bibnamefont {DiCarlo}},\ }\bibfield  {title} {\enquote {\bibinfo
			{title} {Active {Resonator} {Reset} in the {Nonlinear} {Dispersive} {Regime}
				of {Circuit} {QED}},}\ }\href {\doibase 10.1103/PhysRevApplied.6.034008}
	{\bibfield  {journal} {\bibinfo  {journal} {Phys. Rev. Applied}\ }\textbf
		{\bibinfo {volume} {6}},\ \bibinfo {pages} {034008} (\bibinfo {year}
		{2016})}\BibitemShut {NoStop}%
	\bibitem [{\citenamefont {Johnson}\ \emph {et~al.}(2012)\citenamefont
		{Johnson}, \citenamefont {Macklin}, \citenamefont {Slichter}, \citenamefont
		{Vijay}, \citenamefont {Weingarten}, \citenamefont {Clarke},\ and\
		\citenamefont {Siddiqi}}]{Johnson2012}%
	\BibitemOpen
	\bibfield  {author} {\bibinfo {author} {\bibfnamefont {J.~E.}\ \bibnamefont
			{Johnson}}, \bibinfo {author} {\bibfnamefont {C.}~\bibnamefont {Macklin}},
		\bibinfo {author} {\bibfnamefont {D.~H.}\ \bibnamefont {Slichter}}, \bibinfo
		{author} {\bibfnamefont {R.}~\bibnamefont {Vijay}}, \bibinfo {author}
		{\bibfnamefont {E.~B.}\ \bibnamefont {Weingarten}}, \bibinfo {author}
		{\bibfnamefont {John}\ \bibnamefont {Clarke}}, \ and\ \bibinfo {author}
		{\bibfnamefont {I.}~\bibnamefont {Siddiqi}},\ }\bibfield  {title} {\enquote
		{\bibinfo {title} {Heralded state preparation in a superconducting qubit},}\
	}\href {https://doi.org/10.1103/PhysRevLett.109.050506} {\bibfield  {journal}
		{\bibinfo  {journal} {Physical Review Letters}\ }\textbf {\bibinfo {volume}
			{109}} (\bibinfo {year} {2012})}\BibitemShut {NoStop}%
	\bibitem [{\citenamefont {Chow}\ \emph {et~al.}(2010)\citenamefont {Chow},
		\citenamefont {DiCarlo}, \citenamefont {Gambetta}, \citenamefont {Motzoi},
		\citenamefont {Frunzio}, \citenamefont {Girvin},\ and\ \citenamefont
		{Schoelkopf}}]{Chow2010}%
	\BibitemOpen
	\bibfield  {author} {\bibinfo {author} {\bibfnamefont {J.~M.}\ \bibnamefont
			{Chow}}, \bibinfo {author} {\bibfnamefont {L.}~\bibnamefont {DiCarlo}},
		\bibinfo {author} {\bibfnamefont {J.~M.}\ \bibnamefont {Gambetta}}, \bibinfo
		{author} {\bibfnamefont {F.}~\bibnamefont {Motzoi}}, \bibinfo {author}
		{\bibfnamefont {L.}~\bibnamefont {Frunzio}}, \bibinfo {author} {\bibfnamefont
			{S.~M.}\ \bibnamefont {Girvin}}, \ and\ \bibinfo {author} {\bibfnamefont
			{R.~J.}\ \bibnamefont {Schoelkopf}},\ }\bibfield  {title} {\enquote {\bibinfo
			{title} {Optimized driving of superconducting artificial atoms for improved
				single-qubit gates},}\ }\href {https://doi.org/10.1103/physreva.82.040305}
	{\bibfield  {journal} {\bibinfo  {journal} {Physical Review A}\ }\textbf
		{\bibinfo {volume} {82}} (\bibinfo {year} {2010})}\BibitemShut {NoStop}%
	\bibitem [{\citenamefont {McClure}\ \emph {et~al.}(2016)\citenamefont
		{McClure}, \citenamefont {Paik}, \citenamefont {Bishop}, \citenamefont
		{Steffen}, \citenamefont {Chow},\ and\ \citenamefont
		{Gambetta}}]{McClure2016}%
	\BibitemOpen
	\bibfield  {author} {\bibinfo {author} {\bibfnamefont
			{D.{\hspace{0.167em}}T.}\ \bibnamefont {McClure}}, \bibinfo {author}
		{\bibfnamefont {Hanhee}\ \bibnamefont {Paik}}, \bibinfo {author}
		{\bibfnamefont {L.{\hspace{0.167em}}S.}\ \bibnamefont {Bishop}}, \bibinfo
		{author} {\bibfnamefont {M.}~\bibnamefont {Steffen}}, \bibinfo {author}
		{\bibfnamefont {Jerry~M.}\ \bibnamefont {Chow}}, \ and\ \bibinfo {author}
		{\bibfnamefont {Jay~M.}\ \bibnamefont {Gambetta}},\ }\bibfield  {title}
	{\enquote {\bibinfo {title} {Rapid driven reset of a qubit readout
				resonator},}\ }\href {https://doi.org/10.1103/physrevapplied.5.011001}
	{\bibfield  {journal} {\bibinfo  {journal} {Physical Review Applied}\
		}\textbf {\bibinfo {volume} {5}} (\bibinfo {year} {2016})}\BibitemShut
	{NoStop}%
	\bibitem [{\citenamefont {Gambetta}\ \emph {et~al.}(2008)\citenamefont
		{Gambetta}, \citenamefont {Blais}, \citenamefont {Boissonneault},
		\citenamefont {Houck}, \citenamefont {Schuster},\ and\ \citenamefont
		{Girvin}}]{Gambetta2008}%
	\BibitemOpen
	\bibfield  {author} {\bibinfo {author} {\bibfnamefont {Jay}\ \bibnamefont
			{Gambetta}}, \bibinfo {author} {\bibfnamefont {Alexandre}\ \bibnamefont
			{Blais}}, \bibinfo {author} {\bibfnamefont {M.}~\bibnamefont
			{Boissonneault}}, \bibinfo {author} {\bibfnamefont {A.~A.}\ \bibnamefont
			{Houck}}, \bibinfo {author} {\bibfnamefont {D.~I.}\ \bibnamefont {Schuster}},
		\ and\ \bibinfo {author} {\bibfnamefont {S.~M.}\ \bibnamefont {Girvin}},\
	}\bibfield  {title} {\enquote {\bibinfo {title} {Quantum trajectory approach
				to circuit {QED}: Quantum jumps and the zeno effect},}\ }\href
	{https://doi.org/10.1103/physreva.77.012112} {\bibfield  {journal} {\bibinfo
			{journal} {Physical Review A}\ }\textbf {\bibinfo {volume} {77}} (\bibinfo
		{year} {2008})}\BibitemShut {NoStop}%
	\bibitem [{\citenamefont {Kono}\ \emph {et~al.}(2010)\citenamefont {Kono},
		\citenamefont {Koshino}, \citenamefont {Tabuchi}, \citenamefont {Noguchi},\
		and\ \citenamefont {Nakamura}}]{nakamura10}%
	\BibitemOpen
	\bibfield  {author} {\bibinfo {author} {\bibfnamefont {S.}~\bibnamefont
			{Kono}}, \bibinfo {author} {\bibfnamefont {K.}~\bibnamefont {Koshino}},
		\bibinfo {author} {\bibfnamefont {Y.}~\bibnamefont {Tabuchi}}, \bibinfo
		{author} {\bibfnamefont {A.}~\bibnamefont {Noguchi}}, \ and\ \bibinfo
		{author} {\bibfnamefont {Y.}~\bibnamefont {Nakamura}},\ }\bibfield  {title}
	{\enquote {\bibinfo {title} {Quantum non-demolition detection of an itinerant
				microwave photon},}\ }\href {\doibase 10.1038/s41567-018-0066-3} {\bibfield
		{journal} {\bibinfo  {journal} {Nature Phys.}\ }\textbf {\bibinfo {volume}
			{6}},\ \bibinfo {pages} {663} (\bibinfo {year} {2010})}\BibitemShut {NoStop}%
	\bibitem [{\citenamefont {Besse}\ \emph {et~al.}(2018)\citenamefont {Besse},
		\citenamefont {Gasparinetti}, \citenamefont {Collodo}, \citenamefont
		{Walter}, \citenamefont {Kurpiers}, \citenamefont {Pechal}, \citenamefont
		{Eichler},\ and\ \citenamefont {Wallraff}}]{besse18}%
	\BibitemOpen
	\bibfield  {author} {\bibinfo {author} {\bibfnamefont {J.-C.}\ \bibnamefont
			{Besse}}, \bibinfo {author} {\bibfnamefont {S.}~\bibnamefont {Gasparinetti}},
		\bibinfo {author} {\bibfnamefont {M.~C.}\ \bibnamefont {Collodo}}, \bibinfo
		{author} {\bibfnamefont {T.}~\bibnamefont {Walter}}, \bibinfo {author}
		{\bibfnamefont {P.}~\bibnamefont {Kurpiers}}, \bibinfo {author}
		{\bibfnamefont {M.}~\bibnamefont {Pechal}}, \bibinfo {author} {\bibfnamefont
			{C.}~\bibnamefont {Eichler}}, \ and\ \bibinfo {author} {\bibfnamefont
			{A.}~\bibnamefont {Wallraff}},\ }\bibfield  {title} {\enquote {\bibinfo
			{title} {Single-shot quantum nondemolition detection of individual itinerant
				microwave photons},}\ }\href {\doibase 10.1103/PhysRevX.8.021003} {\bibfield
		{journal} {\bibinfo  {journal} {Phys. Rev. X}\ }\textbf {\bibinfo {volume}
			{8}},\ \bibinfo {pages} {021003} (\bibinfo {year} {2018})}\BibitemShut
	{NoStop}%
	\bibitem [{\citenamefont {Ramos}\ and\ \citenamefont
		{García-Ripoll}(2017)}]{ramos17}%
	\BibitemOpen
	\bibfield  {author} {\bibinfo {author} {\bibfnamefont {T.}~\bibnamefont
			{Ramos}}\ and\ \bibinfo {author} {\bibfnamefont {J.J.}\ \bibnamefont
			{García-Ripoll}},\ }\bibfield  {title} {\enquote {\bibinfo {title}
			{Multiphoton scattering tomography with coherent states},}\ }\href {\doibase
		10.1103/PhysRevLett.119.153601} {\bibfield  {journal} {\bibinfo  {journal}
			{Phys. Rev. Lett.}\ }\textbf {\bibinfo {volume} {119}},\ \bibinfo {pages}
		{153601} (\bibinfo {year} {2017})}\BibitemShut {NoStop}%
	\bibitem [{\citenamefont {Lescanne}\ \emph
		{et~al.}(2019{\natexlab{b}})\citenamefont {Lescanne}, \citenamefont
		{Deleglise}, \citenamefont {Albertinale}, \citenamefont {Reglade},
		\citenamefont {Capelle}, \citenamefont {Ivanov}, \citenamefont {Jacqmin},
		\citenamefont {Leghtas},\ and\ \citenamefont {Flurin}}]{lescanne19}%
	\BibitemOpen
	\bibfield  {author} {\bibinfo {author} {\bibfnamefont {R.}~\bibnamefont
			{Lescanne}}, \bibinfo {author} {\bibfnamefont {S.}~\bibnamefont {Deleglise}},
		\bibinfo {author} {\bibfnamefont {E.}~\bibnamefont {Albertinale}}, \bibinfo
		{author} {\bibfnamefont {U.}~\bibnamefont {Reglade}}, \bibinfo {author}
		{\bibfnamefont {T.}~\bibnamefont {Capelle}}, \bibinfo {author} {\bibfnamefont
			{E.}~\bibnamefont {Ivanov}}, \bibinfo {author} {\bibfnamefont
			{T.}~\bibnamefont {Jacqmin}}, \bibinfo {author} {\bibfnamefont
			{Z.}~\bibnamefont {Leghtas}}, \ and\ \bibinfo {author} {\bibfnamefont
			{E.}~\bibnamefont {Flurin}},\ }\bibfield  {title} {\enquote {\bibinfo {title}
			{Detecting itinerant microwave photons with engineered non-linear
				dissipation},}\ }\href {https://arxiv.org/abs/1902.05102} {\bibfield
		{journal} {\bibinfo  {journal} {arXiv:1902.05102}\ } (\bibinfo {year}
		{2019}{\natexlab{b}})}\BibitemShut {NoStop}%
	\bibitem [{\citenamefont {Besse}\ \emph {et~al.}(2019)\citenamefont {Besse},
		\citenamefont {Gasparinetti}, \citenamefont {Collodo}, \citenamefont
		{Walter}, \citenamefont {Remm}, \citenamefont {Krause}, \citenamefont
		{Eichler},\ and\ \citenamefont {Wallraff}}]{besse19}%
	\BibitemOpen
	\bibfield  {author} {\bibinfo {author} {\bibfnamefont {J.-C.}\ \bibnamefont
			{Besse}}, \bibinfo {author} {\bibfnamefont {S.}~\bibnamefont {Gasparinetti}},
		\bibinfo {author} {\bibfnamefont {M.~C.}\ \bibnamefont {Collodo}}, \bibinfo
		{author} {\bibfnamefont {T.}~\bibnamefont {Walter}}, \bibinfo {author}
		{\bibfnamefont {A.}~\bibnamefont {Remm}}, \bibinfo {author} {\bibfnamefont
			{J.}~\bibnamefont {Krause}}, \bibinfo {author} {\bibfnamefont
			{C.}~\bibnamefont {Eichler}}, \ and\ \bibinfo {author} {\bibfnamefont
			{A.}~\bibnamefont {Wallraff}},\ }\bibfield  {title} {\enquote {\bibinfo
			{title} {Parity detection of propagating microwave fields},}\ }\href
	{https://arxiv.org/abs/1912.09896} {\bibfield  {journal} {\bibinfo  {journal}
			{arXiv:1912.0989}\ } (\bibinfo {year} {2019})}\BibitemShut {NoStop}%
	\bibitem [{\citenamefont {Ramos}\ and\ \citenamefont
		{Garcia-Ripoll}(2018)}]{ramos_correlated_2018}%
	\BibitemOpen
	\bibfield  {author} {\bibinfo {author} {\bibfnamefont {T.}~\bibnamefont
			{Ramos}}\ and\ \bibinfo {author} {\bibfnamefont {J.~J.}\ \bibnamefont
			{Garcia-Ripoll}},\ }\bibfield  {title} {\enquote {\bibinfo {title}
			{Correlated dephasing noise in single-photon scattering},}\ }\href {\doibase
		10.1088/1367-2630/aae73b} {\bibfield  {journal} {\bibinfo  {journal} {New J.
				Phys.}\ }\textbf {\bibinfo {volume} {20}},\ \bibinfo {pages} {105007}
		(\bibinfo {year} {2018})}\BibitemShut {NoStop}%
	\bibitem [{\citenamefont {Gardiner}\ and\ \citenamefont
		{Zoller}(2004)}]{QuantumNoise}%
	\BibitemOpen
	\bibfield  {author} {\bibinfo {author} {\bibfnamefont {C.W.}\ \bibnamefont
			{Gardiner}}\ and\ \bibinfo {author} {\bibfnamefont {P.}~\bibnamefont
			{Zoller}},\ }\href@noop {} {\emph {\bibinfo {title} {Quantum Noise}}}\
	(\bibinfo {year} {2004})\BibitemShut {NoStop}%
	\bibitem [{\citenamefont {Schreier}\ \emph {et~al.}(2008)\citenamefont
		{Schreier}, \citenamefont {Houck}, \citenamefont {Koch}, \citenamefont
		{Schuster}, \citenamefont {Johnson}, \citenamefont {Chow}, \citenamefont
		{Gambetta}, \citenamefont {Majer}, \citenamefont {Frunzio}, \citenamefont
		{Devoret}, \citenamefont {Girvin},\ and\ \citenamefont
		{Schoelkopf}}]{Schreier2008}%
	\BibitemOpen
	\bibfield  {author} {\bibinfo {author} {\bibfnamefont {J.~A.}\ \bibnamefont
			{Schreier}}, \bibinfo {author} {\bibfnamefont {A.~A.}\ \bibnamefont {Houck}},
		\bibinfo {author} {\bibfnamefont {Jens}\ \bibnamefont {Koch}}, \bibinfo
		{author} {\bibfnamefont {D.~I.}\ \bibnamefont {Schuster}}, \bibinfo {author}
		{\bibfnamefont {B.~R.}\ \bibnamefont {Johnson}}, \bibinfo {author}
		{\bibfnamefont {J.~M.}\ \bibnamefont {Chow}}, \bibinfo {author}
		{\bibfnamefont {J.~M.}\ \bibnamefont {Gambetta}}, \bibinfo {author}
		{\bibfnamefont {J.}~\bibnamefont {Majer}}, \bibinfo {author} {\bibfnamefont
			{L.}~\bibnamefont {Frunzio}}, \bibinfo {author} {\bibfnamefont {M.~H.}\
			\bibnamefont {Devoret}}, \bibinfo {author} {\bibfnamefont {S.~M.}\
			\bibnamefont {Girvin}}, \ and\ \bibinfo {author} {\bibfnamefont {R.~J.}\
			\bibnamefont {Schoelkopf}},\ }\bibfield  {title} {\enquote {\bibinfo {title}
			{{Suppressing charge noise decoherence in superconducting charge qubits}},}\
	}\href {\doibase 10.1103/PhysRevB.77.180502} {\bibfield  {journal} {\bibinfo
			{journal} {Phys. Rev. B - Condens. Matter Mater. Phys.}\ }\textbf {\bibinfo
			{volume} {77}} (\bibinfo {year} {2008}),\ 10.1103/PhysRevB.77.180502},\
	\Eprint {http://arxiv.org/abs/0712.3581} {arXiv:0712.3581} \BibitemShut
	{NoStop}%
	\bibitem [{\citenamefont {Nigg}\ \emph {et~al.}(2012)\citenamefont {Nigg},
		\citenamefont {Paik}, \citenamefont {Vlastakis}, \citenamefont {Kirchmair},
		\citenamefont {Shankar}, \citenamefont {Frunzio}, \citenamefont {Devoret},
		\citenamefont {Schoelkopf},\ and\ \citenamefont {Girvin}}]{PRL_Nigg_2012}%
	\BibitemOpen
	\bibfield  {author} {\bibinfo {author} {\bibfnamefont {Simon~E.}\
			\bibnamefont {Nigg}}, \bibinfo {author} {\bibfnamefont {Hanhee}\ \bibnamefont
			{Paik}}, \bibinfo {author} {\bibfnamefont {Brian}\ \bibnamefont {Vlastakis}},
		\bibinfo {author} {\bibfnamefont {Gerhard}\ \bibnamefont {Kirchmair}},
		\bibinfo {author} {\bibfnamefont {S.}~\bibnamefont {Shankar}}, \bibinfo
		{author} {\bibfnamefont {Luigi}\ \bibnamefont {Frunzio}}, \bibinfo {author}
		{\bibfnamefont {M.~H.}\ \bibnamefont {Devoret}}, \bibinfo {author}
		{\bibfnamefont {R.~J.}\ \bibnamefont {Schoelkopf}}, \ and\ \bibinfo {author}
		{\bibfnamefont {S.~M.}\ \bibnamefont {Girvin}},\ }\bibfield  {title}
	{\enquote {\bibinfo {title} {Black-box superconducting circuit
				quantization},}\ }\href {\doibase 10.1103/PhysRevLett.108.240502} {\bibfield
		{journal} {\bibinfo  {journal} {Phys. Rev. Lett.}\ }\textbf {\bibinfo
			{volume} {108}},\ \bibinfo {pages} {240502} (\bibinfo {year}
		{2012})}\BibitemShut {NoStop}%
\end{thebibliography}
%

\end{document}